
\input harvmac
\input amssym.def
\input amssym
\baselineskip 14pt
\magnification\magstep1
\parskip 6pt
\newdimen\itemindent \itemindent=32pt
\def\textindent#1{\parindent=\itemindent\let\par=\resetpar%
\indent\llap{#1\enspace}\ignorespaces}

\let\oldpar=\par
\def\resetpar{\oldpar\parindent=20pt\let\par=\oldpar}

\font\ninerm=cmr9 \font\ninesy=cmsy9
\font\eightrm=cmr8 \font\sixrm=cmr6
\font\eighti=cmmi8 \font\sixi=cmmi6
\font\eightsy=cmsy8 \font\sixsy=cmsy6
\font\eightbf=cmbx8 \font\sixbf=cmbx6
\font\eightit=cmti8
\def\eightpoint{\def\rm{\fam0\eightrm}
  \textfont0=\eightrm \scriptfont0=\sixrm \scriptscriptfont0=\fiverm
  \textfont1=\eighti  \scriptfont1=\sixi  \scriptscriptfont1=\fivei
  \textfont2=\eightsy \scriptfont2=\sixsy \scriptscriptfont2=\fivesy
  \textfont3=\tenex   \scriptfont3=\tenex \scriptscriptfont3=\tenex
  \textfont\itfam=\eightit  \def\it{\fam\itfam\eightit}%
  \textfont\bffam=\eightbf  \scriptfont\bffam=\sixbf
  \scriptscriptfont\bffam=\fivebf  \def\bf{\fam\bffam\eightbf}%
  \normalbaselineskip=9pt
  \setbox\strutbox=\hbox{\vrule height7pt depth2pt width0pt}%
  \let\big=\eightbig  \normalbaselines\rm}
\catcode`@=11 %
\def\eightbig#1{{\hbox{$\textfont0=\ninerm\textfont2=\ninesy
  \left#1\vbox to6.5pt{}\right.\n@@space$}}}
\def\vfootnote#1{\insert\footins\bgroup\eightpoint
  \interlinepenalty=\interfootnotelinepenalty
  \splittopskip=\ht\strutbox %
  \splitmaxdepth=\dp\strutbox %
  \leftskip=0pt \rightskip=0pt \spaceskip=0pt \xspaceskip=0pt
  \textindent{#1}\footstrut\futurelet\next\fo@t}
\catcode`@=12 %
\def \bR{{\Bbb R}}
\def \CC{{\Bbb C}}
\def \GG{{\Bbb G}}
\def \TT{{\Bbb T}}
\def \WW{{\Bbb W}}
\def \d{{\rm d}}
\def \de{\delta}
\def \si{\sigma}
\def \Ga{\Gamma}

\def \nab{\nabla}
\def \pr{\partial}
\def \d{{\rm d}}
\def \tr{{\rm tr }}
\def \ta{{\tilde a}}
\def \hI{{\hat{I}}}
\def \hx{{\hat x}}
\def \hy{{\hat y}}
\def \ha{{\hat a}}
\def \bs{{\bar s}}

\def \hG{{\hat G}}

\def \hU{{\hat U}}
\def \bpsi{{\bar \psi}}
\def \rO{{\rm O}}
\def \l{\langle}
\def \r{\rangle}
\def \ep{\epsilon}

\def \cosec{\mathop{\rm cosec}\nolimits}
\def \cosech{\mathop{\rm cosech}\nolimits}
\def \olr#1 {\vbox{\ialign{##\crcr$\leftrightarrow$\crcr\noalign{
\nointerlineskip}$\hfil\displaystyle{#1}\hfil$\crcr}}}
\def \half{{\textstyle {1 \over 2}}}
\def \thir{{\textstyle {1 \over 3}}}
\def \quar{{\textstyle {1 \over 4}}}
\def \ts{\textstyle}
\def \A{{\cal A}}
\def \B{{\cal B}}
\def \C{{\cal C}}
\def \D{{\cal D}}
\def \E{{\cal E}}
\def \F{{\cal F}}
\def \G{{\cal G}}
\def \H{{\cal H}}
\def \I{{\cal I}}
\def \L{{\cal L}}
\def \N{{\cal N}}
\def \O{{\cal O}}
\def \S{{\cal S}}
\def \P{{\cal P}}
\def \Y{{\cal Y}}
\def \hH{{\hat H}}
\def \hK{{\hat K}}
\def \hC{{\hat C}}
\def \hL{{\hat L}}
\def \hhN{{\hat \N}}
\def \hf{{\hat f}}
\def \hhC{{\hat \C}}
\def \hxn{{\xi}}
\def \Gg{K}
\def \down#1{\lower5pt\hbox{$ #1$}}
\def \hde{{\hat \delta}}
\font \bigbf=cmbx10 scaled \magstep1
\def\toinf#1{\mathrel{\mathop{\longrightarrow}\limits_{\scriptscriptstyle {#1}}}}
\def \ite#1{\noindent\llap{#1\enspace}\ignorespaces}

\lref\Cardy{J.L. Cardy, {Phys. Lett.} {B215} (1988) 749.}
\lref\Cbook{J.L. Cardy, Phys. Rev. Lett. 60 (1988) 2709\semi
John Cardy, {\it Scaling and Renormalization in Statistical Physics},
(Cambridge University Press, 1996).}
\lref\Zam{A.B. Zamolodchikov, {JETP Lett.} {43} (1986) 730;
{Sov. J. Nucl. Phys.} {46} (1988) 1090.}
\lref\JO{H. Osborn, Phys. Lett. B222 (1989) 97\semi
I. Jack and H. Osborn, {Nucl. Phys.} {B343} (1990) 647.}
\lref\Weyl{H. Osborn, Nucl. Phys. B363 (1991) 486.}
\lref\Cap{A. Cappelli, D. Friedan and J.I. Latorre, {Nucl. Phys.}
{B352} (1991) 616.}
\lref\Lat{A. Cappelli, J.I. Latorre and X. Vilas\'{\i}s-Cardona, {Nucl. Phys.}
{B376} (1992) 510.}
\lref\Shore{G.M. Shore, {Phys. Lett.} {B253} (1991) 380;
{B256} (1991) 407.}
\lref\ITD{I.T. Drummond and G.M. Shore, Ann. Phys. 117 (1979) 89.}
\lref\Neto{A.H. Castro Neto and E. Fradkin, Nucl. Phys. B400 (1993) 525.}
\lref\Forte{S. Forte and J.I. Latorre, {\it A Proof of the Irreversibility
of Renormalization Group Flows in Four Dimensions}, Nucl. Phys. B535 [FS]
(1998) 709, hep-th/9805015.}
\lref\hughtwo{J. Erdmenger and H. Osborn, Nucl. Phys. {B483} (1997)
431, hep-th/9605009.}
\lref\hughone{H. Osborn and A. Petkou,
    Ann. Phys. {231} (1994) 311, hep-th/9307010.}
\lref\Ent{D. Boyanovsky and R. Holman, Phys. Rev. D40 (1989) 1964, Nucl. Phys.
B332 (1990) 641\semi
N.E. Mavromatos, J.L. Miramontes and J.M. S\'anchez de Santos, Phys. Rev.
D40 (1989) 535\semi
N.E. Mavromatos and J.L. Miramontes, Phys. Lett. B226 (1989) 291\semi
J. Gaite, Phys. Rev. Lett. 81 (1998) 3587, {\it Entropic C theorems
in free and interacting two-dimensional field theories}, hep-th/9810107.}
\lref\Anone{D. Anselmi, D.Z. Freedman, M.T. Grisaru and A.A. Johansen, 
Nucl. Phys. B526 (1998) 543, hep-th/9708042.}
\lref\Anselmi{D. Anselmi, {\it Anomalies, Unitarity and Quantum Irreversibility},
Ann. Phys. 276 (1999) 361, hep-th/9903059.}
\lref\Antwo{D. Anselmi, J. Erlich,  D.Z. Freedman and A.A. Johansen, 
Phys. Rev. D57 (1998) 7570, hep-th/9711035.}
\lref\LO{J.I Latorre and H. Osborn, Nucl. Phys. B511 (1998) 737, 
hep-th/9703196.}
\lref\Hath{S.J. Hathrell, {Ann. Phys.} {139} (1982) 136, {142} (1982) 34\semi
M.D. Freeman, {Ann. Phys.} {153} (1984) 339.}
\lref\grad{ D.J. Wallace and R.K.P. Zia, {Phys. Lett.} {48A} (1974) 327; 
{Ann. Phys.} {92} (1975) 142\semi
B.P. Dolan, Mod. Phys. Lett. 8 (1993) 3103.}
\lref\Banks{T. Banks and A. Zaks, Nucl. Phys. B196 (1982) 189.}
\lref\Aff{H.W.J. Bl\"ote, J.L. Cardy and M.P. Nightingale,
Phys. Rev. Lett. 56 (1986) 273\semi
I. Affleck, Phys. Rev. Lett. 56 (1986) 746.}
\lref\Appel{T. Appelquist, A.G. Cohen and M. Schmaltz, {\it A new constraint on
strongly coupled field theories}, Phys. Rev. D60 (1999) 045003, hep-th/9901109.}
\lref\holo{E. \'Alvarez and C. G\'omez, {\it Geometric Holography, the 
Renormalization Group and the c-Theorem}, Nucl. Phys. B441 (1999) 441,
hep-th/9807226.}
\lref\Holo{D.Z. Freedman, S.S. Gubser, K. Pilch and N. Warner, {\it
Renormalization Group Flows from Holography - Supersymmetry and a c-Theorem},
hep-th/9904017.}
\lref\FO{D.Z. Freedman and H. Osborn, {\it Constructing a c-function for
SUSY Gauge Theories}, Phys. Lett. B432 (1998) 353, hep-th/9804101.}
\lref\allen{B. Allen and T. Jacobson, Comm. Math. Phys. 103 (1986) 669.}
\lref\Allen{B. Allen and C.A. L\"utken,  Comm. Math. Phys. 106 (1986) 201.}
\lref\dewitt{B.S. DeWitt, {\it Dynamical Theory of Groups and Fields},
(Gordon and Breach, 1965).}
\lref\Free{E. D'Hoker and D.Z. Freedman, {\it Gauge boson exchange in 
AdS}${}_{d+1}$, Nucl. Phys. B544 (1999) 612, hep-th/9809179\semi
E. D'Hoker, D.Z. Freedman, S.D. Mathur, A. Matusis and L. Rastelli,
{\it \it Graviton and gauge boson propagators in AdS}${}_{d+1}$, hep-th/9902042.}
\lref\Full{S.A. Fulling, {\it Aspects of Quantum Field Theory in Curved
Space-Time}, (Cambridge University Press, 1989).}
\lref\Burges{C.J.C. Burges, D.Z. Freedman, S. Davis and G.W. Gibbons. Ann. Phys.
167 (1986) 285.}
\lref\Card{J.L. Cardy, Nucl. Phys. B290 (1987) 355.}
\lref\Callan{C.G. Callan and F. Wilczek, Nucl. Phys. B340 (1990) 366.}
\lref\Dus{D.W. D\"usedau and D.Z. Freedman, Phys. Rev. D33 (1986) 389.}
\lref\Adler{S.L. Adler,  Phys. Rev. D6 (1972) 3445; D8 (1973) 2400.}
\lref\DShore{I.T. Drummond and G.M. Shore, Phys. Rev. D19 (1979) 1134\semi
G.M. Shore, Phys. Rev. D21 (1980) 2226.}
\lref\conform{N.K. Nielsen, Nucl. Phys. B65 (1973) 413; B97 (1975) 527\semi
S. Sarker, Nucl. Phys. B83 (1974) 108\semi
R.J. Crewther, Phys. Lett. 397 (1997) 137.}
\lref\posem{J.I. Latorre and H. Osborn, Nucl. Phys. B511 (1998) 737.}
\lref\pope{H. L\"u, C.N. Pope and J. Rahmfeld, {\it A construction of
Killing spinors on $S^n$}, hep-th/9805151.}
\lref\birrell{N.D. Birrell and P.C.W. Davies, {\it Quantum Fields in
Curved Space}, (Cambridge University Press, 1982).}
\lref\Sei{S. Lee, S. Minwalla, M. Rangamani and N. Seiberg, Adv. Theor. Math. Phys. 
2 (1998) 697, hep-th/9806074.}
\lref\Fron{C. Fronsdal, Phys. Rev. D12 (1975) 3819.}
\lref\Camp{R. Camporesi and A. Higuchi, Phys. Rev. D45 (1992) 3591.}
\lref\Mott{E. Mottola, Phys. Rev. D33 (1986) 2136.}
\lref\MottdS{I. Antoniadis and E. Mottola, J. Math. Phys. 32 (1991) 1037.} 

{\nopagenumbers
\rightline{CERN--TH/99-274}
\rightline{DAMTP/99-114}
\rightline{SWAT/99-240}
\rightline{hep-th/9909043}
\vskip 1.5truecm
\centerline {\bigbf Correlation Functions of the Energy Momentum Tensor}
\vskip  6pt
\centerline {\bigbf on Spaces of Constant Curvature}
\vskip 2.0 true cm
\centerline {H. Osborn* ${}^\dagger$ and G.M. Shore** ${}^{\dagger\dagger}$ }

\vskip 12pt
\centerline {*\ Department of Applied Mathematics and Theoretical Physics,}
\centerline {Silver Street, Cambridge, CB3 9EW, England}
\vskip 8pt
\centerline {**\ Theory Division, CERN,}
\centerline {CH-1211,  Gen\` eve 23, Switzerland}

\vskip 0.5 true cm

{\eightpoint
\parindent 1.5cm{

{\narrower\smallskip\parindent 0pt
An analysis of one and two point functions of the energy momentum tensor on
homogeneous spaces of constant curvature is undertaken.  The possibility of 
proving a $c$-theorem in this framework is discussed, in particular in relation to
the coefficients $c,a$, which appear in the energy momentum tensor trace on general 
curved backgrounds in four dimensions. Ward identities relating the correlation
functions are derived and explicit expressions are obtained for free scalar, spinor 
field theories in general dimensions and also free vector fields in dimension four. 
A natural geometric formalism which is independent of any choice of coordinates
is used and the role of conformal symmetries on such constant curvature spaces is
analysed. The results are shown to be constrained by the operator product expansion.
For negative curvature the spectral representation, involving unitary positive
energy representations of $O(d-1,2)$, for two point functions of vector currents
is derived in detail and extended to the energy momentum tensor by analogy. 
It is demonstrated that, at non coincident points, the two point functions are
not related to $a$ in any direct fashion and there is no straightforward 
demonstration obtainable in this framework of irreversibility under renormalisation
group flow of any function of the couplings for four dimensional field theories 
which reduces to $a$ at fixed points.

\narrower}}

\vskip0.5cm
\line{${}^\dagger$ 
address for correspondence: Trinity College, Cambridge, CB2 1TQ, England\hfill}
\line{${}^{\dagger\dagger}$ 
permanent address: Department of Physics, University of Wales Swansea, 
Singleton Park, \hfill}
\line{\hskip3.0cm Swansea, SA2 8PP, Wales\hfill}
\vskip0.2cm
\line{\hskip0.2cm emails:
{\tt \tt ho@damtp.cam.ac.uk} and {\tt G.M.Shore@swansea.ac.uk}\hfill}
}
\vskip0.5cm
\line{CERN--TH/99-274 \hfill}
\line{September, 1999; revised November, 1999 \hfill}

\vfill
\eject}
\pageno=1

\newsec{Introduction}

The energy momentum tensor is a universal probe in any relativistic quantum
field theory. Parameters which are defined through correlation functions
of the energy momentum tensor, or its expectation value when the space on
which the theory is defined has non trivial topology or non zero curvature,
may serve to specify
the theory independent of any particular formulation in terms of elementary
fields. Such parameters should be well defined at any renormalisation group
fixed point where the theory becomes conformal, which should also include
the case of free theories. The cardinal example of course is the Virasoro
central charge $c$ for two dimensional conformal field theories which
may be defined through the trace of the energy momentum tensor on a curved
background, the coefficient of the energy momentum tensor two point function
on flat space $\bR^2$ or a universal term in the dependence of the Casimir energy
on the circumference when the underlying space is compactified on a cylinder 
$S^1 \times \bR$ \Aff. Furthermore away from critical points $c$ 
may be generalised to a function of the couplings $g^i$, $\CC(g)$, 
which monotonically decreases under RG flow as the basic scale of the theory 
is evolved to large distances and the couplings are attracted to any potential 
infra red fixed point. This is the content of the celebrated Zamolodchikov 
$c$-theorem \Zam, where $\CC(g)$ was constructed so as to satisfy
\eqn\flowC{
\beta^i(g) {\pr \over \pr g^i} \CC(g) = \GG_{ij} (g)
\beta^i(g)\beta^j(g) \, ,
}
with $\GG_{ij} (g)$, for unitary theories, positive definite.
Since, in a two
dimensional  conformal field theory $c$ may be interpreted as a measure
of the degrees of freedom, the $c$-theorem incorporates the physical intuition that
RG flow is irreversible as a consequence of loss of information concerning
details at short distances in any infra red limit.\foot{A selection of
papers discussing the $c$-theorem using statistical mechanical methods
is given in \Ent.}

Many efforts have been made to generalise such ideas beyond two dimensions
to realistic four dimensional field theories
\refs{\Cardy,\JO,\Cap,\Lat,\Shore,\Neto,\Forte,\Appel,\Anselmi}, even making 
connections to modern ideas of holography \refs{\holo,\Holo}. 
In particular Cardy \Cardy\ 
discussed a possible generalisation for $c$ to four dimensions in terms of the
energy momentum tensor trace on a curved background. At a conformal fixed point there
are two parameters associated with two independent scalars formed from the curvature
which may apear in the energy momentum tensor trace, and are denoted by $c,a$.
Cardy's conjecture
for a four dimensional generalisation of $c$ involved
$a$, the coefficient of the four dimensional Euler density. An analysis
of this proposal for general four
dimensional renormalisable quantum field theories demonstrated \JO\
irreversibility of the RG flow of a quantity
$\ta$, which is equal to $a$ for vanishing $\beta$-functions, in some 
neighbourhood of weak coupling. Recently non perturbative formulae for the 
flow of both $c$ and $a$ between UV
asymptotically free fixed points and non trivial IR fixed points in
$\N=1$ supersymmetric gauge theories were proposed, on the basis of anomaly 
calculations, which demonstrated that the RG flow of $a$ is monotonic,
so long as the anomalous dimensions of the basic chiral fields
are not too large \refs{\Anone,\Antwo}.  The supersymmetric results were 
subsequently shown to be in accord with the previous
perturbative discussions for general theories in \FO.

It therefore appears natural to try to analyse further how far two dimensional
results for flat space may be generalised to curved backgrounds. 
In this investigation we
restrict attention to homogeneous spaces of constant curvature, for theories continued to
a space with positive definite metric, either the $d$-dimensional sphere $S^d$ or 
the negative curvature hyperboloid $H^d$. For such spaces
\eqn\riem{
R_{\mu\nu\si\rho} = \pm \rho^2 \big ( g_{\mu\si} g_{\nu\rho} -
g_{\mu\rho} g_{\nu\si} \big ) \ \ \Rightarrow \ \ R = \pm d(d-1) \rho^2 \, ,
}
with $1/\rho$ a length scale. The dependence on the metric is then reduced
to just the single parameter $\rho$ and we may write
\eqn\ag{
\rho {\d \over \d \rho} g_{\mu\nu} = - 2g_{\mu\nu} + \L_{u} g_{\mu\nu} \, , 
\quad 
\L_{u} g_{\mu\nu} = \nab_\mu u_{\nu} + \nab_\nu u_{\mu} \, ,
}
where $u^\mu$ is a vector field which depends on the particular choice
of coordinates.   The simplicity of considering homogeneous spaces of 
constant curvature is that the isometry group is as large as on flat space
and therefore  a group theoretic analysis is possible
which, once the appropriate basis functions are introduced, is not essentially 
more complicated than for flat space
\refs{\Adler,\Dus}. The virtues of considering field theories on negative curvature 
spaces were advocated  in \Callan.

For the case of negative curvature an analysis of two point 
correlation functions, where
the positivity properties arising from unitarity are most evident, for the energy
momentum tensor  was first given in \Cap. 
More recently Forte and Latorre \Forte\ have endeavoured to use these results to prove
a four dimensional version of the $c$-theorem by considering as a candidate for $\CC(g)$,
which naturally interpolates between $a$ at fixed points, a function $C(g)$
determined by the one point function of the energy momentum tensor 
on a space of constant negative curvature when $\l T_{\mu\nu}\r $ is proportional
to $g_{\mu\nu}$. It is important to recognise that although we may have $\Delta a <0$
under RG flow between fixed points it is in general necessary to add extra
terms of ${\rm O}(\beta)$ to any interpolating function $C(g)$ to obtain a $\CC(g)$
satisfying \flowC. We show later that there is a freedom of definition for $\CC(g)$
of ${\rm O}(\beta^2)$, and correspondingly for $\GG_{ij} (g)$, which preserves
\flowC\ and which may be
necessary to ensure that $\GG_{ij} (g)$ is positive.

The work in \Forte\ formed part of the stimulus for this investigation although
we attempt to provide a complete set of results for the energy momentum tensor
two point function on constant curvature spaces. The salient 
results obtained here are,

\ite{$\bullet$} An analysis of the Ward identities following from the conservation
equation for the energy momentum tensor and the presence of anomalous terms in its
trace, together with consistency conditions following from renormalisation group
equations, applied to the associated one and two point functions on spaces of constant
curvature. These relations are further restricted for the particular cases of two and 
four dimensions.

\ite{$\bullet$} The derivation of the conditions flowing from the conservation
equations on a general expression of the two point function together with the
calculation of its explicit form for massless scalar, spinor and vector theories.

\ite{$\bullet$} A derivation in detail of the unitary positive energy representations
of $O(d-1,2)$ for a spin one lowest energy state. These are applied to construct
the spectral representation for a vector two point function on a negative curvature 
space. This is extended to the spin two case appropriate to the energy momentum tensor
two point function.

\ite{$\bullet$} An analysis of the implications of these results for the derivation
of a $c$-theorem along similar lines to that in \Forte\ (and also the
extension of the Zamolodchikov proof to curved space). The justification of
positivity conditions through the spectral representation for the two point function
is carefully considered. We are not able to obtain an equation of the form \flowC\ for
$C(g)$ although a related equation of the form $\beta^i \pr_i C = \G - d C$ with $\G>0$
at least in two dimensions is found.

\ite{$\bullet$} A discussion of the possible conditions which would imply
irreversibility of RG flow and the constraints  that a general
$c$-function should satisfy.

\ite{$\bullet$} An application to free massive scalar fields when all contributions
to the various identities may be explicitly calculated.

An analysis of consistency conditions related to the energy momentum tensor
trace anomaly was also previously undertaken in \Weyl\
for general curved space backgrounds. The consistency conditions obtained
subsequently are a subset of those in \Weyl\ but are here directly related to
physical correlation functions. 

In more detail the outline of this paper is as follows. In the next section
a general framework for Ward identities, together with RG equations, for
the energy momentum tensor two point function is described.
This is then specialised to two and four dimensions
in sections 3 and 4. In section 5 we discuss a geometric approach appropriate
for describing two point functions of tensor fields on spaces of constant
curvature. In section 6 the conformal Killing equation is solved, independently
of any choice of coordinates in $d$-dimensions, and the corresponding conformal
group $O(d+1,1)$ identified as well as the appropriate isometry groups.
A scalar function $s$ of two points $x,y$, 
which transforms homogeneously under conformal transformations, is constructed.
An associated bi-vector, related to inversions, which also transforms 
simply under conformal transformations is found. These results allow the construction
of conformally covariant two point functions for any tensor fields.  The 
geometric formalism is extended to spinor fields in section 7 and the corresponding
inversion matrix as well as explicit forms for Killing spinors are obtained. 
The formalism of section 5 is applied in section 8 to determine a general expression 
for the two point function of the energy momentum tensor 
$\l T_{\mu\nu}(x) T_{\alpha\beta}(y) \r$
and the necessary conditions required to satisfy the conservation equation
for the energy momentum tensor are obtained. On $S^2$ or $H^2$ for a traceless energy
momentum tensor these have simple solutions with only an undetermined overall scale
for $S^2$. Expressions which satisfy the conservation equations automatically
are found in terms of two independent scalar functions
which  can be interpreted as corresponding to spin-0 and spin-2 contributions.
The spin-2 function gives a form for the two point function
appropriate to a traceless energy momentum tensor. 
It is shown how to determine generally each of the scalar functions for any
expression for the two point function obeying the conservation equations.
In section 9 the arbitrariness in this decomposition, arising when the
the spin-0 scalar function is a Green function for $-\nab^2-{1\over {d-1}}R$ and
the resulting expression for the energy
momentum tensor two point function is traceless, is discussed. 
In two dimensions, when the spin-2 function is absent,  
$\l T_{\mu\nu}(x) T_{\alpha\beta}(y) \r$ is determined uniquely for conformal
theories  with the overall scale set by the Virasoro central charge $c$. 
In section 10 we calculate the form of the two point function in the
conformal limit for free scalar and spinor fields in general dimensions and
also for free vectors in four dimensions.  For $S^d$ the results are proportional to
the unique conformally covariant form with the same overall coefficient $C_T$ as on
flat space. On $H^d$ the results are not unique but there is a simple expression
for the leading singular term with the same coefficient $C_T$.
Some aspects of these results are
understood in section 11 using the operator product expansion although its
form on spaces of non zero curvature is not yet fully clear.  In section 12
we discuss the positive energy unitary representations of $O(d-1,2)$ and their
significance for unitary quantum field theories on $H^d$. We obtain in detail
the representation for a spin 1 lowest weight state. The technicalities of this
section are then used in section 13 to obtain the spectral representation for
the two point function of a vector field when the intermediate states are 
decomposed into representations of $O(d-1,2)$.
This work motivates a natural extension giving the spectral representation of
the energy momentum tensor two point function. Finally
in section 14 general aspects of the $c$-theorem are discussed and difficulties in
deriving it for a field theory defined on a space of constant curvature
are described. In appendix A the crucial results obtained in section 8 for
positive curvature are listed in the negative curvature case while the calculation
of Green functions in terms of hypergeometric functions is described in appendix B.
Some detailed results for the spin one representation of $O(d-1,2)$, and the
calculation of the norms of the basis states, are deferred to appendix C. This
also contains a summary of some properties of the arbitrary dimension spherical
harmonics used in section 12. In appendix D  we discuss the example of a free
massive scalar field $\phi$, following \Forte, where the mass is the sole coupling.
This involves the explicit construction of the spectral representation for $\phi^2$.

\newsec{Ward Identities and Consistency Conditions}

Our subsequent  discussion depends crucially on the Ward and trace identities
satisfied by correlation functions involving the energy momentum tensor.
These are not necessarily unique since there may be ambiguous
local contact term, involving $\de$-functions, in any two or higher
point function. In consequence it is useful to first give a precise
derivation of the identities in a consistent framework and later take
account of the potential freedom of contact terms. To this end we consider
the vacuum functional $W$ for a quantum field theory defined on an arbitrarily
curved space with metric $g_{\mu\nu}(x)$ and also local sources $g^i(x)$ 
coupled to a set of scalar fields $\O_i(x)$. 
The expectation values of the quantum operator fields in the background for an 
arbitrary metric but with $g^i(x)=g^i$ the physical coupling constants (which
are taken to be dimensionless by introducing an appropriate power of a
scale $\rho$ on which the metric depends) are given by
\eqn\onep{
\sqrt {g(x)} \l T_{\mu\nu}(x)\r =  -2
{\de\over\de g^{\mu\nu}(x)} \, W  \Big |_{g^i(x)=g^i} \, ,\quad
\sqrt {g(x)} \l \O_i(x)\r = - {\de \over \de g^i(x)} W \Big |_{g^i(x)=g^i} \, , 
}
where the functional derivatives, for $d$-dimensions, are defined by
\eqn\deriv{
{\de\over\de g^{\mu\nu}(x)} g^{\alpha\beta}(y) = \de_{(\mu}{}^{\!\alpha}
\de_{\nu)}{}^{\! \beta}\de^d(x-y) \, , \quad
{\de \over \de g^i(x)} g^j(y) = \de_i{}^{\! j} \de^d(x-y) \, .
}
With this prescription the associated two point functions are given by
\eqn\twop{ \eqalign{
\sqrt {g(x)} \sqrt {g(y)}  \l T_{\mu\nu}(x) T_{\alpha\beta}(y) \r = {}& 4
{\de^2\over\de g^{\mu\nu}(x)\de g^{\alpha\beta}(y)} \, W \Big |_{g^i(x)=g^i}
\, ,\cr
\sqrt {g(x)} \sqrt {g(y)} \l  \O_i(x) \O_j(y) \r = {}& 
{\de^2 \over \de g^i(x) \de g^j(y) } \, W  \Big |_{g^i(x)=g^i} \, , \cr}
}
which are manifestly symmetric. We may also similarly define
$ \l T_{\mu\nu}(x) \O_j(y) \r $. In general the correlation functions
$ \l T_{\mu\nu}(x)\dots  \O_j(y) \dots \r $ formed from $T_{\mu\nu}$ and the
scalar fields $\O_j$ form a basic set related by Ward identities and
obeying RG equations which are the subject of discussion here.

The Ward identities may be derived from the condition that $W$ is a scalar
functional, corresponding to the requirement that any regularisation preserves
invariance under diffeomorphisms, which implies
\eqn\diff{
 \int \! d^d x \, \Bigl ( -(\nab^\mu v^\nu + \nab^\nu v^\mu)
{\de\over\de g^{\mu\nu}} + v^\mu \pr_\mu g^i
\, {\de \over \de g^i} \Bigl ) W = 0 \, .
}
It is easy to see that this gives
\eqn\done{
\nab^\mu \l T_{\mu\nu} \r = 0 \, ,
}
and also
\eqna\dtwo
$$\eqalignno{ \!\!\!\!\!\!\!\!\!
\nabla^\mu  \langle T_{\mu\nu} (x)  T_{\alpha \beta}(y) \rangle = {}&
\nabla_\nu \big ( \delta^\sigma_\alpha \delta^\rho_\beta  \, \delta^d(x,y)\big)
\,  \langle  T_{\sigma \rho}(x) \rangle
+ 2 \nabla_\sigma \big ( \delta^\sigma_{(\alpha} \delta^\rho_{\beta)}  \,
\delta^d(x,y) \, \langle  T_{\rho \nu}(x) \rangle \big ) \, , & \dtwo{a} \cr
\!\!\!\!\!\!\!\! \nabla^\mu  \langle T_{\mu\nu} (x)  \O_i(y)  \rangle  = {}&
\pr_\nu \delta^d(x,y) \, \langle \O_i(x)  \rangle \, . & \dtwo{b} \cr}
$$
In these equations, which contain $\delta^d(x,y)=\de^d(x-y)/\sqrt{g(x)}$, 
the appropriate connections to appear in covariant derivatives need to be
clearly specified. Any ambiguities are here resolved by using the
convention that the indices $\nu, \sigma, \rho$ are regarded as `at $x$'
while $\alpha, \beta $ are `at $y$', covariant derivatives such as $\nabla_\nu$
involve differentiation with respect to $x$ and have the required connection
necessary according to the tensorial structure at $x$.

We also consider associated trace identities which take the form
\eqn\onetr{
g^{\mu\nu} \langle T_{\mu\nu} \rangle - \langle \Theta\rangle  = \A \, ,
\qquad \Theta = \beta^i \O_i \, ,
}
where $\A$ is the anomalous contribution to the trace present in field theories
on a curved background. $\A$ is a local scalar formed from the the Riemann
curvature and its derivatives and obeys consistency conditions which
correspond to integrability conditions for $W$. 
In \onetr\ we have assumed that the operators $\O_i$
form a basis for the trace of the energy momentum tensor with coefficients
the $\beta$-functions corresponding to the couplings $g^i$. Derivatives
of lower dimensional operators are thus neglected in \onetr\ but if present they do
not change the essential results of the analysis, as discussed in \Weyl.
As a consequence of \onetr\  with the definitions \twop\ we have
\eqn\trTT{
g^{\mu\nu}(x)  \langle T_{\mu\nu}(x) T_{\alpha \beta}(y) \rangle
- \langle \Theta (x)  T_{\alpha \beta}(y) \rangle - 2  \langle
T_{\alpha \beta}(y) \rangle \, \delta^d(x,y) = \A_{\alpha\beta}(x,y) \, ,
}
where $\A_{\alpha\beta}(x,y)$ is formed from $\de^d(x,y)$ and derivatives
and is explicitly given by
\eqn\derivA{
\A_{\alpha\beta}(x,y) 
= - {2\over \sqrt{g(x)}\sqrt{g(y)}} {\de\over\de g^{\alpha\beta}(y)} \big ( 
\sqrt g \A(x) \big ) \, .
}
We further have
\eqn\trTO{
g^{\mu\nu}(x)  \langle T_{\mu\nu}(x)  \O_i(y)  \rangle -
\langle \Theta (x)  \O_i(y)  \rangle
+ \pr_i \beta^j \langle \O_j(y)  \rangle \, \delta^d(x,y) = 
\B_i(x,y) \, .
}
where $\B_i(x,y)$ also has support only for $x=y$. The appearance of 
$ \pr_i \beta^j$ reflects the fact that this is the anomalous dimension
matrix for the operators $\O_j$. It is easy to see that we
must have for consistency
\eqn\con{
\nab^\alpha \! \A_{\alpha\beta}(x,y) = \pr_\beta \delta^d(x,y) \, \A(y) \, ,
\quad \int \! \d^d y \sqrt g \, \B_i(x,y) = - \pr_i \A(x) \, ,
}
and also
\eqn\conAB{
\H(x,y) = g^{\alpha\beta}(y)\A_{\alpha\beta}(x,y) + \B_i(y,x)\beta^i =
\H(y,x) \, .
}
which is an integrability condition. The necessity of \conAB\ may be seen by
combining \trTT\ and \trTO\ to give
\eqn\trTtheta{ \eqalign{
g^{\mu\nu}(x)  g^{\alpha\beta}(y)& \langle T_{\mu\nu}(x) T_{\alpha \beta}(y) 
\rangle - \langle \Theta (x) \Theta (y) \rangle \cr
& {} =  2 g^{\alpha\beta}(y) \langle T_{\alpha \beta}(y) \rangle \, \delta^d(x,y) 
- \beta^i \pr_i \beta^j \langle \O_j(y)  \rangle \, \delta^d(x,y) 
+ \H (x,y) \, . \cr}
}

To obtain further consistency conditions  it is necessary also to consider the 
linear RG equations for the one
and two point functions. First we define the derivative operator  which
generates constant rescalings of the metric and the associated flow of the
couplings by
\eqn\rgd{
\D = - 2 \int \! \d^d x \, g^{\mu\nu} {\de \over \de g^{\mu\nu}}
+ \beta^i {\pr \over \pr g^i} \, .
}
For a constant curvature metric depending on a single parameter $\rho$ as in
\riem\ we have, up to the effects of the reparameterisation generated by
$u$ as in \ag,
\eqn\dg{
2 \int \! \d^d x \, g^{\mu\nu} {\de \over \de g^{\mu\nu}}
\sim \rho {\d \over \d \rho} \, .
}
With the definition \rgd\ the RG equations for $\l T_{\mu\nu} \r$ and $\l \O_i\r$
are then
\eqna\DT
$$\eqalignno{
( \D + d - 2 )
\l T_{\mu\nu} \r ={}& \C_{\mu\nu} \, , \quad
 \C_{\mu\nu}(x) = \int \! \d^d y \sqrt g \, \A_{\mu\nu}(y,x) \, , & \DT{a} \cr
( \D + d )\l \O_i \r + \pr_i \beta^j \l \O_j \r = {}&\C_i\, , 
\qquad \C_i(x) = \int \! \d^d y \sqrt g \, \B_i(y,x) \, . & \DT{b} \cr}
$$
For the two point functions we have
\eqn\DTT{
\big ( \D + 2d - 4 \big )  \l T_{\mu\nu}(x) T_{\alpha\beta}(y) \r 
= \E_{\mu\nu,\alpha\beta}(x,y) \, ,
}
where
\eqn\defE{
 \E_{\mu\nu,\alpha\beta}(x,y) = {4\over \sqrt {g(x)} \sqrt {g(y)} }
{\de^2\over\de g^{\mu\nu}(x)\de g^{\alpha\beta}(y)} 
\int \! \d^d x \sqrt g \, \A \, ,
}
and
\eqn\DTO{
\big ( \D + 2d -  2 \big ) \l T_{\mu\nu}(x) \O_i(y) \r
+ \pr_i \beta^j  \l T_{\mu\nu}(x) \O_j(y) \r = \F_{\mu\nu,i}(x,y) \, ,
}
for
\eqn\defF{
\F_{\mu\nu,i}(x,y) 
= - {2\over \sqrt {g(x)}\sqrt {g(y)} } {\de \over\de g^{\mu\nu}(x)} 
\big ( \sqrt g \C_i(y)\big )  \, ,
}
together with
\eqn\DOO{ \eqalign{
\big ( \D + 2d \big )& \l \O_i(x) \O_j(y) \r + \pr_i \beta^k  \l \O_k(x)
\O_j(y) \r \cr
{} + {} &  \pr_j \beta^k  \l \O_i(x)  \O_k(y) \r 
- \pr_i \pr_j \beta^k \l \O_k (x) \r \de^d (x,y) = \G_{ij}(x,y) \, . \cr}
}
Given an expression for the trace anomaly $\A$ on a general curved space
$\C_{\mu\nu}$ and $\E_{\mu\nu,\alpha\beta}$, as well as $\A_{\alpha\beta}$,
may be directly calculated. For $d=2$ and $4$ general forms for
$\F_{\mu\nu,i}(x,y)$ and $\G_{ij}(x,y)= \G_{ji}(y,x)$ may be constructed
as a sum of terms involving the curvature and derivatives of $\de^d(x,y)$ 
with the appropriate dimension. From \dtwo{b}\ we must have
\eqn\dF{
\nab^\mu \F_{\mu\nu,i}(x,y) = \pr_\nu \de^d(x,y) \, \C_i(x) \, ,
}
and  \dtwo{a}\ gives a relation between $\E_{\mu\nu,\alpha\beta}$ and 
$\C_{\mu\nu}$.

Requiring  consistency of \DT{a,b}\ and \DTT,\DTO,\DOO\ with \trTT,\trTO\ leads
to 
\eqn\conAF{
( \D + 2d- 2 ) \A_{\alpha\beta}(x,y) = g^{\mu\nu}(x)
\E_{\mu\nu,\alpha\beta}(x,y) - \F_{\alpha\beta,i}(y,x)\beta^i  -
2 \C_{\alpha\beta}(y) \de^d (x,y) \, ,
}
and
\eqn\conBFG{
(\D + 2d) \B_i(x,y) + \pr_i \beta^j \big (\B_j(x,y) - \C_j(y) \de^d(x,y) \big )
= g^{\mu\nu}  \F_{\mu\nu,i}(x,y) -  \G_{ij}(x,y)\beta^j \, .
}

The identities obtained above are explored in the following sections in the
particular cases of two and four
dimensions after restricting to spaces of constant curvature, as given in \riem.
On such homogeneous spaces $\l \O_i \r$ and also the curvature trace anomaly
$\A$ in \onetr\ are just constants and also
\eqn\Thom{
\l T_{\mu\nu}(x) \r =  - {1\over d}C\rho^d g_{\mu\nu}(x) \, ,
}
with $C(g)$ dimensionless.
The result \Thom\ of course trivially satisfies \done.
Furthermore \dtwo{a}\ simplifies in this case to
\eqn\DTr{
\nabla^\mu\langle T_{\mu\nu}(x)T_{\alpha \beta}(y) \rangle =
 - {1\over d}C\rho^d \Big (
\pr_\nu \delta^d(x,y) \,  g_{\alpha\beta}(y)
+ 2 \nabla_\sigma \big ( \delta^\sigma_{(\alpha} \delta^\rho_{\beta)}  \,
 g_{\rho \nu}(x) \, \delta^d(x,y) \big ) \Big ) \, .
}
If we define
\eqn\TTsim{
\langle T_{\mu\nu}(x)T_{\alpha \beta}(y) \rangle_{\rm con} =
\l T_{\mu\nu}(x) T_{\alpha \beta}(y) \r +
{\ts {1\over d}} C\rho^d ( g_{\mu\alpha}g_{\nu\beta} +  g_{\mu\beta}g_{\nu\alpha}
+ g_{\mu\nu} g_{\alpha\beta} ) \de^d(x,y) \, ,
}
then
\eqn\conTT{
\nabla^\mu\langle T_{\mu\nu}(x)T_{\alpha \beta}(y) \rangle_{\rm con} = 0 \, .
}

Using the definition \twop\ and also \trTtheta\ we find in general
\eqna\rhoC
$$\eqalignno{
\rho {\d \over \d \rho} C = {}& {1\over \rho^d} \int \! \d^d y \sqrt g \, 
g^{\mu\nu}(x)  g^{\alpha\beta}(y)\langle T_{\mu\nu}(x) T_{\alpha \beta}(y)
\rangle_{\rm con} - d C & \rhoC{a} \cr
= {}& {1\over \rho^d} \int \! \d^d y \sqrt g \, \Big (
\langle \Theta (x) \Theta (y) \rangle + \H(x,y) \Big )
- {1\over \rho^d} \beta^i \pr_i \beta^j \l \O_j \r \, . & \rhoC{b} \cr} 
$$
Applying the basic definitions in \onep\ with  \deriv, for the
theory defined on a homogeneous space of constant curvature, we have the consistency
conditions
\eqn\consis{
{\pr \over \pr g^i} C = - {\rho} {\d \over \d \rho} \Big ( {1\over \rho^d} \l \O_i \r
\Big ) \, .
}
Since  from \onetr\ $C= - (\beta^i \l \O_i \r + \A ) \rho^{-d}$, \consis\ implies
the RG equation \DT{a}\ which now takes the form
\eqn\DC{
\bigg (- {\rho} {\d \over \d \rho} + \beta^i {\pr \over \pr g^i} \bigg ) C
= {\rho} {\d \over \d \rho} \Big ( {1\over \rho^d} \A \Big ) = 
 - {1\over \rho^d} g^{\mu\nu} \C_{\mu\nu} \, .
}

\newsec{Two Dimensions}

In two dimensions the curved space trace anomaly for an arbitrary metric
is just proportional to the scalar curvature so that in \onetr\ we may
write
\eqn\ctwo{
2\pi \A = {\ts{1\over 12}} c\, R \, .
}
In this case in \trTT\ and \trTO\ we now take
\eqn\two{\eqalign{
2\pi \A_{\alpha\beta}(x,y) = {}& {\ts{1\over 6}}  c \big (
\nabla_\alpha \nabla_\beta - g_{\alpha\beta }\nab^2 \big )\delta^2(x,y)\, , \cr
2\pi \B_i(x,y) ={}&  - {\ts{1\over 12}} \pr_i c \, R \delta^2(x,y) -
{\ts{1\over 6}} w_i \nabla^2 \delta^2(x,y) \, , \cr}
}
which are in accord with \con\ and where $w_i(g)$ is a vector function of the
couplings. Furthermore in two dimensions $\C_{\mu\nu}$ and
$\E_{\mu\nu,\alpha\beta}$ are both zero so that \DT{a}\ and \DTT\ now become
\eqn\DTtwo{
\D \l T_{\mu\nu} \r = 0 \, , \qquad \D \l T_{\mu\nu}(x) T_{\alpha\beta}(y) \r
= 0 \, ,
}
while in \DT{b}\ and \DTO\ we now assume
\eqn\CF{
2\pi \C_i = -  {\ts{1\over 12}}  \pr_i c \, R  \, , \qquad
2\pi  \F_{\mu\nu,i}(x,y) = - {\ts{1\over 6}}  \pr_i c \big (
\nabla_\mu \nabla_\nu - g_{\mu\nu }\nab^2  \big ) \delta^2(x,y) \, ,
}
so that \DT{b}\ now reads
\eqn\DO{
2\pi \big ( ( \D + 2 )\l \O_i \r + \pr_i \beta^j \l \O_j \r \big ) =
-  {\ts{1\over 12}}  \pr_i c \, R  \, .
}
Finally in \DOO\ we take, for $G_{ij}(g)$ a symmetric tensor,
\eqn\Gtwo{
2\pi \G_{ij}(x,y) = {\ts{1\over 6}}  G_{ij} \nab^2 \de^2(x,y) 
+ {\ts{1\over 12}}\pr_i\pr_j c \, R \de^2(x,y)\, .
}

With the expressions  in \two\ and \CF\ the consistency condition \conAF\
is identically satisfied. Inserting the appropriate forms into \conBFG\
leads to the single relation
\eqn\cG{
\pr_i c - G_{ij} \beta^j = - \L_\beta w_i
= - \beta^j \pr_j w_i -  \pr_i \beta^j w_j \, .
}
If we define
\eqn\defC{
{\tilde c} = c + w_i\beta^i \, ,
}
then \cG\ may be rewritten as
\eqn\CGW{
\pr_i {\tilde c} = \big (  G_{ij} + \pr_i w_j - \pr_j w_i \big ) \beta^j \quad 
\Rightarrow \quad \beta^i \pr_i {\tilde c} = G_{ij} \beta^i \beta^j \, .
}

We now show how the Ward and trace identities may be solved, after
restricting to a homogeneous space of constant curvature, to give
explicit forms for the two point functions with relations between them.
Assuming \Thom\ we may take
\eqn\TTtwo{ \eqalign { 
(2\pi)^2 \langle T_{\mu\nu}(x)& T_{\alpha \beta}(y) \rangle_{\rm con} \cr
={}& \big ( \nab_\mu \nab_\nu - g_{\mu\nu} \nab^2 - \half g_{\mu\nu} R \big )
F(x,y) \big ( {\overleftarrow \nab_{\!\alpha}} {\overleftarrow \nab_{\!\beta}}
- {\overleftarrow \nab}{}^2 g_{\alpha\beta} - \half g_{\alpha\beta} R \big )\, ,
\cr}
}
which satisfies \DTr\ identically. In a similar fashion
\eqn\TOtwo{
(2\pi)^2\big ( \langle T_{\mu\nu}(x) \O_i(y) \r - g_{\mu\nu} \l \O_i \r
 \de^2(x,y) \big ) =
\big ( \nab_\mu \nab_\nu - g_{\mu\nu} \nab^2 - \half g_{\mu\nu} R \big )
G_i(x,y) \big ( {- {\overleftarrow \nab}}{}^2 - R \big ) \, , }
automatically satisfies \dtwo{b}. Applying \trTT\  with \two\ gives the
relations
\eqn\relone{
F(x,y) = G_i(x,y)\beta^i + {\ts{1\over 3}} c \, \Gg_0(x,y) \, ,
}
assuming $\Gg_0$ is a solution of
\eqn\Gzero{
(-\nab^2 - R ) \Gg_0(x,y) = 2\pi\de^2(x,y) \, .
}
Explicit solutions of this equation are discussed later. For the case of the
positive curvature sphere \Gzero\ has to be modified although this does
not change \relone\ or \TTtwo. The  RG equations for the two point functions
in \TTtwo\ and \TOtwo\ may now be reexpressed in terms of $F$ and $G_i$,
\eqn\RGG{
\D F(x,y) = 0 \, , \qquad \D G_i(x,y) + \pr_i\beta^j G_j(x,y) =
- {\ts{1\over 6}} \pr_i c \, \Gg_0(x,y) \, .
}
If we write
\eqn\OOtwo{
(2\pi)^2 \l \O_i (x) \O_j (y) \r = \big ( -\nab^2 - R \big )
H_{ij}(x,y) \big ( {- {\overleftarrow \nab}}{}^2 - R \big ) \, ,
}
then we have from \trTO
\eqn\retwo{\eqalign{
G_i(x,y) = {}& H_{ij}(x,y)\beta^j + {\ts{1\over 6}} w_i \Gg_0(x,y) + 
V_i \Gg_1(x,y) \, , \cr
V_i = {}& {\ts{1\over 6}} ( w_i - \half \pr_i c ) R - 2\pi\big ( 2\l \O_i \r 
+ \pr_i \beta^j \l \O_j \r \big ) \, ,\cr}
}
if
\eqn\Gone{
(-\nab^2 - R ) \Gg_1(x,y) =  \Gg_0(x,y) \, .
}
Furthermore \DOO\ gives
\eqn\RGij{
\D H_{ij}(x,y) + \pr_i  \beta^k H_{kj}(x,y) + \pr_j  \beta^k H_{ik}(x,y)
= - {\ts{1\over 6}} G_{ij}  \Gg_0(x,y) - S_{ij} \Gg_1(x,y) \, .
}
for
\eqn\defS{
S_{ij} = {\ts{1\over 6}} (G_{ij} - \half \pr_i\pr_j c ) R 
- 2\pi \pr_i\pr_j \beta^k \l \O_k\r \, .  }
Consistency of \RGij\ with \retwo\ and \RGG\ depends on \cG\ and also
\eqn\VS{
(\D + 2 ) V_i + \pr_i \beta^j V_j = S_{ij} \beta^j \, ,
}
which follows from \cG\ and \DO.

At a fixed point, when $\beta^i=0$, then 
\eqn\Ttwo{
2\pi \langle T_{\mu\nu} \rangle \big |_{\beta^i = 0} =  {\ts{1\over 24}} c R\,
g_{\mu\nu} \, ,
}
so that in \Thom\ $2\pi C |_{\beta^i = 0} = \mp {1\over 6} c$. From
\relone\ the two point function in \TTtwo\ is also determined to be,
\eqn\TTtwoc{ \eqalign {
(2\pi)^2 \langle T_{\mu\nu}(x)& T_{\alpha \beta}(y) \rangle_{\rm con}
\big |_{\beta^i = 0} \cr
={}& {\ts{1\over 6}} c\big ( \nab_\mu \nab_\nu - g_{\mu\nu} \nab^2 - \half g_{\mu\nu} R \big )
\Gg_0(x,y) \big ( {\overleftarrow \nab_{\!\alpha}} {\overleftarrow \nab_{\!\beta}}
- {\overleftarrow \nab}{}^2 g_{\alpha\beta} - \half g_{\alpha\beta} R \big )\, ,
\cr}
}
neglecting contact terms.
The overall coefficient is just $c$ and of course the result
is non local, not just a potentially ambiguous contact term. Explicit
expressions are obtained later in section 8. In the same way
from \retwo\ $G_i = \thir w_i G_0 + V_i G_1$. Substituting in \TOtwo,
disregarding  contact terms and using \cG\ in the expression for $V_i$ gives
\eqn\TOfp{\eqalign{
(2\pi)^2 \langle T_{\mu\nu}(x) \O_i(y) \r \big |_{\beta^i = 0} = {}&
(\nab_\mu\nab_\nu - \half g_{\mu\nu} \nab^2 ) \Gg_0(x,y) \,
\Gamma_i{}^{\! j} ( {\ts{1\over 12}} w_j R - 2\pi \l \O_j \r ) \, , \cr
& \Gamma_i{}^{\! j} = 2 \de_i{}^{\! j} + \pr_i \beta^j \, . \cr}
}
$ \Gamma_i{}^{\! j}$ is the matrix defining the dimensions of the fields 
$\O_i$. The result \TOfp\ demonstrates that
$w_i$ is well defined at a critical point although it should be
recognised that $\l\O_i\r$ is arbitrary  up to terms
$\propto \pr_i f \, R$ for any function $f(g)$ of the couplings which
leads to a corresponding freedom $w_i \sim w_i + \pr_i f$.

\newsec{Four Dimensions}

As is well known in any even dimension beyond two dimensions there are
several possible curvature dependent terms which may contribute to the
energy momentum tensor trace. For four dimensions we take\foot{For free fields 
$c={1\over 120}(12n_V + 6n_F + n_S) , \, a={1\over 360}(62n_V + 11n_F + n_S)$.}
\eqn\Afour{
16\pi^2 \A = c F - a G - b R^2 \, ,
}
neglecting a term $\propto \nab^2 R$ which may be cancelled by a local
redefinition of $W$, and where
\eqn\defGF{
G = {\textstyle{1\over 4}} \epsilon^{\mu\nu\sigma\rho}\epsilon_{
\alpha\beta\gamma\delta} R^{\alpha\beta}{}_{\mu\nu}
R^{\gamma\delta} {}_{\sigma\rho} = 6
R^{\alpha\beta}{}_{[\alpha\beta} R^{\gamma\delta} {}_{\gamma\delta]} \, ,
\qquad F = C^{\alpha\beta\gamma\delta} C_{\alpha\beta\gamma\delta} \, ,
}
with $C$ the Weyl tensor. $G$ is a topological density which is reflected
by its variation being a total divergence,
\eqn\varG{
\delta (\sqrt g G) = \sqrt g \nabla_\gamma V^\gamma \, , \qquad
V^\gamma = 24 R^{[\alpha\beta}{}_{\alpha\beta}\nab^\gamma \de g^{\de]\ep}
g_{\ep\de} \, .
}
For general metrics then \derivA\ gives
\eqn\trA{\eqalign{
16\pi^2 g^{\alpha\beta}(y)\A_{\alpha\beta}(x,y) = {}& - 8a \, \nab_\alpha 
\nab_\beta \big ( G^{\alpha\beta} \de^4(x,y) \big ) + 12b \, R(x) \nab^2
\de^4(x,y) \, , \cr
&  G^{\alpha\beta} = R^{\alpha\beta} - \half g^{\alpha\beta} R \, , \cr}
}
so that, for $R$ not constant, the $b$ term is not symmetric and hence the
condition \conAB\ requires $b=0$ if $\beta^i=0$. On a homogeneous space
with \riem\ then $C_{\alpha\beta\gamma\delta} = 0$ and
\eqn\hom{
G = {\textstyle{1\over 6}} R^2 \, , \qquad
V^\gamma = {\textstyle{1\over 3}} R \big (
- \nabla_\beta \delta g^{\gamma\beta} + \nabla^\gamma
(g_{\alpha\beta}\delta g^{\alpha\beta}) \big ) \, ,
}
so that in this case the anomaly reduces to just
\eqn\Ared{
2\pi^2 \A = - {\ts{1\over 48}} \ha \, R^2 \, , \qquad \ha = a + 6b \, ,
}
and using \hom
\eqn\Ahom{
2\pi^2 \A_{\alpha\beta}(x,y ) =
- {\textstyle{1\over 12}} \ha  \, R \big (\nabla_\alpha \nabla_\beta -
g_{\alpha\beta}\nabla^2 \big ) \delta^4(x,y)  \, .
}
Similarly we require in this case
\eqn\Bhom{ 
2\pi^2 \B_i (x,y) =  {\ts {1\over 48}} \pr_i  \ha  \, R^2 \de^4(x,y)
- Y_i \, R \nab^2 \de^4(x,y) - U_i \, \nab^2 \nab^2 \de^4(x,y) \, .
}

Since now $\C_{\mu\nu}=0$ and
\eqn\eqC{
2\pi^2 \C_i = {\ts {1\over 48}} \pr_i \ha  \, R^2 \, ,
}
the RG eqs. \DT{a,b}\ become
\eqna\DTfour
$$\eqalignno{
( \D + 2 ) \l T_{\mu\nu} \r = {}& 0 \, , & \DTfour{a} \cr
2\pi^2 \big ( ( \D + 4 )\l \O_i \r + \pr_i \beta^j \l \O_j \r \big ) = {}& 
{\ts {1\over 48}} \pr_i \ha  \, R^2 \, . & \DTfour{b} \cr}
$$
With the result \Afour\ the definition \defE\ may be expressed as
\eqnn\Efour
$$\eqalignno{
2\pi^2 \E_{\mu\nu,\alpha\beta}(x,y) = {}& 4c\, \nab^\si \nab^\rho \big (
\E^C{}_{\!\! \mu\si\rho\nu,\alpha\gamma\de\beta} \de^4(x,y) \big )
{\overleftarrow \nab}{}^\gamma {\overleftarrow \nab}{}^\de \cr
{}& - b \,
\big ( \nab_\mu \nab_\nu - g_{\mu\nu} \nab^2 - \quar g_{\mu\nu} R \big )
\de^4(x,y) \big ( {\overleftarrow \nab_{\!\alpha}} {\overleftarrow \nab_{\!\beta}}
- {\overleftarrow \nab}{}^2 g_{\alpha\beta} - \quar g_{\alpha\beta} R \big )\cr
{}& + \half b\, R \, H_{\mu\nu,\alpha\beta}(x,y) \, , & \Efour \cr}
$$
where $\E^C{}_{\!\! \mu\si\rho\nu,\alpha\gamma\de\beta}(x) $ is the projector
at $x$ onto tensors with the symmetry and traceless properties of the Weyl
tensor $C_{\mu\si\rho\nu}$,
\eqn\Csym{
C_{\mu\si\rho\nu}  = C_{[\mu\si][\rho\nu]} \, , \ C_{\mu[\si\rho\nu]} = 0 \, , \
g^{\si\rho} C_{\mu\si\rho\nu} = 0 \, .
}
An explicit form for $\E^C$ is given later by (8.21) for  $y=x$. The term involving
$\E^C$ in \Efour\ is automatically conserved and traceless since,
on spaces of constant curvature when \riem\ holds, for any $C_{\mu\si\rho\nu}$
satisfying \Csym
\eqn\CT{
T_{\mu\nu} = \nab^\si \nab^\rho C_{\mu\si\rho\nu} \quad \Rightarrow
\quad  T_{\mu\nu} = T_{\nu\mu} \, , \
g^{\mu\nu}T_{\mu\nu} = 0 \, , \ \nab^\mu T_{\mu\nu} = 0 \, .
}
The remaining term $H_{\mu\nu,\alpha\beta}(x,y)$ in \Efour\ is then defined by
\eqn\defH{\eqalign{
H_{\mu\nu,\alpha\beta}(x,y) = {}&
\nabla_\mu \nabla_\sigma \big ( \delta^\sigma_{(\alpha} \delta^\rho_{\beta)} \,
g_{\rho\nu}\, \delta^4 (x,y) \big ) +
\nabla_\nu \nabla_\sigma \big ( \delta^\sigma_{(\alpha} \delta^\rho_{\beta)} \,
g_{\rho\mu}\, \delta^4 (x,y)  \cr
{}& - \nabla^2 \big ( \delta^\sigma_{(\alpha} \delta^\rho_{\beta)} \,
g_{\mu\sigma} g_{\nu\rho} \, \delta^4 (x,y) \big ) \cr
{}&\!\!  -  g_{\alpha\beta} \nabla_\mu \nabla_\nu \delta^4 (x,y)
-  g_{\mu\nu} \nabla_\alpha \nabla_\beta \delta^4 (x,y) 
+ g_{\mu\nu} g_{\alpha\beta} \nabla^2 \delta^4 (x,y) \cr
{}& + {\ts{1\over 12}} R \big ( g_{\mu\alpha} g_{\nu\beta}
+ g_{\mu\beta} g_{\nu\alpha} + g_{\mu\nu} g_{\alpha\beta} \big )
\delta^4(x,y) \, , \cr}
}
and, with the conventions on covariant derivatives described above, this is
symmetric
\eqn\sym{
H_{\mu\nu,\alpha\beta}(x,y) = H_{\alpha\beta,\mu\nu}(y,x) \, ,
}
and, for constant curvature, satisfies
\eqn\trH{
\nabla^\mu H_{\mu\nu,\alpha\beta}(x,y) = 0 \, , \quad
H_{\mu\nu,\alpha\beta}(x,y) g^{\alpha\beta}(y) =
- 2 \big (\nabla_\mu \nabla_\nu - g_{\mu\nu} \nabla^2 - \quar g_{\mu\nu} R
\big ) \delta^4 (x,y)\, .
} 
As a consequence of \CT\ and \trH\ the form \Efour\ satisfies
\eqn\trE{\eqalign{
\nab^\mu \E_{\mu\nu,\alpha\beta}(x,y) = {}& 0 \, , \cr
2\pi^2 g^{\mu\nu}(x) \E_{\mu\nu,\alpha\beta}(x,y) = {}& 3 b \, \nab^2 
\de^4(x,y) \big ( {\overleftarrow \nab_{\!\alpha}} {\overleftarrow \nab_{\!\beta}}
- {\overleftarrow \nab}{}^2 g_{\alpha\beta} - \quar g_{\alpha\beta} R \big )\, .
\cr}
}

In addition the corresponding quantities entering into \DTO\ and \DOO\
may now be assumed to be given by (for a general metric then $2\pi^2\C_i$ may
contain a term $\half D_i \nab^2 R$, its variation according to $\defF$
gives rise to the corresponding term below)
\eqn\Ffour{\eqalign{
2\pi^2  \F_{\mu\nu,i}(x,y) ={}& {\ts {1\over 12}} \pr_i  \ha \, R
\big ( \nabla_\mu \nabla_\nu - g_{\mu\nu }\nab^2  \big ) \delta^4(x,y) \cr
{}& + D_i \big ( \nabla_\mu \nabla_\nu - g_{\mu\nu }\nab^2 - \quar  g_{\mu\nu }
R \big )\nab^2 \delta^4(x,y) \, , \cr}
}
and
\eqn\Gfour{
2\pi^2 \G_{ij}(x,y) = G_{ij} \nab^2 \nab^2 \delta^4(x,y) 
+ L_{ij}\, R \nab^2 \delta^4(x,y) - {\ts {1\over 48}} \pr_i \pr_j \ha \, R^2
\delta^4(x,y) \, .
}
Inserting the results \Ahom,  \trE\ and \Ffour\ into \conAF\ then gives
\eqn\bD{
3 b = D_i \beta^i \, .
}
This provides an alternative demonstration of the vanishing of the 
coefficient of the $R^2$ trace anomaly at a fixed point. Imposing now
\conBFG\ leads to two relations, 
\eqn\new{
\L_\beta U_i = G_{ij} \beta^j + 3D_i \, , \qquad
\L_\beta Y_i = L_{ij} \beta^j + D_i + {\ts{1\over 4}} \pr_i \ha \, ,
}
and eliminating $D_i$ then gives an analogous formula to \cG,
\eqn\aG{
{\ts{3\over 4}} \pr_i \ha = \hG_{ij} \beta^j - \L_\beta \hU_i \, , \qquad 
\hG_{ij} = G_{ij} - 3 L_{ij} \, , \quad \hU_i = U_i - 3Y_i \, .
}

In order to apply these results we consider explicit forms for the two
point functions for a homogeneous space of constant curvature given by
\eqn\OOfour{
(2\pi^2)^2 \l \O_i (x) \O_j (y) \r = \big ( -\nab^2 -  \thir R \big )
H_{ij}(x,y) \big ( {- {\overleftarrow \nab}}{}^2 -  \thir R \big ) \, ,
}
and
\eqn\TOfour{
(2\pi^2)^2\big ( \langle T_{\mu\nu}(x) \O_i(y) \r - g_{\mu\nu} \l \O_i \r
 \de^4(x,y) \big ) =
\big ( \nab_\mu \nab_\nu - g_{\mu\nu} \nab^2 - \quar g_{\mu\nu} R \big )
G_i(x,y) \big ( {- {\overleftarrow \nab}}{}^2 - \thir R \big ) \, , }
which satisfies \dtwo{b}. With the result \Bhom\ the relation \trTO\ leads to
\eqn\GHfour{\eqalign{
3G_j(x,y) ={}& \beta^i H_{ij}(x,y)- U_j \, 2\pi^2\de^4(x,y) + 
(Y_j - {\ts{2\over 3}}U_j) R \Gg_0(x,y) + V_j \Gg_1(x,y) \, , \cr
& V_j = \big ( {\ts {1\over 48}} \pr_j \ha - {\ts{1\over 9}} \hU_j \big )
- 2\pi^2 \big ( 4 \l \O_j \r + \pr_j \beta^k \l \O_k \r \big ) \, , \cr}
}
where
\eqn\GGG{
(-\nab^2 - \thir R ) \Gg_0(x,y) = 2\pi^2 \de^4(x,y) \, , \qquad
(-\nab^2 - \thir R ) \Gg_1(x,y) =  \Gg_0(x,y) \, .
}
As in two dimensions the equation for $\Gg_0$ must be modified in the positive
curvature case as demonstrated for its explicit solution found later.
The RG equations \DOO\ and \DTO\ now become
\eqn\DOOfour{\eqalign{
(\D + 4 )H_{ij}(x,y)+ \pr_i & \beta^k H_{kj}(x,y) + \pr_j \beta^k H_{jk}(x,y)\cr
= {}& G_{ij}\, 2\pi^2\de^4(x,y) - (L_{ij} - {\ts{2\over 3}} G_{ij} ) R \Gg_0(x,y) 
- S_{ij} \Gg_1(x,y) \, , \cr 
S_{ij} = {}& \big ( {\ts{1\over 48}} \pr_i \pr_j \ha  - {\ts{1\over 9}} 
\hG_{ij} \big ) R^2 - 2\pi^2 \pr_i \pr_j \beta^k  \l \O_k \r \, , \cr}
}
and
\eqn\DTOfour{
(\D + 4 ) G_i(x,y)+ \pr_i \beta^j G_j (x,y) = - D_i \, 2\pi^2 \de^4(x,y) 
+ \thir ({\ts{1\over 4}}\pr_i \ha - D_i) R \Gg_0(x,y ) \, .
}
Compatibility of \DTOfour\ and \DOOfour\ with \GHfour, which requires
\eqn\DD{
(\D + 4 ) V_j + \pr_j \beta^k V_k = S_{jk}\beta^k \, ,
}
is guaranteed as a consequence of \new. 

It remains to find a general form for the energy momentum tensor two point
function. For homogeneous spaces in general dimensions other than two this 
involves two independent tensor structures which represent spin 2 and spin 0.
For the conserved two point function defined by \conTT
\eqn\TTfour{\eqalign {
(2\pi^2)^2 & \langle T_{\mu\nu}(x) T_{\alpha \beta}(y) \rangle_{\rm con}
=  \nab^\si \nab^\rho \big (
\E^C{}_{\!\! \mu\si\rho\nu,\alpha\gamma\de\beta}(x,y) F_2 (x,y) \big )
{\overleftarrow \nab}{}^\gamma {\overleftarrow \nab}{}^\de \cr
& \ {}+ \big ( \nab_\mu \nab_\nu - g_{\mu\nu} \nab^2 - \quar g_{\mu\nu} R \big )
F_0(x,y) \big ( {\overleftarrow \nab_{\!\alpha}} {\overleftarrow \nab_{\!\beta}}
- {\overleftarrow \nab}{}^2 g_{\alpha\beta} - \quar g_{\alpha\beta} R \big )\, ,
\cr}
}
where $\E^C{}_{\!\! \mu\si\rho\nu,\alpha\gamma\de\beta}(x,y)$ is a bi-tensor,
constructed explicitly later, which satisfies the symmetry and traceless
conditions of the Weyl tensor separately at $x$ and $y$ and which for $x=y$
reduces to the projector $\E^C{}_{\!\! \mu\si\rho\nu,\alpha\gamma\de\beta}(x)$
introduced above. Both expressions on the right hand side of \TTfour\ are
automatically conserved, the first term, involving $F_2$, using \CT\
is in addition traceless. It is important to recognise that the decomposition 
in \TTfour\ is not unique. If $F_0 \to \Gg_0$, satisfying \GGG, then this 
term is also both conserved and traceless for non coincident $x,y$ and, as shown in
section 8, any conserved traceless two point function can always be written in terms
of an appropriate $F_2$ for $x\ne y$. This therefore leads to the relation
\eqn\Hfour{\eqalign{\!\!\!\!\!\!
2\pi^2 H_{\mu\nu,\alpha\beta}(x,y) {}& =
 8 \nab^\si \nab^\rho \big (
\E^C{}_{\!\! \mu\si\rho\nu,\alpha\gamma\de\beta}(x,y) {\Gg}_2 (x,y) \big )
{\overleftarrow \nab}{}^\gamma {\overleftarrow \nab}{}^\de \cr
{}-{\ts{2\over 3}} &
\big ( \nab_\mu \nab_\nu - g_{\mu\nu} \nab^2 - \quar g_{\mu\nu} R \big )
\Gg_0(x,y) \big ( {\overleftarrow \nab_{\!\alpha}} {\overleftarrow \nab_{\!\beta}}
- {\overleftarrow \nab}{}^2 g_{\alpha\beta} - \quar g_{\alpha\beta} R \big )\, ,
\cr}
}
for a suitable ${\Gg}_2 (x,y)$
where $H_{\mu\nu,\alpha\beta}(x,y)$ is purely a local contact term. Due to
the derivatives the right hand side of \Hfour\ is well defined as a distribution.
By considering the divergence and trace of both sides of this relation
$H_{\mu\nu,\alpha\beta}(x,y)$ may be identified with the previous definition in 
\defH\ since it is the unique form satisfying \sym\ and \trH.
On flat space $\E^C$ is a constant tensor and $\Gg_0$ and $\Gg_2$ are identical.

Applying the decomposition in \TTfour\ we may first note,
by using the general relation \trTT\ with \TOfour\ and \Ahom, that 
\eqn\trfour{ 
3 F_0(x,y) = G_i(x,y) \beta^i - {\ts{1\over 12}} \ha \, R \Gg_0 (x,y) \,.
}
In order to implement the RG equation \DTT\ it is necessary to rewrite
the result for $\E_{\mu\nu,\alpha\beta}(x,y)$, which is given by \Efour, by
using \Hfour\ in the more convenient form,
\eqn\Efourn{\eqalign{\!\!\!\!\!\!
 2 & \pi^2\E_{\mu\nu,\alpha\beta}(x,y) = 4\, \nab^\si \nab^\rho \Big (
\E^C{}_{\!\! \mu\si\rho\nu,\alpha\gamma\de\beta}(x,y)
\big (c \, 2\pi^2\de^4(x,y) + b \, R {\Gg}_2 (x,y)\big ) \Big )
{\overleftarrow \nab}{}^\gamma {\overleftarrow \nab}{}^\de \cr
\!\!\!\!\!\! {}& - b \,
\big ( \nab_\mu \nab_\nu - g_{\mu\nu} \nab^2 - \quar g_{\mu\nu} R \big )
\big ( \de^4(x,y) + \thir R \Gg_0(x,y) \big )
\big ( {\overleftarrow \nab_{\!\alpha}} {\overleftarrow \nab_{\!\beta}}
- {\overleftarrow \nab}{}^2 g_{\alpha\beta} - \quar g_{\alpha\beta} R \big )\,.
\cr}
}
With this result the RG equations reduce to
\eqna\rgfour
$$\eqalignno{
(\D + 4 ) F_2(x,y) = {}& 4
\big ( c \, 2\pi^2\de^4(x,y) + b \, R {\Gg}_2 (x,y)\big ) \, , &\rgfour{a} \cr
(\D + 4 ) F_0(x,y) = {}& - b \big ( 2\pi^2 \de^4(x,y) + \thir R \Gg_0(x,y) \big )
\, . &\rgfour{b} \cr}
$$
It is important to note that \rgfour{b} follows directly from the
expression \trfour\ and \DTOfour\ so long as \bD\ holds. 

The general results simplify if we restrict to a fixed point where $\beta$-functions
vanish. The expectation value of a single energy momentum tensor becomes
\eqn\Tfour{
16\pi^2 \langle T_{\mu\nu} \rangle \big |_{\beta^i = 0} = - {\ts{1\over 24}} a R^2\,
g_{\mu\nu} \, ,
}
or in \Thom\ $2\pi^2 C |_{\beta^i = 0} = 3a$.  From \GHfour\ we must also have
\eqnn\TOfpfour
$$\eqalignno{  \!\!\!\!\!\!\!\!
(2\pi^2)^2 \langle T_{\mu\nu}(x) \O_i(y) \r \big |_{\beta^i = 0} = {}&
- (\nab_\mu\nab_\nu - \quar g_{\mu\nu} \nab^2 ) \Gg_0(x,y) \, \thir
\Gamma_i{}^{\! j}\big ( {\ts {1\over 36}} \hU_j R^2 + 2\pi^2\l \O_j\r \big) \, , \cr
& \Gamma_i{}^{\! j} = 4 \de_i{}^{\! j} + \pr_i \beta^j \, , & \TOfpfour \cr}
$$
which is the extension of \TOfp\ to four dimensions. This provides a
definition of $\hU_i$ at a fixed point.  In the energy momentum tensor 
two point function $F_0 = - {\ts{2\over 9}} a \, R \Gg_0 $,
from \trfour, and using \Hfour\ we may write
\eqn\TTcon{\eqalign {
(2\pi^2)^2 \langle T_{\mu\nu}(x) T_{\alpha \beta}(y) \rangle_{\rm con} 
\big |_{\beta^i = 0} ={}&   \nab^\si \nab^\rho \big (
\E^C{}_{\!\! \mu\si\rho\nu,\alpha\gamma\de\beta}(x,y) {\hat F}_2 (x,y) \big )
{\overleftarrow \nab}{}^\gamma {\overleftarrow \nab}{}^\de  \cr
{}& + {\ts {1\over 24}} a R \, 2\pi^2 H_{\mu\nu,\alpha\beta}(x,y)  \, , \cr
{\hat F}_2 (x,y)= {}& F_2(x,y) 
- {\ts{1\over 3}} a \, R {\Gg}_2 (x,y)\, .\cr}
}
Hence in this case, apart from a contact term, there is only the manifestly
conserved and traceless spin 2 contribution to this two point function. 

\newsec{Geometrical Results for Spaces of Constant Curvature}

In order to find more explicit expressions for the two point functions 
considered above we discuss here some geometrical results which allow
natural expressions for the two point functions on homogeneous spaces
of constant curvature to be found. A related discussion was given by
Allen and Jacobsen \allen\ but there are differences in derivation and
also application.

Our starting point is the geodetic interval $\si(x,y)$ which is a bi-scalar
defined for any curved manifold which is unique for $x$ in the neighbourhood
of $y$ and which satisfies \dewitt,
\eqn\bis{
g^{\mu\nu}\pr_\mu \si \pr_\nu \si = 2 \si \, .
}
On flat space $\si(x,y)= \half (x-y)^2$. 
For a homogeneous space of constant curvature then any $T_{\mu\nu\dots}(x,y)$, which 
transforms covariantly under all isometries as a tensor at $x$ and a scalar at $y$,
may be expanded in a basis formed by $\pr_\mu \si(x,y)$ and $g_{\mu\nu}(x)$
with coefficients functions of $\si(x,y)$ \allen. Thus we may write
\eqn\nabsi{
\nab_\mu \nab_\nu \si = g_{\mu\nu}\,  f(\si) +  \pr_\mu \si \pr_\nu \si
\, g(\si) \, .
}
Since, from \bis, $\nab_\mu \nab_\nu \si \pr^\nu \si = \pr_\mu \si$, 
we have
\eqn\fg{
 f(\si) + 2 \si g(\si) = 1 \, .
}
Imposing $[\nab_\si , \nab_\mu] \nab_\nu \si = - R^\rho{}_{\nu\si\mu}\pr_\rho
\si$ on \nabsi\ gives using \riem
\eqn\dfg{
f'(\si) -  f(\si)g(\si) = \mp \rho^2 \, ,
}
or defining $\si = \half \theta^2 /\rho^2$, so that $\theta/\rho$ is the geodesic
distance from $x$ to $y$,
\eqn\df{
\theta {\d \over \d \theta} f - f(1-f) = \mp \theta^2 \, .
}
With the boundary condition $f(0)=1$ the solutions are
\eqn\solf{
f = {\theta \over \tan \theta} \, , \ \ {\theta \over \tanh \theta} \, , \qquad
g = {\rho^2\over \theta^2} \big ( 1 - f \big ) \, .
}
For the sphere $\theta$ is the angular separation of $x$ and $y$ and we may
restrict $0\le \theta\le \pi$, $\theta=\pi$ corresponds to $x$ and $y$ being
antipodal points.\foot{If we define the bi-scalar $\Delta(x,y) = \det(
- \pr_\mu \si(x,y) \! {\overleftarrow \pr}_{\! \alpha} )/\sqrt{g(x)}\sqrt{g(y)}$
then solving the equations in \dewitt\ gives $\Delta=(\theta/\tan \theta)^{d-1},
\ (\theta/\tanh \theta)^{d-1}$. For the spherical case the divergence at
$\theta=\pi$ is a reflection of this being a caustic.}

For any bi-tensor, $T_{\mu\nu\dots,\alpha\beta\dots}(x,y)$, then the
basis for expansion may be extended to include $\si(x,y) \!
{\overleftarrow \pr}_{\! \alpha}$, which transforms as a covariant vector at
$y$, and also the bi-vector $g_{\mu\alpha}(x,y) = - \pr_\mu \si(x,y) \!
{\overleftarrow \pr}_{\! \alpha}$ as well as $g_{\alpha\beta}(y)$. 
In practice it is more convenient to
introduce $I^\mu{}_{\alpha}(x,y)$ which gives parallel transport of
vectors along the geodesic from $y$ to $x$. This is defined by
\eqn\defI{
\si^\rho \nab_\rho I^\mu{}_{\alpha} = 0 \, , \qquad
 I^\mu{}_{\alpha}(x,x) = \de^\mu{}_{\!\alpha} \, .
}
For homogeneous spaces as considered here we may write the general form
\eqn\homI{
I_{\mu\alpha} = g_{\mu\alpha} \, a(\si) - \pr_\mu \si \, \si \!
{\overleftarrow \pr}_{\! \alpha} \, b(\si) \, .
}
Inserting this into \defI\ and using \bis,\nabsi,\fg\ gives
\eqn\diffI{
a'(\si) + g(\si)a(\si)= 0 \, , \qquad {\d \over \d \si} \big (
a(\si) + 2 \si b(\si) \big ) = 0 \, .
}
Solving these with the boundary condition $a(0)=1$ gives
\eqn\solab{
a = {\sin \theta \over \theta} \, , \ \ {\sinh \theta \over \theta} \, , \qquad
b = {\rho^2\over \theta^2} \big ( 1 - a \big ) \, .
}
An important consistency check, which follows directly from \defI, is that
\eqn\II{
I^\mu{}_\alpha I_{\mu\beta} = g_{\alpha\beta} \, , \qquad
I_{\mu\alpha}I_\nu{}^\alpha = g_{\mu\nu} \, ,
}
and further we have
\eqn\Isi{
\pr_\mu \si(x,y) I^\mu{}_\alpha(x,y) = - \si(x,y) \!
{\overleftarrow \pr}_{\! \alpha} \, .
} 

In the following sections these results are used to to obtain a natural form
for the tensorial expansions of two point functions whose coefficients are
functions of the biscalar $\theta(x,y)$. It is convenient to use, as well as
the parallel transport bi-vector $I_{\mu\alpha}$, a basis formed by
$\hx_\mu$ and $\hy_\alpha$, which are unit vectors at $x, \, y$, given by
\eqn\defxy{
\pr_\mu \theta = \rho \hx_\mu , \quad  \theta
{\overleftarrow \pr}_{\! \alpha} = \rho \hy_\alpha \, , \qquad
\hx_\mu I^\mu {}_{\! \alpha} = - \hy_\alpha \, , \quad
I_\mu{}^{\! \alpha} \hy_\alpha = - \hx_\mu \, ,
}
using \II\ and \Isi. With these definitions \nabsi\ and \solf\ then become
\eqn\diffx{
\nab_\mu \hx_\nu = \rho \cot \theta ( g_{\mu\nu} - \hx_\mu \hx_\nu ) \, , \
\rho \coth \theta ( g_{\mu\nu} - \hx_\mu \hx_\nu ) \, ,
}
and also from \homI\ and \solab\ we have,
\eqn\diffy{\eqalign{
\pr_\mu \hy_\alpha = {}& 
 - \rho \cosec\theta ( I_{\mu\alpha} + \hx_\mu \hy_\alpha) \, ,  
\  - \rho \cosech\theta ( I_{\mu\alpha} + \hx_\mu \hy_\alpha) \, ,  \cr
\nab_\mu I_{\nu\alpha} = {}& \rho \tan \half \theta
( g_{\mu\nu} \hy_\alpha + \hx_\nu I_{\mu\alpha} ) \, , \
- \rho \tanh \half \theta ( g_{\mu\nu} \hy_\alpha + \hx_\nu I_{\mu\alpha} ) \, .\cr}
}
For the spherical case  $\hx_\mu, \hy_\alpha$ and $I_{\mu\alpha}$ are singular 
when $\theta=\pi$ and $x,y$ are antipodal points.
Using \diffx\ it is easy to verify
\eqn\nabsq{
\nab^2 F(\theta) = \rho^2 \Big ( {\d^2 \over \d \theta^2} + 
{d-1 \over \tan \theta} \, {\d \over \d \theta} \Big ) F(\theta) \, , \ \
\rho^2 \Big ( {\d^2 \over \d \theta^2} +
{d-1 \over \tanh \theta} \, {\d \over \d \theta} \Big ) F(\theta) \, \, .
}

\newsec{Conformal Symmetries}

At a renormalisation group fixed point quantum field theories are additionally
constrained by conformal invariance. Conformal symmetry may be extended to
spaces of non zero curvature by seeking conformal Killing vectors which 
satisfy
\eqn\kill{
\nab_\mu v_\nu + \nab_\nu v_\mu = 2w \, g_{\mu\nu} \, , \qquad 
w = {1\over d} \nab{\cdot v} \, .
}
On flat space the solutions are well known. We show here how it is similarly possible
to solve \kill\ on homogeneous spaces of constant curvature without any restriction
to particular choices of coordinates.\foot{However writing the metric
in a conformally flat form $g_{\mu\nu}=\Omega^2 \de_{\mu\nu}$ then if
$v^\mu$ is a solution of the flat space conformal Killing equation it
remains a solution for the metric $g_{\mu\nu}$.} From \kill\ and \riem\ we first
derive an expression for two covariant derivatives of $v$
\eqn\ddv{
\nab_\si \nab_\mu v_\nu = \mp \rho^2 (g_{\si\mu}v_\nu - g_{\si\nu}v_\mu )
- g_{\si\mu} \pr_\nu w + g_{\si\nu} \pr_\mu w + g_{\mu\nu} \pr_\si w \, .
}
Also by contracting with $\nab^\mu \nab^\nu$ we may obtain\foot{This demonstrates
that there is no solution with $w=1$ if $R\ne 0$ so that there is no
dilation current $j^\mu = T^{\mu\nu}v_\nu$ satisfying $\nab_\mu j^\mu
= g_{\mu\nu} T^{\mu\nu}$ except in flat space.}
\eqn\delw{
(d-1) \nab^2 w + R w =0 \, ,
}
and by using this with \kill\ together with $(d-2) \nab_\nu w = -(\nab^2 +
{1\over d} R ) v_\nu$, which also follows from \kill, we find for $d\ne 2$
\eqn\nabw{
\nab_\mu \nab_\nu w = \mp \rho^2 g_{\mu\nu} w \, .
}
This is easily extended to arbitrarily many derivatives
\eqn\dddw{ \eqalign{
\nab_{(\mu_1}\dots \nab_{\mu_{2n})} w
={}& (\mp \rho^2)^n g_{(\mu_1\mu_2} \dots g_{\mu_{2n-1}\mu_{2n})} w \, , \cr
\nab_{(\mu_1}\dots \nab_{\mu_{2n+1})} w = {}&
(\mp \rho^2)^n g_{(\mu_1\mu_2} \dots g_{\mu_{2n-1}\mu_{2n}} 
\pr_{\mu_{2n+1})} w \, , \cr}
}
and then applying the covariant Taylor expansion, which for any scalar $F$
takes the form \Full
\eqn\taylor{
\!\! F(x) = \sum_{n=0} \, {1\over n!} \si^{\alpha_1} \dots \si^{\alpha_n}
(-1)^n \nab_{(\alpha_1} \dots \nab_{\alpha_n)} F(y) \, , \quad
\si^\alpha(x,y) = g^{\alpha\beta}(y) \,\si(x,y) \! {\overleftarrow \pr}_{\! \beta} \, ,
}
leads to
\eqn\wsol{
w(x) 
= \cos \theta \, w(y) - {1\over \rho} \sin \theta \, \hy{\cdot \pr}w(y)\, , \ 
\cosh \theta \, w(y) - {1\over \rho} \sinh \theta \, \hy{\cdot \pr}w(y)  \, .
}
This result demonstrates that $w(x)$ is determined completely by the values of
$w$ and $\pr_\alpha w$ at any arbitrary $y$ so that there are $d+1$ 
independent forms for $w$ (for $R>0$ from \delw\ these correspond to the $d+1$
normalisable eigenvectors of $-\nab^2$ with eigenvalue $d\rho^2$). If we define
\eqn\defom{
\nab_{[\mu} v_{\nu]} = - \omega_{\mu\nu} \, ,
}
then from \ddv\ and \nabw\ we get
\eqn\dddv{ \eqalign{
\!\!\!\nab_{(\mu_1}\dots \nab_{\mu_{2n})} v_\nu
={}& (\mp \rho^2)^n \Big \{\big ( g_{(\mu_1\mu_2} \dots g_{\mu_{2n-1}\mu_{2n})} v_\nu 
- v_{(\mu_1} g_{(\mu_2\mu_3} \dots g_{\mu_{2n})\nu} \big ) \cr
{}& \pm {1\over \rho^2}\big (g_{(\mu_1\mu_2}\dots g_{\mu_{2n-1}\mu_{2n})} \pr_\nu w
- 2 \pr_{(\mu_1}\! w \,g_{(\mu_2\mu_3} \dots g_{\mu_{2n})\nu} \big )\Big \} \, , \cr
\!\!\!\nab_{(\mu_1}\dots \nab_{\mu_{2n+1})} v_\nu = {}&
- (\mp \rho^2)^n \big ( g_{(\mu_1\mu_2} \dots g_{\mu_{2n-1}\mu_{2n}}
\omega_{\mu_{2n+1})\nu} - g_{(\mu_2\mu_3} \dots g_{\mu_{2n+1})\nu} w \big )\, , \cr}
}
Applying the Taylor expansion \taylor\ to $I_\mu{}^\alpha(x,y)v_\alpha(y)$ leads to
\eqnn\vsol
$$ \eqalignno{
v_\mu (x) = {}& I_\mu{}^\alpha \bigg \{ {\cos \theta\atop \cosh \theta}
\, v_\alpha(y) \pm {(1-\cos \theta)\atop (\cosh \theta - 1)} \Big (
\hy{\cdot v}(y) \, \hy_\alpha \mp {1\over \rho^2} \pr_\alpha w(y) \pm {2\over \rho^2}
\hy{\cdot \pr}w(y) \, \hy_\alpha \Big ) \cr
& \qquad  - {1\over \rho}\, {\sin \theta\atop \sinh \theta}
\, \big ( \omega_{\alpha\beta}(y)\hy^\beta + w(y)\hy_\alpha \big ) \bigg \} \, , 
& \vsol \cr}
$$
It is straightforward to verify, using \diffx\ and \diffy, that \vsol\ and \wsol\
satisfy \kill. Besides $w$ and $\pr_\alpha w$ the general expression for $v_\mu(x)$
is determined by $v_\alpha$ and $\omega_{\alpha\beta}$ at some arbitrary $y$,
giving $\half (d+1)(d+2)$ linearly independent vectors. 
Taking $y=0$ it is easy to see that \vsol\ and \wsol\
reduce to the standard results for flat space with $v_\mu(x)$
quadratic and $w(x)$ linear in $x$. From the definition \defom\ we may also
derive
\eqn\omsol{
\omega_{\mu\nu} (x) =  I_\mu{}^\alpha I_\nu{}^\beta
\bigg \{  \omega_{\alpha\beta}(y) \mp {\sin \theta\atop \sinh \theta}
2 \hy_{[\alpha} \Big ( \rho v_{\beta]}(y) \pm {1\over \rho} \pr_{\beta]} w(y)\Big ) 
\pm {(1-\cos \theta)\atop (\cosh \theta - 1)} 2 \hy_{[\alpha}
\omega_{\beta]\gamma}\hy^\gamma \bigg \} \, .
}

Using the above solutions for conformal Killing vectors, which are specified
by $v^\alpha, \omega^\alpha{}_{\!\beta}, w , \pr^\alpha \! w$, it is straightforward
to calculate the Lie algebra of the associated vector fields,
\eqn\Lie{
[v_1 , v_2]^\mu = - v_3{}^{\! \mu} \quad \Rightarrow \quad
v_3{}^{\! \mu} = \omega_1{}^{\! \mu}{}_{\!\nu} v_2{}^{\! \nu}  - v_1{}^{\! \mu} w_2
- ( 1 \leftrightarrow 2 ) \, .
}
Using \vsol, \wsol\ and \omsol\ this is identical with the algebra of matrices $W$,
\eqn\LieW{
[W_1 , W_2 ] = W_3 \, ,
}
where
\eqn\defW{
W^A{}_{B} = \pmatrix{\omega^\alpha{}_{\!\beta} & \rho v^\alpha \pm {1\over \rho}
\pr^\alpha w & \pm {1\over \rho} \pr^\alpha w \cr \mp  \rho v_\beta - {1\over \rho}
\pr_\beta w & 0 & w \cr {1\over \rho} \pr_\beta w & w & 0 \cr} \, .
}
Since, for
\eqn\defG{
G_{AB} =  \pmatrix{ \de_{\alpha\beta} & 0 & 0\cr 0 & \pm 1 & 0 \cr 0 & 0 & \mp 1}\, ,
}
we have
\eqn\anti{
W_{AB} = G_{AC} W^C{}_{B} = - W_{BA} \, ,
}
it is clear that \LieW\ corresponds in both cases to the Lie algebra of the
$d$-dimensional conformal group $O(d+1,1)$. 
If we restrict to Killing vectors for which $w=0$ it
is also evident that the algebra reduces to that for the isometry groups
$O(d+1)$ or $O(d,1)$ for $S^d$ or $H^d$ respectively as expected.

For construction of conformally covariant expressions for correlation functions
it is necessary to construct bi-scalar functions  of $x,y$ which transform
homogeneously under such conformal transformations. In consequence we consider
the generalisation $s$ of the flat space $(x-y)^2$ which is required to satisfy
\eqn\defs{
\big ( v^\mu(x) \pr_\mu + v^\alpha(y)\pr_\alpha \big ) s = 
\big ( w(x) + w(y) \big ) s \, .
}
Writing $s(\theta)$ the left hand side of \defs\ involves
\eqn\vsi{ 
\rho \big (v^\mu(x) \hx_\mu + v^\alpha(y) \hy_\alpha\big) =  \tan \half \theta
\big ( w(x) + w(y) \big ) \, , \ \ \tanh \half \theta
\big ( w(x) + w(y) \big ) \, ,
}
from \vsol\ and \wsol. Hence \defs\ becomes
\eqn\eqf{
{\d \over \d \theta} s = \cot \half \theta \, s \, , \ \
\coth  \half \theta \, s \, .
}
Imposing the boundary condition that $s\sim \theta^2/\rho^2$ as $x\to y$ gives the
solution
\eqn\sols{
\rho^2 s = 2 (1-\cos \theta ) \, , \ \  2 (\cosh \theta - 1 ) \, ,
}
so that $\sqrt s $ may be interpreted as the chordal distance between $x$ 
and $y$.  For the hyperbolic case it is useful also to define
\eqn\sbar{
\rho^2 \bs = \rho^2 s + 4 = 2(\cosh \theta + 1 ) \, .
}
Any power $s^{-\lambda}$ also transforms homogeneously under
conformal transformations, as in \defs, and \nabsq\ gives
\eqn\nabss{
\big ( \nab^2 \pm \rho^2 \lambda (\lambda - d +1 ) \big ) s^{-\lambda}
= \lambda (2\lambda + 2 - d) s^{-\lambda -1 } \, .
}

Using the bi-scalar $s$ we may further define a bi-vector by
\eqn\definv{
- \pr_\mu \ln s(x,y) \! {\overleftarrow \pr}_{\! \alpha} = {2\over
s(x,y)} \, \I_{\mu\alpha}(x,y) \, , 
}
which gives
\eqn\invxy{
\I_{\mu\alpha} = I_{\mu\alpha} + 2 \hx_\mu \hy_\alpha \, .
}
$\I_{\mu\alpha}$ generalises the inversion tensor to spaces of constant curvature 
and, from \II\ and \Isi, we have $\I^\mu{}_\alpha \I_{\mu\beta} 
= g_{\alpha\beta}$. From its definition \definv\ and \defs, with \defom, we have
\eqn\tinv{
\big ( v^\mu(x) \pr_\mu + v^\alpha(y)\pr_\alpha \big ) \I_{\mu\alpha}(x,y)
= \omega_\mu{}^\nu(x) \I_{\nu\alpha}(x,y) + \omega_\alpha{}^\beta(y)
\I_{\mu\beta}(x,y) \, .
}
For the positive curvature case since
$s(x,y)$ is single valued for arbitrary $x,y$ \definv\ also ensures that  
$\I_{\mu\alpha}$ is well defined at $\theta=\pi$.

For later reference it is useful also to define related bi-vector
\eqn\defhI{
\hI_{\mu\alpha} = - \half \pr_\mu s {\overleftarrow \pr_{\!\alpha}}  = I_{\mu\alpha}
+ (1-\cos \theta) \hx_\mu \hy_\alpha \, , \quad
I_{\mu\alpha} - (\cosh \theta - 1 ) \hx_\mu \hy_\alpha \, ,
}
which satisfies
\eqn\DhI{
\nab_{\mu}\hI_{\nu\alpha} =  g_{\mu\nu} \sin\theta \, \hy_\alpha \, , \
- g_{\mu\nu} \sinh \theta \, \hy_\alpha \, .
}

\newsec{Spinors}

For spinor fields we may define, using vielbeins as usual, Dirac matrices
$\gamma_\mu(x)$ satisfying $\{ \gamma_\mu, \gamma_\nu \} = 2g_{\mu\nu}$. The
essential geometrical object for our purposes is the bispinor $I(x,y)$ which
acting on spinor at $y$ parallel transports it along the geodesic  from $y$ to $x$.
This satisfies
\eqn\pI{
\hx^\mu \nab_\mu I = 0 \, , \qquad I(x,x) = 1 \, ,
}
where $\nab_\mu = \pr_\mu + \omega_\mu$ is the spinor covariant derivative.
For a homogeneous space of constant curvature we follow a similar approach to
that of Allen and L\"utken in four dimensions \Allen\ and express the covariant
derivative in a form compatible with \pI,
\eqn\derI{
\nab_\mu I(x,y) = - \half \rho \, \alpha(\theta) ( \gamma_\mu \gamma{\cdot \hx}
- \hx_\mu ) I(x,y) \, .
}
Using \diffx\ we may then find
\eqn\curI{\eqalign{ \!\!\!\!\!
\nab_{[\mu}\nab_{\nu]} I(x,y) = {}& \half \alpha ( \alpha + \cot \theta )
\gamma_{[\mu} \gamma_{\nu]} I(x,y) - \half ( \alpha' - \alpha^2 - \alpha \cot \theta)
\hx_{[\mu} \gamma_{\nu]}  \gamma{\cdot \hx} I(x,y) \, , \cr
{}& \!\! \half \alpha ( \alpha + \coth \theta )
\gamma_{[\mu} \gamma_{\nu]} I(x,y) - \half ( \alpha' - \alpha^2 - \alpha \coth \theta)
\hx_{[\mu} \gamma_{\nu]}  \gamma{\cdot \hx} I(x,y) \, . \cr}
}
For a spinor $\psi$ the commutator of covariant derivatives is given by
\eqn\dpsi{
[\nab_\mu , \nab_\nu ] \psi = \half R_{\si\rho \mu\nu} s^{\si\rho} \psi
= \pm \rho^2 s_{\mu\nu} \psi \, , \qquad s_{\mu\nu} = \half \gamma_{[\mu}
\gamma_{\nu]} \, ,
}
using \riem. Applying this integrability condition to \curI\ leads to equations for 
$\alpha$ which are readily solved giving
\eqn\aI{
\alpha = \tan \half \theta \, , \ \tanh \half \theta \, ,
}
and hence \derI\ becomes
\eqn\derIa{ \eqalign{
\nab_\mu I(x,y) = {}&
- \rho \tan\half \theta \,\quar [ \gamma_\mu , \gamma{\cdot \hx} ]
I(x,y) \, , \ - \rho \tanh\half \theta \, 
\quar [ \gamma_\mu ,\gamma{\cdot \hx} ] I(x,y)\, , \cr
I(x,y) {\overleftarrow \nab}{}_{\! \alpha} 
= {}&\rho \tan\half \theta \, I(x,y) \quar [\gamma_\alpha, \gamma{\cdot \hy} ]\, , \
\rho \tanh\half \theta \, I(x,y) \quar [ \gamma_\alpha, \gamma{\cdot \hy} ] \, . \cr}
}
The spinor parallel transport $I(x,y)$ is easily seem to satisfy
\eqn\invI{
I(x,y) I(y,x) = 1 \, ,
}
and also
\eqn\tgamma{
I(x,y) \gamma_\alpha  I(y,x) = \gamma_\nu I^\nu{}_{\! \alpha}(x,y) \, ,
}
which may be verified my applying a covariant derivative to both sides, using
\derIa\ with \aI\ as well as \diffy.

It is also useful to define
\eqn\Iinv{
\I(x,y) =  \gamma{\cdot \hx} I(x,y) = - I(x,y) \gamma{\cdot \hy} \, ,
}
which plays the role of the inversion matrix on spinors analogous to the
inversion tensor \invxy. It is easy to see, using \tgamma, that
\eqn\twoI{
\tr\big (\gamma_\mu I(x,y) \gamma_\alpha I(y,x) \big )  =
2^{{1\over 2}d}I_{\mu\alpha}(x,y)\, , \quad
\tr\big (\gamma_\mu \I(x,y) \gamma_\alpha \I(y,x) \big ) = 
2^{{1\over 2}d} \I_{\mu\alpha}(x,y)\, .
}
and from \Iinv\
\eqn\II{
\I(x,y) \I(y,x) = - 1 \, .
}
{}From \derIa\ and  \aI\ we may also easily obtain
\eqn\derII{ \eqalign{
\nab_\mu \I(x,y) = {}&
\rho \cot\half \theta \,\quar [ \gamma_\mu , \gamma{\cdot \hx} ]
\I(x,y) \, , \ \rho \coth\half \theta \,
\quar [ \gamma_\mu ,\gamma{\cdot \hx} ] \I(x,y)\, , \cr
\I(x,y) {\overleftarrow \nab}{}_{\! \alpha}
= {}&- \rho \cot\half \theta \, I(x,y) \quar [\gamma_\alpha, \gamma{\cdot \hy} ]\, , \
- \rho \coth\half\theta \, I(x,y)\quar [ \gamma_\alpha, \gamma{\cdot \hy} ] \, . \cr}
}
The importance of $\I$ is that it transforms homogeneously under conformal
transformations similarly to \tinv\ since from \derII, using \vsol\ and \omsol\
and the spin matrices defined in \dpsi,
\eqn\conII{
v^\mu(x) \nab_\mu \I(x,y) + \I(x,y) {\overleftarrow \nab}{}_{\! \alpha}v^\alpha(y)
= {\ts{1\over 2}} \omega_{\mu\nu}(x) s^{\mu\nu} \I(x,y)
-  \I(x,y) {\ts{1\over 2}} \omega_{\alpha\beta}(y) s^{\alpha\beta}  \,.
}

Using the above results it is easy to construct a spinor Green function satisfying
\eqn\GS{
\gamma^\mu \nab_\mu S(x,y) = \de^d(x,y) \, ,
}
which are then given by
\eqn\Stwo{\eqalign{
S(x,y)_+ = {}& {1\over S_d} \, {1\over s^{{1\over 2}(d-1)}}\, \I(x,y) \, , \cr
S(x,y)_- = {}&  {1\over S_d} \bigg (  {1\over s^{{1\over 2}(d-1)}} \, \I(x,y)
\pm  {1\over \bs^{{1\over 2}(d-1)}} \, I(x,y) \bigg ) \, , \cr}
}
where 
\eqn\Sd{
S_d = {2\pi^{{1\over 2}d}\over \Gamma(\half d)} \, .
}
In the negative curvature case there are two inequivalent Green functions
which correspond to two alternative boundary conditions and are appropriate for 
different spinor representations \Allen\foot{In \Callan\ the $\pm$ is replaced
in four dimensions by an arbitrary phase $e^{i\xi\gamma_5}$.}. 
The boundary conditions in either case violate chiral symmetry.

Following \Burges\ and \Allen\ we may also construct a Killing bispinor which
satisfies the Killing spinor equation at $x$ and at $y$. Consistent with \dpsi\
a Killing spinor $\ep$ satisfies
\eqn\Kills{
\nab_\mu \ep^{\pm} = \pm \rho \half i \gamma_\mu \ep^{\pm} \, , \
\pm \rho \half \gamma_\mu \ep^{\pm} \, .
}
Such spinors allow the construction of solutions of the Dirac equation in terms of
those for a scalar field since if
\eqn\pp{
\psi = \big ( (- \gamma{\cdot \pr} + \lambda^\pm \rho)\phi \big ) \ep^\pm \, , 
}
then
\eqn\ppp{
\Big (-\nab^2 + {d-2\over 4(d-1)}R \Big ) \phi = 0 \ \Rightarrow \ \gamma{\cdot \nab}
\psi = 0 \quad \hbox{if} \quad \lambda^\pm = \mp \half i (d-2) , \
\mp \half (d-2) \,.
} 
Writing the  Killing bispinor in the general form
$\S(x,y) = p(\theta) I(x,y) + q (\theta) \I(x,y)$ and using \derIa,\aI\ and \derII\
then leads to differential equations for $p,q$ which are easily solved,
\eqn\solS{
\S^\pm = \cos \half \theta I \pm i \sin \half \theta \I \, , \
\cosh \half \theta I \pm \sinh \half \theta \I \, .
}
These satisfy
\eqn\nabS{
\nab_\mu \S^\pm = \pm \rho \half i \gamma_\mu \S^\pm \, , \
\pm \rho \half \gamma_\mu \S^\pm \, , \qquad 
\S^\pm {\overleftarrow \nab_{\!\alpha}} = \mp \rho \half i \S^\pm \gamma_\alpha 
\, , \ \mp \rho \half S^\pm \gamma_\alpha \, ,
}  
and hence for any solution of \Kills\ we may write
\eqn\ssol{
\ep^{\pm} (x) = \S^\pm (x,y) \ep^{\pm} (y) \, ,
} 
so that the general solution is determined by a fixed spinor at any arbitrary
point $y$. This gives solutions coinciding with those found with particular
coordinate choices \pope. It is of course possible to form Killing vectors
from Killing spinors, this is exemplified here by
\eqn\SKill{ \eqalign{
\nab_\mu \S^-(y,x) \gamma_\nu \S^+(x,y)  = {}& g_{\mu\nu}\, \rho i
\S^-(y,x) \S^+(x,y) \, , \ g_{\mu\nu}\, \rho \S^-(y,x) \S^+(x,y) \, , \cr
\nab_{(\mu} \S^+(y,x) \gamma_{\nu)} \S^+(x,y) = {}& 0 \, . \cr}
}
Furthermore  from \solS
\eqn\SSS{\eqalign{
\!\!\! \S^-(y,x) \gamma_\mu \S^+(x,y) = {}& I_\mu{}^\alpha \bigg \{
\gamma_\alpha \mp  {(1-\cos \theta)\atop (\cosh \theta - 1)} 
\gamma{\cdot \hy} \, \hy_\alpha  - {i \sin \theta \atop \sinh \theta} \,
\hy_\alpha \bigg \} \, , \cr
\!\!\! \S^+(y,x)\gamma_\mu \S^+(x,y) = {}& I_\mu{}^\alpha \bigg \{{\cos \theta \atop
\cosh \theta } \, \gamma_\alpha
\pm {(1- \cos \theta )\atop (\cosh \theta - 1)}
\gamma{\cdot \hy} \, \hy_\alpha -  {  i \sin \theta \atop \sinh \theta}\, 
\half [\gamma_\alpha, \gamma_\beta]\hy^\beta \bigg \} \, , \cr}
}
which are in accord with the general form exhibited in \vsol.

\newsec{Two Point Functions on Spaces of Constant Curvature}

In this section we obtain a general decomposition for the two point function
of the energy momentum tensor for arbitrary dimension $d$ using the
geometrical results of the previous section. For simplicity we confine our
attention here to the positive curvature sphere $S^d$ although they are
easily extended to the negative curvature case using the usual correspondence
between trigonometric and hyperbolic functions. The critical formulae are displayed
in appendix A.

It is convenient  first to consider $\Gamma_{\mu\nu}(x,y)$ which is a symmetric
tensor at $x$ and has the form
\eqn\GTO{
\Gamma_{\mu\nu}(x,y) = \hx_\mu \hx_\nu A(\theta) + g_{\mu\nu}B(\theta) \, .
}
Imposing the conservation equation $\nab^\mu \Gamma_{\mu\nu} (x,y) = 0$, using \diffx\
easily gives
\eqn\DE{
A' + B ' + (d-1) \cot \theta \, A = 0 \, .
}
Imposing also the traceless condition $g^{\mu\nu}(x) \Gamma_{\mu\nu}(x,y) = 0$,
or $A + d B = 0$, leads to a solution
\eqn\solD{
\Gamma_{\mu\nu}(x,y) = K (\sin \theta )^{-d} ( d  \hx_\mu \hx_\nu - g_{\mu\nu} ) \, .
}
However such a solution is unacceptable due to the singularity at $\theta=\pi$,
although there is no difficulty with the corresponding solution in the  hyperbolic
case when $\sin \theta \to \sinh \theta$.

The corresponding results for the two point
function of the energy momentum tensor are less trivial. The  general form for
a bi-tensor $\Gamma_{\mu\nu,\alpha\beta}(x,y)$, symmetric
in $\mu\nu$ and $\alpha\beta$,  may be reduced to six independent functions
of $\theta$
\eqn\twoG{\eqalign{
\Gamma_{\mu\nu,\alpha\beta} = {}& \hx_\mu\hx_\nu \hy_\alpha \hy_\beta \, R
+ \big ( I_{\mu\alpha} \hx_\nu \hy_\beta + \mu \leftrightarrow \nu ,
\alpha \leftrightarrow \beta \big ) S + \big (  I_{\mu\alpha} I _{\nu\beta}
+ I_{\mu\beta}  I_{\nu\alpha} \big ) T \cr
{}& + \big (  \hx_\mu\hx_\nu g_{\alpha\beta} \, U_1 + \hy_\alpha \hy_\beta
g_{\mu\nu} \, U_2 \big ) + g_{\mu\nu}  g_{\alpha\beta} \, V \, . \cr}
}
For symmetry, $\Gamma_{\mu\nu,\alpha\beta}(x,y) = \Gamma_{\alpha\beta,\mu\nu}(y,x)$,
it is clearly necessary that
\eqn\UU{
U_1 = U_2 = U \, ,
}
and imposing tracelessness, $g^{\mu\nu} \Gamma_{\mu\nu,\alpha\beta}
= \Gamma_{\mu\nu,\alpha\beta} g^{\alpha\beta} = 0$, further requires
\eqn\trace{
P_1 \equiv R - 4S + d U = 0 \, , \qquad P_2 \equiv 2T + U + dV = 0 \, .
}
In two dimensions the basis used in \twoG\ is overcomplete as a consequence
of the identity
\eqn\twocom{
\big ( I_{\mu\alpha} \hx_\nu \hy_\beta + \mu \leftrightarrow \nu ,
\alpha \leftrightarrow \beta \big ) + I_{\mu\alpha} I _{\nu\beta}
+ I_{\mu\beta}  I_{\nu\alpha} + 2 \big ( \hx_\mu\hx_\nu g_{\alpha\beta}
+ \hy_\alpha \hy_\beta g_{\mu\nu} 
-  g_{\mu\nu}  g_{\alpha\beta} \big ) = 0 \, .
}

In order to impose the conservation equation $\nab^\mu 
\Gamma_{\mu\nu,\alpha\beta} (x,y) = 0 $ we make use of \diffx\ as well as \diffy\
giving in terms of the expansion \twoG,
\eqn\conG{\eqalign{
& R'-2S'+ U_2{\!}' + (d-1)(\cot \theta \, R + 2 \tan  \half \theta \, S )
+ 2 \cot \half \theta \, S - 2\cosec  \theta \, U_2 =  0 \, , \cr
& S'-T'  + d \cot \theta \, S + d \tan  \half \theta \, T - 
\cosec  \theta \, U_2 = 0 \, , \cr
& U_1{\!}' + V' + (d-1) \cot \theta \, U_1 - 2 \cosec  \theta \, S + 2
\tan  \half \theta \, T =  0 \, . \cr}
}
Requiring the symmetry condition \UU\ of course guarantees that
$\Gamma_{\mu\nu,\alpha\beta} {\overleftarrow \nab}{}^\alpha = 0$ as well.
For subsequent use it is convenient to rewrite \conG. With the definitions \trace,
and assuming \UU, we have
\eqn\PPcon{
P_1\!{}' +  P_2\!{}'  =  - (d-1) \cot \theta P_1 \, .
}
In addition defining
\eqn\defQ{
Q = 2T + {d-1\over d}(R - 4 S) \, ,
}
then gives (equivalent equations for $d=4$ were found in \MottdS),
\eqna\conQ
$$\eqalignno{
Q' + d \cot \theta \, Q + {1\over d}\,  P_1\!{}' ={}& - 
2d  \cosec \theta \, (S-T) \, , 
&\conQ{a} \cr
(S-T)' + d \cot \theta \, (S-T) = {}& - \cosec \theta \, {1\over d-1}
\big (Q +(d-2)(d+1)T \big ) \cr
{} & + \cosec \theta \, {1\over d}\, P_1  \, . &\conQ{b} \cr}
$$
Hence, if $P_1 = P_2 =0$ there are just two independent equations and in this
case knowing $Q$ in general determines $R,S,T$.

In two dimensions these relations provide stronger constraints since on the right 
hand side of \conQ{b} $4T+R-4S = 2Q$ (it should be noted that $R$ and $T-S$ are
independent of ambiguities that arise from \twocom). 
Taking the sum and difference of  \conQ{a,b} then gives 
\eqna\RSTtwo
$$\eqalignno{
R' + 2\cot \half \theta \, R + P_1\!{}' ={}& 2\cosec\theta \, P_1 \, , 
& \RSTtwo{a} \cr
(R-8S+8T)' - 2\tan \half \theta \, (R-8S+8T) + P_1\!{}'
= {}& - 2 \cosec\theta \, P_1  \, . & \RSTtwo{b} \cr}
$$
If $P_1=0$ there is a straightforward solution
\eqn\Rtwo{
R = C\, {\rho^4 \over \sin^4 \! \half \theta} \, , \qquad R-8S+8T = 0 \, ,
}
where we have imposed the condition that there are no singularities at $\theta=\pi$.

Explicit forms satisfying \conG\ may be found in terms of the two
independent terms displayed for $d=4$ in \TTfour. Initially we consider the
contribution corresponding to spin zero intermediate states
\eqn\TTzero{
\Gamma_{0,\mu\nu,\alpha\beta} (x,y) =
\big ( \nab_\mu \nab_\nu - g_{\mu\nu} \nab^2 - (d-1)\rho^2 g_{\mu\nu} \big )
F_0(\theta) \big ( {\overleftarrow \nab_{\!\alpha}}\! {\overleftarrow \nab_{\!\beta}}
- {\overleftarrow \nab}{}^2 g_{\alpha\beta} - (d-1)\rho^2 g_{\alpha\beta}\big )
\, .
}
To evaluate this we first write
\eqn\AB{
F_0(\theta) \big ( {\overleftarrow \nab_{\!\alpha}}\! {\overleftarrow \nab_{\!\beta}}
- {\overleftarrow \nab}{}^2 g_{\alpha\beta} - (d-1)\rho^2 g_{\alpha\beta}\big )
=  \hy_\alpha \hy_\beta A + g_{\alpha\beta} B  \, ,
}
where $A(\theta), B(\theta)$ are given by
\eqn\ABF{
A = \rho^2 \big ( F_0{\!}'' - \cot \theta \,  F_0{\!}'\big ) \, , \qquad
B = - \rho^2 \big ( F_0{\!}'' +  (d-2)\cot \theta \, F_0{\!}'  + (d-1) F_0 \big )\, .
}
These expressions automatically satisfy \DE.
Inserting \AB\ into \TTzero, for general $A,B$, further gives
\eqnn\ABzero
$$\eqalignno{
R_0 = {}& \rho^2 \big ( A'' - ( \cot \theta + 4 \cosec  \theta ) A' + 4 \cosec  \theta
\cot \half \theta \, A \big ) \, , \cr
S_0 = {}& \rho^2 \big ( - \cosec  \theta A' + \cosec \theta \cot \half \theta \, A 
\big ) \, , \qquad
T_0 = \rho^2 \cosec^2 \theta \, A \, , \cr
\!\!\!\! U_{0,1} = {}& \rho^2 \big ( B'' - \cot \theta \, B' \big ) \, , \quad 
U_{0,2} = - \rho^2 \big ( A'' + (d-2)
\cot \theta \, A' - (d-1) ( 2\cosec^2  \theta -1 ) A \big ) \, , \cr
V_0 ={}& - \rho^2 \big ( B''+(d-2)\cot \theta \, B'+(d-1) B +  2\cosec^2 \theta \, A 
\big ) \, . & \ABzero \cr}
$$
It is easy to verify that these satisfy \conG\ and, given the results for
$A,B$ in \ABF, that the symmetry condition \UU\ also holds. With the definitions
in \trace\ and also \UU\ we may note that
\eqn\PP{
P_1 + P_2 = - \rho^2 (d-1) \big ( \cot \theta (A' + dB') + A +dB \big ) \, ,
}
which may easily solved for $A+dB$ up to $A+dB \sim A+dB + k \cos \theta$,
reflecting the freedom of adding a solution of the homogeneous equation.
Consistency of \PP\ with \ABzero\ and \UU\ requires \PPcon.
Furthermore we may obtain from \ABF
\eqn\solF{
A + dB = - (d-1) \big ( \nab^2 F_0 + d \rho^2 F_0 \big ) \, ,
}
which may be inverted to determine $F_0$ if the constant $k$ in the solution
for $A + dB$ is chosen to make this orthogonal to $\cos \theta$ as this is a
zero mode for $\nabla^2 + d \rho^2$. Hence from \ABF\ we may find $A,B$
separately and then from \ABzero\ obtain $R_0,S_0,T_0,U_0,V_0$ satisfying 
the conservation equations. Since $P_1$ is determined by \PPcon\ these functions
reproduce the same $P_1,P_2$ in \trace\ as obtained from the original $R,S,T,U,V$.

Besides \TTzero\ we may use \CT\ to obtain a second solution of the conservation 
equations by taking, as in \TTfour,
\eqn\TTC{
\Gamma_{2,\mu\nu,\alpha\beta} (x,y) = \nab^\si \nab^\rho \big (
\E^C{}_{\!\! \mu\si\rho\nu,\alpha\gamma\de\beta}(x,y) F_2 (\theta) \big )
{\overleftarrow \nab}{}^\gamma {\overleftarrow \nab}{}^\de  \, ,
}
where for general $d$ the bi-tensor $\E^C$ may now be expressed explicitly
in terms of the parallel transport matrix $I$ defined in \defI\ by (if $d=3$ then
$\E^C=0$),
\eqnn\EC
$$\eqalignno{ \!\!\!\!\!\!
\E^C{}_{\!\! \mu\si\rho\nu,\alpha\gamma\de\beta} {}& = 
{\ts {1\over 12}}\big( I_{\mu\alpha}I_{\nu\beta}I_{\si\gamma}I_{\rho\de}
+ I_{\mu\de}I_{\si\beta}I_{\rho\alpha}I_{\nu\gamma}
- {\mu \leftrightarrow \si}, {\nu \leftrightarrow \rho} \big ) \cr
& \ + {\ts{1\over 24}}\big( I_{\mu\alpha}I_{\nu\gamma}I_{\rho\de}I_{\si\beta}
- {\mu \leftrightarrow \si}, {\nu \leftrightarrow \rho},
{\alpha \leftrightarrow \gamma} , {\de \leftrightarrow \beta}  \big ) \cr
\!\!\!\! \!\! & \ - {1\over d-2} \, {\ts {1\over 8}}\big(
g_{\mu\rho}g_{\alpha\delta}I_{\si\gamma}I_{\nu\beta}
+ g_{\mu\rho}g_{\alpha\delta}I_{\si\beta}I_{\nu\gamma}
- {\mu \leftrightarrow \si}, {\nu \leftrightarrow \rho},
{\alpha \leftrightarrow \gamma} , {\de \leftrightarrow \beta}  \big ) \cr
& \ + {1\over (d-2)(d-1)}\, {\ts {1\over 2}}\big(
g_{\mu\rho} g_{\nu\si} - g_{\mu\nu}g_{\si\rho} \big ) \big (
g_{\alpha\de}g_{\beta\gamma} - g_{\alpha\beta}g_{\gamma\de} \big )\, . & \EC \cr}
$$
For $x=y$, when $I_{\mu\alpha} = g_{\mu\alpha}$,
this reduces to the projector onto tensors with the symmetries and
traceless conditions of the Weyl tensor for general $d$ given \Csym. 
To evaluate \TTC\ we may first obtain
\eqn\TF{
\F_{\mu\si\rho\nu,\alpha\beta}(x,y) G(\theta) = \big (
\E^C{}_{\!\! \mu\si\rho\nu,\alpha\gamma\de\beta}(x,y) F_2 (\theta) \big )
{\overleftarrow \nab}{}^\gamma {\overleftarrow \nab}{}^\de  \, , 
}
where the bi-tensor $ \F_{\mu\si\rho\nu,\alpha\beta} $, which is a
symmetric traceless tensor at $y$, is given by
\eqn\FF{ \eqalign{
\F_{\mu\si\rho\nu,\alpha\beta} = {}& \hx_\mu \hx_\nu
\Big ( I_{\si\alpha} I_{\rho\beta} +  I_{\si\beta} I_{\rho\alpha} -
{2\over d}\, g_{\si\rho}g_{\alpha\beta} \Big ) - 
{\mu \leftrightarrow \si}, {\nu \leftrightarrow \rho}  \cr
{}& - {1\over d-2} \big ( g_{\mu\nu} X_{\si\rho,\alpha\beta}
- {\mu \leftrightarrow \si}, {\nu \leftrightarrow \rho} \big )  \cr
{}& + {4\over (d-2)(d-1)d } \, \big(
g_{\mu\rho} g_{\nu\si} - g_{\mu\nu}g_{\si\rho} \big ) \big ( d \, \hy_\alpha
\hy_\beta - g_{\alpha\beta} \big ) \, , \cr
X_{\si\rho,\alpha\beta} = I_{\si\alpha}I_{\rho\beta} {}& + I_{\si\beta}
I_{\rho\alpha} + \big ( I_{\si\alpha} \hx_\rho \hy_\beta +
{ \si \leftrightarrow \rho}, {\alpha \leftrightarrow \beta} \big ) +
{2\over d} \big ( 2 \hx_\si \hx_\rho - g_{\si\rho} \big ) g_{\alpha\beta} \, ,
\cr}
}
and
\eqn\Gres{
8 G = \rho^2 \big ( F_2{\!}''- \cot \theta \, F_2{\!}' - 2(d-1) \tan \half \theta \,  
F_2{\!}' + (d-1)(d-2) \tan^2 \! \half \theta \, F_2 \big ) \, .
}
Using \TF\ in  \TTC\ gives an expression which automatically satisfies the
symmetry and traceless conditions \UU\ and \trace\ for any $G$ and
corresponding to the general form \twoG\ we have solution of \conG\ given in
terms of
{\eqnn\RStwo
$$\eqalignno{
R_2 = {}&\rho^2 2{d-3\over d-1}\,\bigg \{ G'' + (2d-3)\cot \theta \, G' - 2(d+1)\cosec 
\theta \, G' \cr
{}& \ {}+ \Big ( (d-1)(d-2) \cot^2 \!\theta - 2(d+1)(d-2) \cot \theta \,
\cosec \theta  + (d^2 + d +2 ) \cosec^2 \!\theta \Big ) G \bigg \} \, , \cr
S_2 = {}& \rho^2{d-3\over d-2} \, \bigg \{ G'' + (2d-3)\cot \theta \, G' - {(d+1)(d-2)
\over d-1} \cosec \theta \, G' \cr
{}& \ {}+ \Big ( (d-1)(d-2) \cot^2 \!\theta - (d-2)^2{d+1\over d-1}\cot \theta \,
\cosec\theta - {d^2 - d +2\over d-1} \cosec^2 \!\theta \Big ) G \bigg \} \, , \cr
T_2 = {}&\rho^2 {d-3\over d-2} \, \bigg \{ G'' + (2d-3)\cot \theta \, G' \cr
{}& \ {}+ \Big ( (d-1)(d-2) \cot^2 \!\theta 
- {d^2 - d +2 \over d-1} \cosec^2 \!\theta \Big ) G \bigg \} \, . & \RStwo \cr}
$$}

If the traceless conditions \trace\ are satisfied then the expressions given by
\RStwo\ are generally valid. This is easily seen since, with the definition
\defQ\ in terms of $R_2,S_2,T_2$, we have
\eqn\QG{
Q = \rho^2 2(d-3)d\,{d+1 \over d-1} \cosec^2 \! \theta \, G \, ,
}
and \conQ{a,b}, with $P_1=0$, are equivalent to \RStwo. Conversely for arbitrary 
$R,S, \dots$ satisfying \conG\ we may first solve  \PP\ and \solF\ to obtain 
$R_0, S_0, \dots$
and then define $R_2 = R-R_0 , S_2 = S - S_0 , \dots $ which must necessarily
satisfy \conG\ as well as $P_1=P_2=0$. Hence defining $Q$ in \defQ\ in terms of these
$R_2, S_2 , \dots$ ensures through \QG\ that they may all be expressed as in \RStwo.
Furthermore $F_2$ may be found by solving \Gres\ in conjunction with \QG\
which may alternatively be expressed as
\eqn\GF{
\rho^4 {\d^2\over \d w^2} \big ( w^{d-1} F_2)
= {16(d-1)\over (d-3)d(d+1)}\,  w^{d-1} Q \, , \qquad w = \half ( 1+\cos \theta ) \, .
}
Assuming $Q$ and $F_2$ are non singular at $w=0$ ensures a unique solution.
In consequence the  general bi-tensor $\Gamma_{\mu\nu,\alpha\beta}$ as given by
\twoG\ subject to the conservation equation $\nab^\mu \Gamma_{\mu\nu,\alpha\beta}
=0$ may be decomposed into spin 0 and spin 2 pieces,
\eqn\TTgen{
\Gamma_{\mu\nu,\alpha\beta} (x,y) = 
\Gamma_{0,\mu\nu,\alpha\beta} (x,y) + \Gamma_{2,\mu\nu,\alpha\beta} (x,y) \, ,
}
where the two independent expressions are given by \TTzero\ and \TTC.

As an illustration which is relevant for the case of conformally invariant 
theories, for which $P_1=P_2 = 0$, we may consider
\eqn\GRST{
\quar R = S = 2T = Q = C \, {1\over s^d}  = C (\quar \rho^2 )^d (1-w)^{-d} \, ,
}
when \GF\ may be integrated to give
\eqn\Ftwoc{
F_2 = C (\quar \rho^2 )^{d-2} {d-1\over (d-3)d^2(d+1)^2} \, w^2F(d,d;d+2;w) \, . 
}

\newsec{Ambiguities in Spin 0 - Spin 2 Decomposition}

In the previous section we showed how the energy momentum tensor can be
decomposed into a spin 2 traceless part and a spin 0 contribution determined
by the trace, as in  \TTgen\ with \TTC\ and \twoG. However the spin 0
part may also result in a traceless expression for 
$\langle T_{\mu\nu}(x) T_{\alpha \beta}(y) \rangle$ if $F_0$ is proportional
to the Green function $\Gg_0$ for $-\nab^2 - {1\over d-1} R$,
\eqn\FG{
F_0(\theta) = C_0 \, \rho^{d-2} \Gg_0(\theta) \, ,
}
This Green function is constructed in appendix B. In the positive curvature
case, since $d\rho^2$ is an eigenvalue of $-\nab^2$, this satisfies 
\eqn\Kzero{
(- \nab^2 - d \rho^2 ) \Gg_0(\theta)_+ = S_d\de^d(x,y) -  k_d \, \rho^d \cos \theta \, .
}
Determining $A,B$ from \AB\ and \ABF\ ensures they satisfy \DE\ and from \Kzero
\eqn\Abeq{
A + dB = -(d-1) k_d \cos \theta \, .
}
Although the traceless condition $A + dB = 0$ is not satisfied the corresponding
expression for $\Gamma_{0,\mu\nu,\alpha\beta}$ is since
$\big ( \nab_\mu \nab_\nu - g_{\mu\nu} ( \nab^2 + \rho^2 ) \big ) \cos \theta =0$.
For general $d$ $A,B$ are not expressible in terms of elementary functions
but for $d=2,4$, either by solving \Abeq\ with $k_2 = 3/2, \ k_4 = 15/4$
or by using the explicit form for $\Gg_0$ given in appendix B, we have
\eqn\ABsol{\eqalign{
A = {}& C_0 \Big ( {2 \over s} + \half \rho^2 \! \cos \theta \Big ) \, , 
\qquad d=2 \, , \cr
A = {}& C_0 \Big ( {2\over s^2} (2 + \rho^2 s) +
{\ts {3\over 4}}\rho^4 \! \cos \theta \Big ) \, , \qquad d=4 \, . \cr}
}
Applying \ABzero\ for $d=2$ then gives, if we
make use of \twocom\ to set $U_0=0$,
\eqn\Rplus{
{\ts{1\over 8}} R_0 = \half S_0 = T_0 = - V_0 = {6C_0\over s^2} \, .
}
Assuming \TTtwoc\ requires taking $ C_0 = {1\over 6} c$ and, using \invxy,
gives the following simple form at a fixed point
\eqn\TTc{
4\pi^2 \langle T_{\mu\nu}(x) T_{\alpha \beta}(y) \rangle_+ \big |_{\beta^i =0}
= {c\over s^2 } \Big (
\I_{\mu\alpha} \I_{\nu\beta} + \I_{\mu\beta} \I_{\nu\alpha} 
- g_{\mu\nu} g_{\alpha\beta} \Big ) \, .
}
In two dimensions this the unique expression, as expected from the solution
obtained in \Rtwo, since the contribution 
corresponding to $ \Gamma_{2,\mu\nu,\alpha\beta}$ is absent.
When $d=4$ we may also obtain
\eqn\Rplusf{ \eqalign{
T_0 ={}&  C_0 {\rho^2 \over s^3 } \big ( 4 + 3\rho^2 s + {\ts{3\over 2}}
(\rho^2s)^2 \big ) \, , \quad
S_0 = C_0{\rho^2 \over s^3 } \big ( 24 + 8\rho^2 s + {\ts{3\over 2}}
(\rho^2s)^2 \big ) \, , \cr
R_0 ={}& C_0 {2\rho^2 \over s^3 } \big ( 96 + 12\rho^2 s + 
(\rho^2s)^2 \big ) \, , \cr}
}
In this case if we use \defQ\ we find
\eqn\Qfour{
Q_0 = 80 C_0 \, {\rho^2 \over s^3 }\,,
}
and inserting in \GF\ gives 
\eqn\Ftwo{
\rho^2 {\d^2\over \d w^2} \big ( w^3 F_2) = 3 C_0  \, 
{w^3 \over (1-w)^3} \, .
}
In consequence the two point function may be equivalently be expressed in terms of
$F_2$, i.e. for $F_2$ determined by \Ftwo\ inserted in \Gres\ and hence \RStwo\ we
have $R_2=R_0$ etc. This corresponds to the existence of $\Gg_2$ in \Hfour\ and
by solving \Ftwo\ we may find
\eqn\Ggtwo{
\Gg_2 = {\rho^2 \over 8 w^3} \Big ( {w\over 1-w} + 11 w
- w^2 + 6(2-w)\ln(1-w) \Big ) \, .
}

For the negative curvature case there are no complications due to zero modes
as reflected in \Kzero. In this case $A,B$ satisfy the traceless condition
$A+dB=0$ and the dependence on $\theta$ is simple for any $d$,
\eqn\AH{
A = C_0 \rho^{2d-2}\, {d\over (\sinh \theta)^d} \, .
}
For $d=2$ the results,
setting $U_0 = 0$ as before, involve essentially two independent terms
\eqn\Rminus{
\quar R_0 = S_0 = 12 C_0\, {1 \over  s^2} \, , \qquad T_0 = - V_0 = 
6C_0\bigg ( {1\over s^2} + {1\over \bs^2} \bigg ) \, .
}
Inserting these expressions into \TTtwoc\ with $C_0$ determined in terms of
$c$ as above now gives
\eqn\TTcc{\eqalign{
4\pi^2 \langle T_{\mu\nu}(x) T_{\alpha \beta}(y) \rangle_- \big |_{\beta^i =0}
= {}&{c\over s^2 }\Big ( \I_{\mu\alpha} \I_{\nu\beta} + \I_{\mu\beta} \I_{\nu\alpha}
- g_{\mu\nu} g_{\alpha\beta} \Big ) \cr
{}& + {c\over \bs^2 } \Big ( I_{\mu\alpha} I_{\nu\beta} + I_{\mu\beta} I_{\nu\alpha}
- g_{\mu\nu} g_{\alpha\beta} \Big ) \, . \cr}
}
{}From \AH\ for any $d$ we may find using \defQ\
\eqn\Qd{
Q_0 = C_0 \rho^{2d} d(d+1) \, 
{d + (d-1) \sinh^2 \theta \over (\sinh \theta)^{d+2}}  \, ,
}
and for $d=4$ the complete expression for $\Gamma_{2,\mu\nu,\alpha\beta}$ is
given by 
\eqn\Rminusf{ \!\!\!\!\!\!\!
T_0 = 4C_0\rho^8 \, {1\over \sinh^6 \! \theta} \, , \
S_0 = 4C_0\rho^8 \, {5\cosh \theta + 1 \over \sinh^6 \! \theta} \, , \
R_0 = 16C_0\rho^8 \, {5\cosh^2\! \theta + 5\cosh \theta + 2
\over \sinh^6 \! \theta} \, .
}
As in the positive curvature case we may 
solve the equivalent equation to \GF, given in (A.10), to determine
a corresponding $F_2$ which gives identical forms for $R,S,T$ with $U,V$
determined by \trace. In our later discussion of the spectral representation
we show that the resulting $F_2$ cannot be accommodated by imposing the
unitarity bound on possible spin 2 intermediate states.

\newsec{Free Fields}

It is important in order to understand what results may be found from an analysis
of the two point function on spaces of constant curvature to calculate the possible
forms at renormalisation group fixed points. To this end we first consider the
results for conformally invariant free field theories, free scalar and fermion
theories in general dimension $d$ and abelian gauge fields in dimension four. In
each case we initially consider the theory on a sphere, with $R>0$, avoiding
boundary conditions which are necessary for the hyperbolic case when $R<0$.

The free conformally coupled scalar field $\phi$ satisfies, on a space of constant 
curvature where \riem\ holds, $\Delta \phi \equiv
(-\nab^2 \pm  \quar d(d-2) \rho^2 ) \phi = 0$. The
associated  energy momentum tensor may be written in various equivalent forms but
here we take
\eqn\Tphi{
T_{\mu\nu} = \pr_\mu \phi \pr_\nu \phi - {1\over 4} \, {1\over d-1} \Big (
(d-2) \nab_\mu \nab_\nu + g_{\mu \nu} \nab^2 \mp (d-1)(d-2) \rho^2 g_{\mu\nu}
\Big ) \phi^2 \, ,
}
which is conserved and traceless on the equations of motion.\foot{$g^{\mu\nu}T_{\mu\nu}
= \phi\, \Delta \phi$, $\nab^\mu T_{\mu\nu} = \half \phi \pr_\nu \Delta \phi
- \half \Delta \phi \, \pr_\nu \phi$.} 
Correlation functions for the energy momentum tensor and other composite fields
are determined in terms of the basic scalar field two point function which
for the sphere may be taken, with $s$ the chordal distance defined in \sols\
and $S_d$ given in \Sd, as
\eqn\twophi{
\l \phi(x) \phi(y) \r_+ = {1\over (d-2) S_d} \, {1\over s^{{1\over 2}d - 1}} \, . 
}
It is first useful to verify
\eqn\TphiS{
\l T_{\mu\nu}(x) \phi^2 ( y) \r_+ = 0 \, ,
}
and then we may obtain for the two point function of the energy momentum tensor
$S_d^{\, 2} \l T_{\mu\nu}(x) T_{\alpha\beta}(y) \r_+$
the form given in \twoG\ with
\eqn\RST{
2 T_+ = S_+ = \quar R_+  = - {1\over d }\, V_- = {d\over d-1}  \,
{1\over s^d} \, , \quad U_+ = 0 \, .
}
The actual result may be written simply, using \invxy, in the form
\eqn\TTconS{
S_d^{\, 2} \l T_{\mu\nu}(x) T_{\alpha\beta}(y) \r_{+,{\rm conformal}} = C_T \, 
{1\over s^d} \,
\Big ( \half \big (\I_{\mu\alpha} \I_{\nu\beta} + \I_{\mu\beta} \I_{\nu\alpha} \big )
- {1\over d} \, g_{\mu\nu} g_{\alpha\beta} \Big ) \, ,
}
where
\eqn\CTphi{
C_{T,\phi} = {d\over d-1} \, .
}
Comparing with \TTc\ for $d=2$ gives $c=1$ as expected. the expression \TTconS\
is just what would be expected by conformal transformation from flat space.

When the scalar fields are considered on the negative curvature
hyperboloid $H^d$ boundary conditions are necessary. For Dirichlet
boundary conditions the basic two point function of the conformally coupled
scalar field becomes
\eqn\twophiH{
\l \phi(x) \phi(y) \r_- = {1\over (d-2) S_d} \bigg ( {1\over s^{{1\over 2}d - 1}} 
- {1\over \bs^{{1\over 2}d - 1}} \bigg ) \, .
}
Unlike \twophi\ this is well defined as $d\to 2$.
In this case, in contrast to \TphiS, we have
\eqn\TphiH{
S_d^{\, 2} \l T_{\mu\nu}(x) \phi^2 (y) \r_- = {4\over d-1}  \,
{1\over \rho^2 (s\bs)^{{1\over 2}d}} \big (d \,\hx_\mu \hx_\nu - g_{\mu\nu} \big ) \, .
}
Using this result then we may find, after some calculation, the corresponding 
expressions to \RST\ for $\l T_{\mu\nu}(x) T_{\alpha\beta}(y) \r_-$ in this
case which are given by
\eqn\RSTH{ \eqalign{
(d-1) T_-  {}& = \half d \bigg ( {1\over s^d} + {1\over \bs^d}
- {2\over  (s\bs)^{{1\over 2}d}} \bigg )  - d \, {d-2\over d-1} \,
{4\over  \rho^4 (s\bs)^{{1\over 2}d + 1}} \, , \cr
(d-1) S_-  {}& = d \bigg ( {1\over s^d} - {1\over  (s\bs)^{{1\over 2}d}} 
\bigg ) - d \, {d-2\over d-1} \big ( (d+1) \cosh \theta + 1 \big )
{4\over  \rho^4 (s\bs)^{{1\over 2}d + 1}} \, , \cr
(d-1) R_-  {}& = d \bigg ( {4\over s^d} + (d^2 - 4) 
{1\over  (s\bs)^{{1\over 2}d}} \bigg ) \cr
&\ {} - d \, {d-2\over d-1} \big ( 
d(d+1) \cosh^2 \! \theta + 4 (d+1) \cosh \theta + d + 4 \big )
{4\over  \rho^4 (s\bs)^{{1\over 2}d + 1}} \, , \cr
(d-1) U_- {}&  ={}  - d^2 {1\over  (s\bs)^{{1\over 2}d}} 
+ d \, {d-2\over d-1} \big ( (d+1) \cosh^2 \! \theta + 1 \big )
{4\over  \rho^4 (s\bs)^{{1\over 2}d + 1}} \, , \cr
(d-1) V_- {}&  = - \bigg ( {1\over s^d} + {1\over \bs^d}
- (d+2) {1\over  (s\bs)^{{1\over 2}d}} \bigg ) -  {d-2\over d-1} 
\big ((d+1)\!\cosh^2 \!\theta - 1 \big ) {4\over  \rho^4 (s\bs)^{{1\over 2}d + 1}} \, .
\cr} 
}
These results satisfy the traceless conditions \trace, as expected, and further
for $d=2$ the terms involving $s\bs$ disappear as a consequence of \twocom\ and
the result is compatible with \TTcc\ for $c=1$. The expression written in \RSTH\
is such that the different terms separately obey the conservation equations. The 
resulting expression is no longer in the simple form given by \TTconS\ in the
positive curvature case but if the terms involving $s\bs$, which vanish if $d=2$,
are dropped we have
\eqn\TTconH{\eqalign{
S_d^{\, 2} \l T_{\mu\nu}(x) T_{\alpha\beta}(y) \r_{-,{\rm conformal}} = {}& 
C_T \bigg \{ {1\over s^d}
\Big ( \half \big (\I_{\mu\alpha} \I_{\nu\beta} + \I_{\mu\beta} \I_{\nu\alpha}
\big ) - {1\over d}\,  g_{\mu\nu} g_{\alpha\beta} \Big ) \cr
& \qquad {} + {1\over \bs^d}
\Big ( \half \big (I_{\mu\alpha} I_{\nu\beta} + I_{\mu\beta} I_{\nu\alpha}
\big ) - {1\over d}\,  g_{\mu\nu} g_{\alpha\beta} \Big ) \bigg \} \, , \cr}
}
with the same result for $C_T$ as in \CTphi. The remaining terms present in \RSTH\
can be understood in terms of the operator product expansion but it is clear that
the form given in \TTconH\ cannot be the unique expression for conformal field
theories.

For massless vector fields, satisfying the free equations $\nab^\mu F_{\mu\nu}
+ \pr_\nu \nab^\mu A_\mu / \xi=0$ with ${ F_{\mu\nu}= \pr_\mu A_\nu - \pr_\mu A_\nu}$
and $\xi$ a gauge fixing parameter, we restrict to $d=4$ when the theory is
conformally invariant on flat space. Neglecting terms which are irrelevant in
gauge invariant correlation functions the energy momentum tensor is
\eqn\VT{
T_{\mu\nu} = F_{\mu\si} F_\nu{}^{\si} - \quar g_{\mu\nu}\, F_{\si\rho}F^{\si\rho} \, .
}
{}From the two point functions of the gauge field $A_\mu$, which have
been obtained in \allen\ and more recently in \Free\ and are discussed in
appendix B, the two point correlation
functions of the field strength $F_{\mu\nu}$ are simple
\eqn\FFV{\eqalign{
\l F_{\mu\nu}(x) F_{\alpha\beta}(y) \r_+ = {}&{1\over \pi^2 s^2}
\big (\I_{\mu\alpha} \I_{\nu\beta} - \I_{\mu\beta} \I_{\nu\alpha}  \big ) \, ,\cr
\l F_{\mu\nu}(x) F_{\alpha\beta}(y) \r_-  = {}&{1\over \pi^2 s^2}
\big (\I_{\mu\alpha} \I_{\nu\beta} - \I_{\mu\beta} \I_{\nu\alpha}  \big ) +
{1\over \pi^2 \bs^2}
\big (I_{\mu\alpha} I_{\nu\beta} - I_{\mu\beta} I_{\nu\alpha}  \big ) \, . \cr}
}
With the explicit form \VT\ it is straightforward to calculate in this case that
the energy momentum tensor two point function is just as in \TTconS\ or \TTconH\
for $d=4$ with 
\eqn\CTV{
C_{T,V} = 8 \,  .
}
We may also easily show that
\eqn\TFF{
\l T_{\mu\nu}(x) F_{\alpha\beta}F^{\alpha\beta}(y) \r_+ = 0 \, , \quad
(2\pi^2)^2\l T_{\mu\nu}(x) F_{\alpha\beta}F^{\alpha\beta}(y) \r_- = 
{32\over  s^2\bs^2} ( 4 \hx_\mu \hx_\nu - g_{\mu\nu} ) \, .
}

For massless spinor fields, satisfying $\gamma^\mu \nab_\mu \psi = 0$, the
energy momentum tensor may be taken as
\eqn\Tpsi{
T_{\mu\nu} = \bpsi \gamma_{(\mu} \olr \nab_{\nu)} \psi \, ,
}
which is conserved and traceless for any $d$. The formalism for spinors
on constant curvature spaces was described in section 7 and the 
basic two point functions were obtained in \Stwo\ in terms of the inversion
bispinor $\I(x,y)$ and parallel transport bispinor $I(x,y)$ giving
\eqn\twopsi{\eqalign{
\l  \psi(x) \bpsi(y) \r_+ 
= {}& {1\over S_d} \, {1\over s^{{1\over 2}(d-1)}}\, \I \, , \cr
\l  \psi(x) \bpsi(y) \r_- 
= {}&  {1\over S_d} \bigg (  {1\over s^{{1\over 2}(d-1)}} \, \I
\pm  {1\over \bs^{{1\over 2}(d-1)}} \, I \bigg ) \, . \cr}
}
{}From this we may obtain
\eqn\twode{\eqalign{\!\!\!
\l &  \nab_\mu \psi(x) \bpsi(y)\overleftarrow\nabla_{\!\alpha} \r_+  \cr
& = {1\over S_d} \, {1\over s^{{1\over 2}(d+1)}}\bigg \{
\gamma_\mu \I\gamma_\alpha  + d\I_{\mu\alpha}\, \I
+ \cos^2\! \half \theta ( \gamma_\mu \gamma{\cdot \hx} - d \hx_\mu ) \I
( \gamma{\cdot \hy} \gamma_\alpha - d \hy_\alpha ) \bigg \} \, , \cr
\!\!\!\l & \nab_\mu  \psi(x) \bpsi(y) \overleftarrow\nabla_{\!\alpha} \r_- \cr
& = {1\over S_d} \, {1\over s^{{1\over 2}(d+1)}}\bigg \{
\gamma_\mu \I\gamma_\alpha  + d\I_{\mu\alpha} \, \I
+ \cosh^2\! \half \theta ( \gamma_\mu \gamma{\cdot \hx} - d \hx_\mu ) \I
( \gamma{\cdot \hy} \gamma_\alpha - d \hy_\alpha ) \bigg \} \, , \cr
& \ {}\mp {1\over S_d} \, {1\over \bs^{{1\over 2}(d+1)}}\bigg \{
\gamma_\mu I\gamma_\alpha  - d I_{\mu\alpha} \, I
- \sinh^2\! \half \theta ( \gamma_\mu \gamma{\cdot \hx} - d \hx_\mu ) I
( \gamma{\cdot \hy} \gamma_\alpha - d \hy_\alpha ) \bigg \} \, , \cr}
}
and then straightforwardly calculate expressions for $\l T_{\mu\nu}(x) T_{\alpha\beta}
(y) \r_\pm$ which are just as in \TTconS\ or \TTconH\ with
\eqn\CTpsi{
C_{T,\psi} =  2^{{1\over 2}d-1 }d \, .
}
It is also useful to note that
\eqn\Tpsitwo{
\l T_{\mu\nu}(x) \bpsi\psi(y) \r_+ = 0 \, , \quad
S_d^{\, 2} \l T_{\mu\nu}(x) \bpsi\psi ( y) \r_- =  \pm 2^{{1\over 2}d+1}
\, {1\over \rho (s\bs)^{{1\over 2}d}} \big (d \,\hx_\mu \hx_\nu -
g_{\mu\nu} \big ) 
\, .
}

\newsec{Operator Product Expansion}

The singular behaviour of the above results for two point functions for free field
theories may be understood in terms of the operator product expansion.
This may also be applied for non trivial interacting theories to determine 
the various possible contributions to two, and higher, point functions. We restrict 
here to the situation of theories at a fixed point, so that the flat space theory
is conformally invariant, and further consider just the operator
product expansions involving the energy momentum tensor and a scalar field $\O$
with an arbitrary dimension $\eta$. On curved space the leading term 
in any operator product expansion coefficient is determined by the
corresponding results on flat space \refs{\Card,\hughone}.  Using
the notation of section 5 we consider first the operator product expansion of 
$T_{\mu\nu}$ and $\O$ which may be written as
\eqn\OPETO{
T_{\mu\nu}(x) \O(y) \sim - {\eta \over d-1} \, {1\over S_d} \, 
{1\over s^{{1\over 2}d}}
\big ( d \hx_\mu \hx_\nu - g_{\mu\nu} \big ) \O(y) \, .
}
The overall coefficient is determined through Ward identities
\refs{\Card,\hughone}. Secondly the contribution of the scalar $\O$ to the
operator product expansion of two energy momentum tensors has the form
\eqn\OPETT{ \eqalign{
T_{\mu\nu}(x) T_{\alpha\beta}(y) \sim {1\over s^{d-{1\over 2}\eta}} \Big (
 &  a \,\hx_\mu \hx_\nu \hy_\alpha \hy_\beta 
+ b \big ( \hx_\mu \hy_\alpha I_{\nu\beta}  +  \mu \leftrightarrow \nu ,
\alpha \leftrightarrow \beta \big ) \cr
& {} + c \big (  I_{\mu\alpha} I _{\nu\beta}
+ I_{\mu\beta}  I_{\nu\alpha} \big ) - \hbox{traces} ( \mu\nu, \alpha\beta) \Big )
\O(y) \, , \cr}
}
where $a,b,c$ satisfy the relations, derived essentially by imposing the 
conservation equations on the expansion coefficient,
\eqn\abc{
(a-4b) \big ( 1 - \half (d-\eta)(d-1) \big ) + d\eta b = 0 \, , \quad
a-4b - d(d-\eta) b + d(2d-\eta) c = 0 \, .
}
In this case there remains a single undetermined scale reflecting the arbitrariness 
in the normalisation of $\O$.

The operator product \OPETO\ is in accord with the the leading singular
behaviour as $s\to 0$ of the expressions \TphiH, \TFF\ and
\Tpsitwo\ in the hyperbolic case taking then
\eqn\onep{
\!\!\!\!\l \phi^2 \r_- = - {1\over (d-2) S_d} \, \big (\half \rho)^{d-2} , \quad
\l F_{\alpha\beta}F^{\alpha\beta} \r_- = - {3\over 4\pi^2} \, \rho^4 \, , \quad
\l \bpsi\psi \r_- = \mp {2^{{1\over 2}d} \over S_d} \, \big (\half \rho)^{d-1}\, .
} 
These results for the one-point functions follow directly from \twophiH, \FFV\
and \twopsi\ when the coincident limit is regularised by dropping the singular
contributions in $s$ (which is consistent with the vanishing of these one-point
functions in the spherical case). As usual with dimensional regularisation
this prescription is essentially unambiguous for $d\ne$ integer, but for the fermion
case the result for $\l \bpsi\psi \r $ is unique and well defined even at $d=4$
since it is non zero only
if chirality is violated so that the first term in \twopsi\ cannot contribute.
The form of the two point function for the energy momentum tensor in the case
of scalar fields given by \RSTH\ may also be understood by using the operator
product expansion in \OPETT. For a free scalar theory taking $\O \to \phi^2$,
which has dimension $d-2$, we have
\eqn\abcs{
c = {d(d-2)^2\over 4(d-1)^2} \, {1\over S_d} \, , \qquad b=(d+2) c \, , \qquad
a=(d+2)(d+4)c \, ,
}
which satisfy \abc. With the result \onep\ for $\l \phi^2 \r$ \OPETT\ and \abcs\
generate the terms in \RSTH\ $\propto s^{-{1\over 2}d-1}$ as $s\to 0$. There are no
corresponding contributions in the case of free vectors or fermions from 
$\l F_{\alpha\beta}F^{\alpha\beta} \r$ or $\l \bpsi\psi \r$ since these
operators do not appear in the operator product expansion of two energy
momentum tensors, for $\bpsi\psi$ by chirality and for 
$F_{\alpha\beta}F^{\alpha\beta}$ for free vector theories in $d=4$ by direct
calculation \hughone. Although the overall normalisation of $a,b,c$ in \abc\
is arbitrary such freedom is absent from the associated contributions to
$\l T_{\mu\nu}(x) T_{\alpha\beta}(y) \r$ since it is cancelled by a
corresponding factor in $\l \O\r$.

Although the above results demonstrate the relevance of the operator product
expansion to understanding the form of the two point function on spaces
of constant curvature there are also areas where its role is less transparent.
It was our initial hope, based on the fact that the operator product
expansion of two energy momentum tensors always contains the energy momentum
tensor itself and that its one point function, which has the form shown in \Thom,
is in general non zero, that expression for the two point function for the energy
momentum tensor on such spaces would involve its coefficient $C$. In particular
in four dimensions at a fixed point, as a consequence of \Tfour, this would
require a dependence on the parameter $a$, which is defined initially  through 
the energy momentum tensor trace on general curved backgrounds, in some evident
fashion. Such a result would be significant since $a$ is the favourite candidate
for a four dimensional generalisation of the $c$-theorem \Cardy\ (there is now
strong theoretical support for this conjecture in supersymmetric theories \Anone)
and it could be hoped that a proof of the  $c$-theorem in four dimensions
might be found in terms of the two point functions of the energy momentum tensor
on curved space following a similar approach to the original discussion of
Zamolodchikov \Zam. Nevertheless, despite the arguments based on the operator
product expansion for the appearance of the parameter $a$ in the two point
function at non coincident points,
this expectation is not supported by the explicit free field expressions obtained 
in section 10. This is most apparent from the results for
fermions and vectors which are have the identical form \TTconH\ when $d=4$
with the coefficient $C_T$ determined by $c$ in the trace anomaly \Afour. On the other
hand $a$, which is also initially defined in terms of \Afour, has no direct
relation to $c$ as demonstrated by results for free fields.

It appears therefore that there may be some unresolved questions when applying the
operator product expansion on spaces with non zero curvature. For a conformal
field theory on flat space all one point functions vanish (except trivially
for the identity operator). For quasi-primary operators, which transform homogeneously
under conformal transformations, two point functions, which are non zero only if both
operators have the same spin and scale dimension, are
unique up to an overall constant and three point functions are also given by a finite
number of linearly independent forms.  
Together these determine completely the operator product expansions for any pair
of quasi-primary operators in terms of all quasi-primary operators and their
descendents or derivatives for which the associated three point functions are non zero
and also the identity operator if their two point function is non zero. If
the operator product expansion is extended to a curved space background the
leading singular contributions to the coefficient functions in the expansion
are of course determined by those for flat space, but there may now be 
less singular contributions involving  the curvature which can naturally appear
associated with derivative terms.
The contribution of the identity operator to the product of two operators 
may also be expected in general now to be a function 
of the separation rather than just a power determined by the scale dimension of the
operators, since a non zero curvature introduces a scale. Moreover such a function 
cannot just be identified with the associated two point function if other
operators present in the expansion have a non zero expectation value on curved space.
In the particular case of the contribution of the energy momentum tensor to the
operator product expansion of two energy momentum tensors the coefficient functions
found from flat space calculations correspond to purely traceless energy momentum 
tensors \refs{\hughone,\hughtwo} so that the straightforward generalisation
to curved space does not generate terms associated with its expected trace anomaly. 
The necessity of a non zero trace for $\l T_{\mu\nu}\r$ on a general curved space for
a flat space conformal theory arises in $d=2,4,\dots$ dimensions as a consequence
of the requirement that the regularisation procedure necessary to determine
$\l T_{\mu\nu}\r$ should maintain the conservation equation \done, \birrell. 
The trace is then unambiguous, independent of boundary conditions and non zero. 
However, on a homogeneous space of constant curvature, the conservation equation
is trivially satisfied as a consequence of \Thom. Hence, as remarked in \Burges,
there is no calculation uniquely determining the anomaly, or the coefficient
$C$ in \Thom,  which is entirely intrinsic to the
theory on the constant curvature space (although calculations adapted to such
spaces are described in \refs{\DShore,\Burges,\Camp}).

When $d=4$ the contribution of the energy momentum tensor, of dimension 4, to the
operator product expansion of two energy momentum tensors will in general
mix with terms quadratic in the Riemann tensor and such terms will contribute
to the two point function. This may be illustrated by the associated example
of the two point function of the energy momentum tensor and the scalar field
$\O$. \TOfpfour\ for the hyperbolic case gives
\eqn\TOdd{
(2\pi^2)^2 \langle T_{\mu\nu}(x) \O_i(y) \r_- \big |_{\beta^i = 0} = 
- {1\over 3 (\quar \rho^2 s \bs)^2}\big ( 4 \hx_\mu \hx_\nu - g_{\mu\nu} \big ) 
\Gamma_i{}^{\! j}\big ( {\ts {1\over 36}} \hU_j R^2 + 2\pi^2\l \O_j \r_- \big ) \, . 
}
Since $\Gamma_i{}^{\! j}$ is the dimension matrix for the fields $\O_j$ this
is exactly in accord with \OPETO\ but the presence of a non zero $\hU_j$ in
general demonstrates such a mixing with curvature terms.

\newsec{Unitary  Representations of $O(d-1,2)$ for Spin One}

In a quantum field theory with a unique vacuum the correlation functions may be
expressed in terms of the vacuum state expectation values of products of local
field operators. For unitarity the operators are assumed to act on a Hilbert
space with positive norm. A further essential requirement is that there is
a hermitian Hamiltonian operator $\hH$, which annihilates the vacuum but
otherwise with positive spectrum, generating translations in a coordinate $\tau$.
This ensures that the theory may be analytically continued, by letting 
$\tau \to i t$, from an underlying $d$-dimensional  with positive metric
to define a unitary quantum field theory on the associated space with a metric with 
signature $(-,+,\dots)$. We here consider field theories on the homogeneous
constant curvature spaces $S^d$ and $H^d$.
In the positive curvature case the analytic continuation of $S^d$,
when $\tau$ is periodic, leads to de Sitter space $dS_d$ and the correlation
functions of the resulting quantum field theory are interpreted as corresponding
to a thermal bath of non zero temperature \birrell, and so are not given by
vacuum matrix elements of field operators.\foot{For a relevant discussion see \Mott.} 
In consequence we restrict our attention here to the negative
curvature case when the homogeneous space $H^d$ is continued to
$AdS_d$. Hence we consider states which form unitary representations of the isometry
group $O(d-1,2)$ where the vacuum $|0\r$ is a unique state forming a trivial
singlet representation. The general unitary positive energy representations 
are constructed from lowest weight states which form a basis for a representation 
space for $O(d-1)$. The representations are used subsequently to obtain spectral
representations for the two point correlation functions.
For scalar fields the unitary representations are formed from a spin 0 or singlet
lowest weight state and the spectral representations are well known \refs{\Dus,\Cap}.
However the extension to fields with spin involves further 
complications so in this section, and appendix C, we construct in detail the
representation for a spin one lowest weight state.

In order to identify convenient coordinates, with  a privileged choice for $\tau$,
for $H^d$ we consider first its embedding in $\bR^{d+1}$ as the hypersurface given
 by the constraint 
$\rho^2 g_{ab} \eta^a \eta^b= - 1$ where with $\eta^a = (\eta^0 , \eta^i ,
\eta^d)$ we have $g_{00}= - g_{dd}=1,\,  g_{ij}= \delta_{ij}$. Global coordinates 
$x= (\tau, r , \hxn_i)$ for
$H^d$ are then given by $\rho \eta^0 = \sinh \tau \, \sec r , \, \rho \eta^i
= \tan r \, \hxn_i , \, \rho \eta^d = - \cosh \tau \, \sec r$, where $\hxn_i \hxn_i =1,
\, \hxn_i \in S^{d-2}$ and $\tau,r$ are in the ranges
$-\infty < \tau < \infty, \, 0\le r\le {\pi\over 2}$. On continuing $\tau \to i t$
these coordinates then cover the simply connected covering space for $AdS_d$.
For two points, represented
by $\eta^a,\eta^{\prime a}$, $\cosh \theta = - \rho^2 g_{ab} \eta^a \eta^{\prime b}
= \cosh(\tau-\tau')\sec r \sec r' - \tan r \tan r' \xi{\cdot \xi'}$.
The associated metric $\d s^2 = g_{ab} \d \eta^a \d \eta^b$ becomes
\eqn\metric{
\rho^2 \d s^2 = \sec^2 r \big ( \d \tau^2 + \d r^2 + \sin^2 r \, \d s_{S^{d-2}}^2
\big ) \, .
}
The isometry group $O(d,1)$ generators $L_{ab} = \eta_a \pr_b - \eta_b \pr_a$ obey
the algebra
\eqn\alg{
[L_{ab} , L_{cd} ] = g_{bc} \, L_{ad} - g_{ac} \, L_{bd} - g_{bd} \, L_{ac} 
+ g_{ad} \, L_{bc} \, .
}
Writing
\eqn\newgen{
H = L_{0d} \, , \qquad L_{\pm i} = L_{0i} \mp L_{di} \, ,
}
the algebra \alg\ decomposes as
\eqn\algn{ \eqalign{
[ H , L_{\pm i}] =  \pm L_{\pm i}{}&  \, , \quad [ L_{-i} , L_{+j} ] 
= 2 \de_{ij} H - 2 L_{ij}  \, , \quad [ L_{+i} , L_{+j} ] = 
[ L_{-i} , L_{-j} ] = 0 \, , \cr
 & [ L_{ij} , H ]  = 0 \, , \quad [ L_{ij} , L_{\pm k }] = 
\de_{jk}  L_{\pm i} - \de_{ik}  L_{\pm j} \, , \cr}
}
with $L_{ij}$ the generators of $O(d-1)$ obeying an algebra of the same form as
\alg\ with $g_{ab} \to \de_{ij}$. With these coordinates we have
\eqn\rep{
H = - {\pr \over \pr \tau} \, , \quad
L_{\pm i} = - e^{\mp \tau} \bigg (  \hxn_i \Big ( \sin r \, {\pr \over \pr \tau}
\pm \cos r \, {\pr \over \pr r } \Big )  \pm \cosec r \, D_i \bigg ) \, ,
\quad L_{ij} = \hxn_i D_j - \hxn_j D_i \, ,
}
where $D_i$ is the tangential derivative
\eqn\Sd{
[ D_i , D_j ] = L_{ij} \, , \qquad \hxn_i D_i = 0 \, ,  \qquad
D_i \hxn_j = \de_{ij} - \hxn_i \hxn_j \, ,
}
so that we may write $D_j = \hxn_i L_{ij}$ and also $ [ L_{ij} , D_k ]
= \de_{jk} D_i  - \de_{ik}  D_j$.

In a quantum field theory which is unitary on continuation to $AdS_d$ the
isometry group is represented by operator generators $\hL_{ab}$ obeying
\alg\ which on decomposing as in \newgen\ obey the hermeticity properties
\eqn\herm{
\hH^\dagger = \hH \, , \qquad \hL_{-i}{}^{\! \dagger} = \hL_{+i} \, , \qquad
\hL_{ij}{}^{\! \dagger} = - \hL_{ij} \, .
}
These conditions correspond to restricting to unitary representations of
the algebra of $O(d-1,2)$.
The vacuum state $|0\r$ of course satisfies $ \hL_{ab} |0\r = 0$, forming the
trivial singlet representation. The quadratic
Casimir  has the form
\eqn\Cas{
\hC_2 = - \half \hL^{ab} \hL_{ab} = -\half \hL_{ij} \hL_{ij} -
 \hL_{\pm i} \hL_{\mp i} + \hH^2 \mp (d-1) \hH \, ,
}
while
\eqn\Casd{\eqalign{
- \half L^{ab} L_{ab} = {1\over \rho^2} \nab^2_{H^d} ={}&
\cos^2 r \Big ( {\pr^2 \over \pr \tau^2} + {\pr^2 \over \pr r^2} \Big )
+ (d-2) \cot r {\pr \over \pr r} + \cot^2 r \nab^2_{S^{d-2}} \, , \cr
& \nab^2_{S^{d-2}} = D_i D_i = \half L_{ij} L_{ij} \, . \cr}
}

For a scalar field $\phi(x)$ the action of the generators is simply
\eqn\Lphi{
[ \hL_{ab} , \phi ] = - L_{ab} \phi \, .
} 
To determine the extension to vector fields we
consider the transformation of a vector field $\A_a(\eta)$ on $\bR^{d+1}$
for which the corresponding action of the generators is 
$[ \hL_{ab} , \A_c ] = - L_{ab} \A_c - g_{ac} \A_b + g_{bc} \A_a$. This may be
reduced to a $d$-component field on the embedded hypersurface
$H^d$ through the invariant constraint
$\eta^a \A_a (\eta) =0$. In terms of the coordinates $x= (\tau, r , \hxn_i)$ it
is natural to consider instead the field $A_a(x)$ where $A_\pm = \A_0 \mp \A_d $ 
and $A_i= \A_i - \hxn_i \hxn_j \A_j $ is the component of $\A_i$ tangential to 
$S^{d-2}$ so that $\hxn_i A_i = 0$. Using the constraint to eliminate 
$\hxn{\cdot \A}$ the action 
of the generators then becomes, with $H, \, L_{\pm i}, \, L_{ij}$ given by \rep,
\eqn\LHA{\eqalign{
[ \hH , A_\pm ] ={}& - H A_\pm \pm A_\pm \, , \quad [\hH , A_i ] = - H A_i \, , \cr
[ \hL_{ij} , A_\pm ] = {}& - L_{ij}  A_\pm \, , \quad 
[ \hL_{ij} , A_k ] = - L_{ij}  A_k - \de_{ik} A_j + \de_{jk} A_i \, , \cr}
}
and
\eqnn\LLA
$$\eqalignno{
[\hL_{\pm i}, A_\pm ] = {}& -  L_{\pm i} A_{\pm} \, , \cr
[\hL_{\pm i}, A_j ] = {}& -  L_{\pm i} A_j \pm e^{\mp \tau} \cosec r \, \hxn_j A_i
+ ( \de_{ij} - \hxn_i \hxn_j ) 
\big ( A_\pm - \half \cosec^2 r ( A_\pm - e^{\mp 2\tau} A_\mp ) \big )  \, , \cr
[\hL_{\pm i}, A_\mp ] = {}& -  L_{\pm i} A_{\mp} - 2 A_i 
+ \hxn_i \cosec r \big ( e^\tau A_+ - e^{-\tau} A_- \big ) \, . & \LLA \cr}
$$
For a scalar field $\phi(\eta)$ there is a corresponding vector $\A_a = \eta^b
L_{ab} \phi$ satisfying $\eta^a\A_a = 0$. In the same fashion as previously when
$\A_a(\eta) \to A_a(x)$ on $H^d$ we may therefore define the vector field
$\nab_a \phi(x) = e_a{}^\mu(x) \pr_\mu \phi(x)$ where explicitly in terms of the
coordinates $x= (\tau, r , \hxn_i)$
\eqn\nabphi{
\nab_\pm \phi = e^{\mp \tau} \big ( \cos r \pr_\tau \phi \mp \sin r \pr_r \phi)
\, , \qquad \nab_i \phi = \cot r \,  D_i \phi \, .
}
Similarly for a vector $\A_a$ there is an associated scalar $ - \eta^a L_{ab}
\A^b$ which may be used to defined the divergence $\nab{\cdot A}$ and which
is then given by,
\eqn\nabA{\eqalign{
\nab{\cdot A} = {}&\half \cos r \big ( e^\tau \pr_\tau A_+ + e^{-\tau} \pr_\tau A_-
\big ) \cr
{}& - \half \cot r \big ( \cos r \pr_r + (d-3)\cosec r \big )\big (
e^\tau A_+ - e^{-\tau} A_- \big )  + \cot r \, D_i A_i \, . \cr}
}

For scalar fields the appropriate representation is defined in terms of a
spin-0 lowest weight state $|\lambda \r $, $\l \lambda | \lambda \r$=1, satisfying
\eqn\wei{
\hH |\lambda \r = \lambda |\lambda \r \, ,\qquad \hL_{-i}|\lambda \r = 
\hL_{ij} |\lambda \r = 0 \, .
}
The representation space is then spanned by  linear combinations of states 
of the form $\prod_i \hL_{+i}{}^{\! r_i} |\lambda\r$. These may be decomposed
into representations of $O(d-2)$ by  using symmetric traceless rank $\ell$
tensors $\C_{i_1 \dots i_\ell}$, $\ell = 1,2,\dots$,  satisfying
\eqn\tens{
\C_{i_1 \dots i_\ell} = \C_{(i_1 \dots i_\ell)} \, , \qquad
\C_{i i i_1 \dots i_{\ell-2} } =0 \, ,
}
to define, taking for $\ell=0, \, \C=1$,
\eqn\basis{
| n \, \ell , \C \r = \hK_+{}^{\!\! n} \hL_{+ i_1} \dots \hL_{+ i_\ell} |\lambda \r \,
\C_{i_1 \dots i_\ell} \, ,
\quad \hK_+ = \hL_{+i} \hL_{+i} \, , \qquad n,\ell = 0 ,1, 2, \dots \, .
}
It is easy to see that $\hH | n \, \ell , \C \r = (\lambda + 2n +\ell)
| n \, \ell , \C \r$ and for all states the Casimir operator, given by \Cas,
takes the value $\lambda(\lambda - d +1)$ while for each $\ell$ they transform
irreducibly under $O(d-2)$. These states  satisfy the orthogonality
condition
\eqn\orthog{
\l n' \, \ell' , \C'  | n \, \ell , \C \r =  \de_{n'n} 
\de_{\ell'\ell}\,  \C'\!\cdot \C \,\N_{n\ell} \, , \qquad \C'\!\cdot \C =
\C'{}_{\!i_1 \dots i_\ell} \C_{i_1 \dots i_\ell} \,  ,
}
where, as shown in appendix C, the norms are
\eqn\norm{
\N_{n\ell} = 2^{4n+\ell}n! \ell!(\lambda)_{n+\ell}(\mu+\ell)_n(\lambda+1-\mu)_n 
\, , \quad \mu= \half(d-1) \, , \quad (\lambda)_n =
{\Gamma(\lambda+n)\over \Gamma(\lambda)} \, .
}
A unitary representation therefore requires as usual
$\lambda >0 ,\ge \half(d-3)$ or $\lambda=0$ when there is just the singlet vacuum
state $|0\r$. An complete orthogonal basis of states, $\{ |n\, \ell, I\r\}$, 
of the form given by \basis\ may be obtained by introducing for each $\ell$ a 
basis of symmetric traceless
tensors $\C^I{}_{\!\!i_1 \dots i_\ell}$, satisfying \tens, such that
\eqn\tenC{
\C^{I'}\!\!\cdot \C^I = \de_{I'I} \, ,
\qquad \sum_I \C^I{}_{\!\!j_1 \dots j_\ell}\C^I{}_{\!\!i_1 \dots i_\ell} =
\P^{(\ell)}{}_{\!\!j_1 \dots j_\ell, i_1 \dots i_\ell} \, ,
}
where $\P^{(\ell)}$ is here the projector onto symmetric traceless tensors of rank 
$\ell$.

The action of $\hL_{+i}$ on the basis states \basis\ takes $(n\, \ell) \to
(n{+1}\, \ell{-1}),\, (n\, \ell{+1})$. It is convenient to define the $\ell\pm 1$
rank symmetric traceless tensors
\eqn\Cpm{ \eqalign{
\C^+{}_{\!\!i, i_1 \dots i_{\ell+1}} = {}& \de_{i(i_1}
\C_{i_2 \dots i_{\ell+1})} - {\ell \over d+2\ell -3} \,
\de_{(i_1i_2} \C_{i_3 \dots i_{\ell+1})i} \, , \quad \ell = 0,1,\dots \, , \cr
\C^-{}_{\!\!i, i_1 \dots i_{\ell-1}} = {}& \C_{ i i_1 \dots i_{\ell-1}} \, , \qquad
\ell = 1,2,\dots \, , \cr}
}
so that
\eqn\lplus{
\hL_{+i} | n \, \ell , \C \r = | n\, \ell{+1}, \C^+{}_{\!\!i} \r  +
 {\ell \over d+2\ell -3} \, | n{+1}\, \ell{-1} ,  \C^-{}_{\!\!i} \r \, .
}

To obtain the spectral representation of the two point function of the
scalar field $\phi$ we need to determine the matrix elements  of $\phi$ between
$|0\r$ and arbitrary states in the representation. For the lowest weight state
\eqn\phil{
\l 0 | \phi (x) | \lambda \r = e^{-\lambda \tau} f(r) \, ,
}
and $\hL_{-i} | \lambda \r =0$ leads to, from \rep, to $\cos r f'(r) + \lambda
\sin r f(r) = 0$ so that
\eqn\Req{
f(r) = N (\cos r )^\lambda \, .
}
A general matrix element has the form
\eqn\mat{
\l 0 | \phi (x) | n \, \ell , \C \r = e^{-\lambda_{n\ell}\tau}
f_{n\ell}(r) \, Y_\ell^\C (\hxn) \, , \quad \lambda_{n\ell} = \lambda + 2n + \ell \, .
}
where we define appropriate spherical harmonics (for further properties see
appendix C, a useful summary of spherical harmonics in arbitrary dimensions is 
given in
\Sei) by
\eqn\har{
Y_\ell^\C (\hxn) = \C_{i_1 \dots i_\ell} \hxn_{i_1} \dots \hxn_{i_\ell} \, .
}
{}From \Cas, \Casd\ and \Lphi\ with $\hL_{ab}|0\r =0$ it is easy to see that
the matrix elements \mat\  satisfy
$\nab^2_{H^d}(e^{-\lambda_{n\ell}\tau}f_{n\ell}(r) \, Y_\ell^\C (\hxn)) = 
\rho^2\lambda(\lambda-d+1)e^{-\lambda_{n\ell}\tau}f_{n\ell}(r) \, 
Y_\ell^\C (\hxn)$ which may be reduced to a simple equation for $f_{n\ell}$. In order 
to find  $f_{n\ell}$ with the overall scale determined we use instead \Lphi\ for
$\hL_{+i}$, with the expression \rep\ for $L_{+i}$, and from \Cpm,
\eqn\Ypm{\eqalign{ 
\hxn_i Y_\ell^\C (\hxn) ={}&  Y_{\ell+1}^{\C^+{}_{\!\!i}} (\hxn) 
+ {\ell \over d+2\ell -3} 
Y_{\ell-1}^{\C^-{}_{\!\!i}} (\hxn) \, , \cr 
D_i Y_\ell^\C (\hxn) = {}& \ell \, Y_{\ell-1}^{\C^-{}_{\!\!i}} (\hxn) 
- \ell \hxn_i Y_\ell^\C (\hxn) =
- \ell \, Y_{\ell+1}^{\C^+{}_{\!\!i}} (\hxn) + \ell \, {d + \ell-3 \over d+2\ell -3}
Y_{\ell-1}^{\C^-{}_{\!\!i}} (\hxn) \, , \cr}
}   
to obtain
\eqn\feq{\eqalign{
-\lambda_{n\ell} \sin r f_{n\ell}(r) + \cos r f'{}_{\!\! n\ell}(r)
- \ell \cosec r f_{n\ell}(r) = {}& - f_{n \, \ell{+1}}(r)\, , \quad \ell=0,1,\dots \, ,
\cr
-\lambda_{n\ell} \sin r f_{n\ell}(r) + \cos r f'{}_{\!\! n\ell}(r)
+ ( d+\ell-3) \cosec r f_{n\ell}(r) = {}& - f_{n{+1} \, \ell{-1}}(r)\, , \,
\ell = 1,2,\dots \, , \cr}
}
relating $f_{n\ell}$ for differing $n,\ell$.
The solutions are well known, involving Jacobi polynomials 
$P_n^{(\alpha,\beta)}$ which satisfy the crucial, for our purposes, recurrence
relations, 
\eqn\Jac{\eqalign{
\cos r {\d \over \d r} P_n^{(\alpha,\beta)}& (\cos 2r) - 2(n+\alpha+\beta+ 1)
\sin r \, P_n^{(\alpha,\beta)}(\cos 2r) \cr = {}& - 2(n+\alpha+\beta+ 1) \sin r \,
P_n^{(\alpha+1,\beta)}(\cos 2r) \cr
 = {}& - 2 \alpha \cosec r \, P_n^{(\alpha,\beta)}(\cos 2r) + 2 (n+1) \cosec r
\, P_{n+1}^{(\alpha-1,\beta)}(\cos 2r)  \, . \cr}
}
Since \Req\ with $P_0^{(\alpha,\beta)}(x)=1$ gives
$f_{00}(r) = N (\cos r )^\lambda$ we may then find from \feq
\eqn\sol{
f_{n\ell}(r) = N \, (-1)^n2^{2n+\ell} n!(\lambda)_{n+\ell} (\cos r )^\lambda \,
(\sin r)^\ell P_n^{(\ell+\mu-1,\lambda-\mu)}(\cos 2r ) \, .
}
For subsequent application it is sufficient to
consider only $r=0$ when, using $P_n^{(\alpha,\beta)}(1) = (\alpha+1)_n/n!$, the
results reduce to just
\eqn\zero{
f_{n\ell}(0) = N \, (-1)^n 2^{2n} (\lambda)_n (\mu)_n \, \de_{\ell 0} \, .
}

The above results are designed to set the scene for the more involved
analysis when spin is involved. For spin one we consider a lowest weight
state $|\lambda,i\r$,  $\l \lambda,i|\lambda,j\r= \de_{ij}$, with \wei\
replaced by,
\eqn\weiV{
\hH |\lambda ,i\r = \lambda |\lambda ,i\r \, ,\qquad \hL_{-i}|\lambda ,i\r = 0\, ,
\qquad
\hL_{ij} |\lambda , k \r =  \de_{jk} |\lambda ,i\r - \de_{ik} |\lambda ,j\r \, .
}
In order to build a basis of states analogous to \basis\ it is necessary
besides \tens\ to introduce tensors $\hhC_{j i_1 \dots i_\ell}$, $\ell = 1,2,\dots$
belonging to mixed symmetry,
described by $(\ell,1,0 \dots)$ Young tableaux, representations which satisfy
the properties
\eqn\tensm{
\hhC_{j i_1 \dots i_\ell} = \hhC_{j(i_1 \dots i_\ell)} \, , \qquad
\hhC_{(j i_1 \dots i_\ell)} = 0 \, , \qquad 
\hhC_{j j i_1 \dots i_{\ell-1} } = \hhC_{j i i i_1 \dots i_{\ell-2} } =0 \, .
}
Corresponding to \basis\ we now define the states, 
\eqn\basisV{ \eqalign{
| n \, \ell \, - , \C \r = {}&\hK_+{}^{\!\! n} \hL_{+ i_1} \dots 
\hL_{+ i_{\ell-1}} |\lambda,k \r \, \C_{k i_1 \dots i_{\ell-1} } \, , \qquad \,
n=0,1, \dots \, , \ell=1,2,\dots  \, ,\cr
| n \, \ell \, + , \C \r = {}&\hK_+{}^{\!\! n} \hL_{+ i_1} \dots 
\hL_{+ i_{\ell}} \hL_{+k} |\lambda,k \r \, \C_{i_1 \dots i_{\ell} } \, , \qquad  \,
n,\ell =0,1, \dots  \, ,\cr
| n \, \ell , \hhC \r = {}&\hK_+{}^{\!\! n} \hL_{+ i_1} \dots 
\hL_{+ i_{\ell}} |\lambda,k \r \, \hhC_{k i_1 \dots i_{\ell} } \, , \qquad \qquad \
n=0,1, \dots \, , \ell=1,2,\dots  \, , \cr }
}
with the associated eigenvalues of $\hH$ taking the values,
\eqn\eigen{
\lambda_{n\ell\mp} = \lambda+ 2n+\ell\mp 1 \, , \qquad 
\lambda_{n\ell} = \lambda+ 2n+\ell \, .
}
For $\ell =1,2,\dots$ the scalar products of the states $| n \, \ell \, \mp , \C \r$
are described by a $2\times 2$ matrix,
\eqn\orthov{ 
\pmatrix{ \l n'{+1} \, \ell' \, - , \C'  | n{+1} \, \ell \, -, \C \r &
\l n'{+1} \, \ell' \, - , \C'  | n \, \ell \, + , \C \r  \cr
\l n' \, \ell' \, + , \C'  | n{+1} \, \ell \, -, \C \r &
\l n' \, \ell' \, + , \C'  | n \, \ell \, + , \C \r \cr}
=  \de_{n'n} \de_{\ell'\ell}\,  \C'\!\cdot  \C \,
{\underline \N}_{n\ell} \,  ,
}
while also
\eqn\orthoh{\eqalign{
\l n' \, \ell' , \hhC'  | n \, \ell , \hhC \r =  {}& \de_{n'n}
\de_{\ell'\ell}\,   \hhC' \! \cdot \hhC \, {\hat \N}_{n\ell} \, , \qquad
\hhC' \! \cdot \hhC = \hhC'{}_{\!ji_1 \dots i_\ell} \hhC_{ji_1 \dots i_\ell} \, , \cr
\l n' \, 0 \, +  , 1 | n \, 0\, +,  1 \r  = {}& \de_{n'n} \, \N_{n+} \, . \cr}
}
{}From appendix C
\eqn\normV{\eqalign{
{\hat \N}_{n\ell} = {}& 2^{4n+\ell}n! \ell!(\lambda+1)_{n+\ell-1}(\mu+\ell)_n
(\lambda+1-\mu)_n \, (\lambda -1) \, , \cr
\N_{n+} = {}& 2^{4n+2} n! (\lambda+1)_{n}(\mu)_{n+1}(\lambda+1-\mu)_n
(\lambda - d +2)  \, , \cr}
}
so that unitarity requires $\lambda \ge 1,d-2$. The expressions for the elements
of ${\underline \N}_{n\ell}$ are more involved and so are deferred to appendix C.

In a similar fashion to \Cpm\ we define
\eqnn\Cpm
$$\eqalignno{
\hhC^+{}_{\!\!i, j i_1 \dots i_{\ell+1}} = {}& \de_{i(i_1}
\hhC_{j|i_2 \dots i_{\ell+1})} 
-  {\ell -1 \over d + 2\ell -3}\Big (
\de_{(i_1i_2} \hhC_{j|i_3 \dots i_{\ell+1})i} -
\de_{(i_1i_2} \hhC_{i_3 \dots i_{\ell+1})ji} \Big ) \, , \cr
{} + &{1 \over d+\ell -3}\Big ( \ell \, \de_{j(i_1} \hhC_{i_2 \dots i_{\ell+1})i} - 
\de_{(i_1i_2} \hhC_{j|i_3 \dots i_{\ell+1})i} -
(\ell -1) \de_{(i_1i_2} \hhC_{i_3 \dots i_{\ell+1})ji} \Big ) \, ,\cr
\hhC^-{}_{\!\!i, j i_1 \dots i_{\ell-1}} = {}& \hhC_{j i_1 \dots i_{\ell-1}i}
 - \hhC_{(i_1 \dots i_{\ell-1})ji} \, , \qquad
\hhC_{i, i_1 \dots i_{\ell}} = \hhC_{ii_1 \dots i_{\ell}} \, , & \Cpm \cr}
$$
where $\hhC^\pm{}_{\!\!i,j i_1 \dots i_\ell}$ are mixed symmetry tensors satisfying
\tensm\ while $\hhC_{i, i_1 \dots i_{\ell}}$ is a symmetric tensor obeying \tens. 
Furthermore from $\C_{i_1 \dots i_\ell}$ we may also define a mixed symmetry tensor
satisfying \tensm\ by
\eqn\mix{
\C_{i,j i_1 \dots i_\ell} = {1\over \ell+1}\bigg (
\de_{i(i_1} \C_{j|i_2 \dots i_\ell)} - \de_{ij} \C_{i_1 \dots i_\ell}
- {\ell-1 \over d+\ell-4} \Big ( \de_{(i_1 i_2} \C_{j|i_3 \dots i_\ell)i}
- \de_{j(i_1} \C_{i_2 \dots i_\ell)i}  \Big ) \bigg ) \, .
}
With the states defined by \basisV\ we then have
\eqn\raiseV{\eqalign{ \!\!\!\!
\hL_{+i} | n \, \ell \, + , \C \r = {}& | n\, \ell{+1}\, + , \C^+{}_{\!\!i} \r  +
 {\ell \over d+2\ell -3} \, | n{+1}\, \ell{-1}\, +  ,  \C^-{}_{\!\!i} \r \, , \cr
\!\!\!\!
\hL_{+i} | n \, \ell \, - , \C \r = {}& | n\, \ell{+1}\, - , \C^+{}_{\!\!i} \r  
+ | n \, \ell , \C_i \r +
{(\ell-1)(d+\ell-3) \over (d+2\ell -3)(d+\ell-4)} \, 
| n{+1}\, \ell{-1}\, -  ,  \C^-{}_{\!\!i} \r \cr
{}& + {d-3 \over (d+2\ell -3)(d+\ell-4)} \, 
| n\, \ell{-1}\, +  ,  \C^-{}_{\!\!i} \r \, , \cr
\!\!\!\!
\hL_{+i} | n \, \ell , \hhC \r = {}& | n\, \ell{+1} , \hhC^+{}_{\!\!i} \r 
+ {\ell-1 \over d+2\ell-3} \, | n{+1}\, \ell{-1} , \hhC^-{}_{\!\!i} \r \cr
{}& + {1\over d+\ell-3} \Big ( | n\, \ell\, +  ,  \hhC_{i} \r   
- | n{+1}\, \ell\, -  ,  \hhC_{i} \r \Big ) \, .\cr}
}

Following a similar route to the case of scalar fields we consider now
the matrix elements of a vector current $A_a(x)$ between the singlet $|0\r$
and the states of this spin one representation.\foot{In four dimensions the spin
one representation was discussed by Fronsdal \Fron\ who obtained 
equivalent results for $\l 0 | A_a | \lambda ,k \r$ to those obtained below.}
For the lowest weight state we take, consistent with $\hxn_j A_j =0$,
\eqn\Al{ \eqalign{
\l 0 | A_- (x) | \lambda ,k \r = {}& 0 \, , \qquad
\l 0 | A_+ (x) | \lambda ,k \r = \hxn_k e^{-(\lambda+1)\tau} g(r) \, , \cr
\l 0 | A_j (x) | \lambda ,k \r = {}&  (\de_{jk} - \hxn_j \hxn_k )
e^{-\lambda \tau} f(r) \, . \cr}
}
Imposing the equations following from $\hL_{-i} | \lambda ,k \r = 0$, which also
requires the vanishing of the $A_-$ matrix element, and using \rep\ we get
\eqn\fgeq{
\lambda \sin r f + \cos r f' =0 \, , \quad f + \half \cosec r g = 0 \, , \quad
(\lambda + 1 ) g + \cos r g' - \cosec r g = 0 \, ,
}
for which the solution is
\eqn\fgsol{
f(r) = N(\cos r )^\lambda \, , \qquad g(r) = -2 N(\cos r )^\lambda \sin r \, .
}
It is easy to verify from \nabA\ that this solution implies\foot{Alternatively
for a spin 0 lowest weight state we find $\l 0 |  A_- (x) | \lambda \r =
N e^{-(\lambda-1)\tau} (\cos r )^{\lambda -1}$, 
$ \l 0 | A_j (x) | \lambda \r = 0 , \ \l 0 |  A_+ (x) | \lambda \r =
N e^{-(\lambda+1)\tau} (\cos r )^{\lambda -1} \cos 2r$ but in this case
$\l 0 | A_a (x) | \lambda \r = - \nab_a Ne^{-\tau}(\cos r)^\lambda/\lambda$. It
is useful to also note that $\l 0 | \nab{\cdot A}(x) | \lambda \r = -(\lambda-d+1)
N e^{-\tau}(\cos r)^\lambda$ so the current is conserved if $\lambda=d-1$.}
\eqn\div{
\l 0 | \nab{\cdot A} | \lambda ,k \r = 0 \, .
}
For other states the matrix elements may be determined in a similar fashion
to previously. To write expressions for matrix elements we need besides \har\
vector spherical harmonics which may be defined in terms of tensors satisfying
\tensm,
\eqn\harV{
\Y_{\ell,j}^\hhC(\hxn) 
= \hhC_{j i_1 \dots i_\ell} \hxn_{i_1} \dots \hxn_{i_\ell} \, .
}
Like $D_j Y_\ell^\C(\hxn)$ this satisfies $n_j \Y_{\ell,j}^\hhC(\hxn)=0$ and
also we have $D_j \Y_{\ell,j}^\hhC(\hxn)=0$. Using the above spherical harmonics
the matrix elements may then be written in general as
\eqnn\matV
$$\eqalignno{
\l 0 | A_j (x) | n \, \ell \, \pm , \C \r = {}& e^{- \lambda_{n\ell\pm}\tau}
f_{n\ell\pm}(r) D_j Y_\ell^\C(\hxn) \, , \quad
\l 0 | A_j (x) | n \, \ell , \hhC \r =  e^{- \lambda_{n\ell}\tau}
{\hat f}_{n\ell}(r) \Y_{\ell,j}^\hhC(\hxn) \, , \cr
\l 0 | A_+ (x) | n \, \ell \, \pm , \C \r  = {}& e^{- (\lambda_{n\ell\pm}+1)\tau}
g^+{}_{\!\! n\ell\pm}(r)  Y_\ell^\C(\hxn) \, , \qquad 
\l 0 | A_+ (x) | n \, \ell , \hhC \r  = 0 \, , \cr
\l 0 | A_- (x) | n \, \ell \, \pm , \C \r  = {}& e^{- (\lambda_{n\ell\pm}-1)\tau}
g^-{}_{\!\! n\ell\pm}(r)  Y_\ell^\C(\hxn) \, , \qquad 
\l 0 | A_- (x) | n \, \ell , \hhC \r  = 0 \, . & \matV
\cr}
$$
In order to use \raiseV\ we need besides \Ypm
\eqn\DY{\eqalign{
\hxn_i D_j Y_\ell^\C(\hxn) = {}& D_j\big ( \hxn_i Y_\ell^\C(\hxn) \big ) 
 - (\de_{ij} - \hxn_i \hxn_j ) Y_\ell^\C(\hxn) \cr
={} & \ell \, \Y_{\ell,j}^{\C_i}(\hxn) +
D_j \Big ( {\ell \over \ell+1}\,Y_{\ell+1}^{\C^+{}_{\!\!i}} (\hxn) + 
{\ell (d+\ell-3)\over (d+2\ell -3)(d+\ell-4)} 
Y_{\ell-1}^{\C^-{}_{\!\!i}} (\hxn) \Big ) \, , \cr
D_i D_j Y_\ell^\C(\hxn) {}& + \hxn_j D_i Y_\ell^\C(\hxn) 
= D_j  D_i Y_\ell^\C(\hxn) + \hxn_i D_j Y_\ell^\C(\hxn)   \cr
={} & \ell \, \Y_{\ell,j}^{\C_i}(\hxn) -
D_j \Big ( {\ell^2 \over \ell+1}\,Y_{\ell+1}^{\C^+{}_{\!\!i}}  (\hxn)
- {\ell (d+\ell-3)^2\over (d+2\ell -3)(d+\ell-4)} 
Y_{\ell-1}^{\C^-{}_{\!\!i}} (\hxn) \Big ) \, , \cr}
}
using
\eqn\nnY{
\!\!\! (\de_{ij} - \hxn_i \hxn_j ) Y_\ell^\C(\hxn)  
=  D_j \Big ( {1\over \ell+1}\,Y_{\ell+1}^{\C^+{}_{\!\!i}}  (\hxn) 
- {\ell \over (d+2\ell -3)(d+\ell-4)} Y_{\ell-1}^{\C^-{}_{\!\!i}} (\hxn) \Big ){} -
\ell \, \Y_{\ell,j}^{\C_i}(\hxn) \, ,
}
and furthermore
\eqn\Ypm{\eqalign{
\!\!\!\hxn_i \Y_{\ell,j}^\hhC (\hxn) ={}& \Y_{\ell+1,j}^{\hhC^+{}_{\!\!i}} (\hxn) + 
{\ell - 1 \over d+2\ell -3} \, \Y_{\ell-1,j }^{\hhC^-{}_{\!\!i}} (\hxn) 
- {1 \over d+\ell -3}  {1\over \ell}\, D_j Y_\ell^{\hhC_i}(\hxn) \, , \cr
\!\!\! D_i \Y_{\ell,j}^\hhC (\hxn) + \hxn_j  \Y_{\ell,i}^{\hhC} (\hxn) 
= {}& - \ell \, \Y_{\ell+1,j}^{\hhC^+{}_{\!\!i}} (\hxn) +
(\ell-1){d + \ell -3 \over d+2\ell -3} \, \Y_{\ell-1,j }^{\hhC^-{}_{\!\!i}} (\hxn)\cr
{}& - {d-3 \over d+\ell -3} {1\over \ell}\, D_j Y_\ell^{\hhC_i}(\hxn) \, . \cr}
}

With these results we may obtain equations determining $f_{n\ell\pm}(r)$,
${\hat f}_{n\ell}(r)$, $g^+{}_{\!\! n\ell\pm}(r)$, $g^-{}_{\!\! n\ell\pm}(r)$ by
using the results for the commutators of $\hL_{+i}$ in \LLA\ with, from \Al\ and
\fgsol,
\eqn\init{
f_{01-}(r) = N(\cos r)^\lambda \, , \qquad g^+{}_{\!\! 01-}(r)  = 
- 2 N(\cos r)^\lambda \sin r \, , \qquad g^-{}_{\!\! 01-}(r) = 0 \, .
}
The details are quite lengthy so we restrict
ourselves here to the main results. The essential equations are similar in general
form to \feq\ along with various algebraic relations necessary for consistency.
{}From $\l 0 | A_\pm (x) | n \, \ell , \hhC \r  = 0 $ we may obtain,
using $\Y_{\ell,i}^\hhC(\hxn) =  Y_\ell^{\hhC_i}(\hxn) $, for $\ell=1,2,\dots$,
\eqn\ggf{
g^+{}_{\!\! n{+1}\, \ell -}(r) -g^+{}_{\!\! n\ell +} (r) = 0 \, , \qquad
g^-{}_{\!\! n{+1}\, \ell -}(r) -g^-{}_{\!\! n\ell +} (r) = 
- 2(d+\ell-3){\hat f}_{n\ell}(r) \, ,
}
and from the analysis of $[\hL_{+i}, A_j ]$  we may obtain
\eqn\fff{
\hf_{n\, \ell{+1}}(r) + \hf_{n{+1}\, \ell{-1}}(r) = \ell \big ( f_{n{+1}\, \ell -}(r) -
f_{n\ell +}(r) \big ) \, , \qquad \ell =2,3,\dots \, .
}
The results, apart from $g^+{}_{\!\! n\ell +}(r)$ which is easily obtained
from \ggf, are then
\eqn\solv{\eqalign{ 
\!\!\!\!\!\!\!\!\! g^+{}_{\!\! n\ell-}(r) = {}& - N \, (-1)^n2^{2n+\ell} n!
(\lambda+1)_{n+\ell-1} 
(\cos r )^\lambda \, (\sin r)^\ell P_n^{(\ell+\mu-1,\lambda-\mu)}(\cos 2r ) \, , \cr
\!\!\!\!\!\!\!\!\! \hf_{n\ell} (r) = {}& N \, (\lambda-1)(-1)^n2^{2n+\ell} n!
(\lambda+1)_{n+\ell-1} 
(\cos r )^\lambda \, (\sin r)^\ell P_n^{(\ell+\mu-1,\lambda-\mu)}(\cos 2r ) \, , \cr 
\!\!\!\!\!\!\!\!\! f_{n\ell-}(r) ={}& N \, (-1)^n2^{2n+\ell-1} n!(\lambda)_{n+\ell-1}
(\cos r )^\lambda \,(\sin r)^{\ell-1} \cr
\!\!\!\!\!\!\!\!\! & \ \times {1\over \lambda\ell}\Big ( \ell
P_n^{(\ell+\mu-1,\lambda-\mu-1)}(\cos 2r ) + (\lambda-1)
P_n^{(\ell+\mu-2,\lambda-\mu)}(\cos 2r ) \Big ) \, , \cr
\!\!\!\!\!\!\!\!\! f_{n\ell+}(r) ={}& - N \, (-1)^n2^{2n+\ell} n!(\lambda+1)_{n+\ell-1}
(\cos r )^\lambda \,(\sin r)^{\ell-1} \cr
\!\!\!\!\!\!\!\!\! & \ \times  \Big ( 2(n+1)
P_{n+1}^{(\ell+\mu-1,\lambda-\mu-1)}(\cos 2r ) + (\lambda-1)
P_n^{(\ell+\mu-1,\lambda-\mu)}(\cos 2r ) \Big ) \, , \cr
\!\!\!\!\!\!\!\!\! g^-{}_{\!\! n\ell-}(r) = {}& - N \, (-1)^n2^{2n+\ell} n!
(\lambda+1)_{n+\ell-2} (\cos r )^\lambda \, (\sin r)^\ell\cr
\!\!\!\!\!\!\!\!\! & \ \times (\lambda + \mu + n + \ell -2)\,  
P_{n-1}^{(\ell+\mu-1,\lambda-\mu)}(\cos 2r ) \, , \cr
\!\!\!\!\!\!\!\!\! g^-{}_{\!\! n\ell+}(r) = {}& N \, (-1)^n2^{2n+\ell+1} n!
(\lambda+1)_{n+\ell-1} (\cos r )^\lambda \, (\sin r)^\ell\cr
\!\!\!\!\!\!\!\!\! & \ \times \big(2(\lambda + n )(\mu + n+ \ell )- 
(\lambda-1)\ell \big ) \, P_{n}^{(\ell+\mu-1,\lambda-\mu)}(\cos 2r ) \, , \cr}
}
where also $g^-{}_{\!\! 0\ell -}(r)=0$. Otherwise, except for $f_{n\ell+}(r)$
which is defined only for $\ell \ge 1$, the formulae given by \solv\
are valid for all $n,\ell$ required by the definition of the basis in \basisV. As a
consequence of \div\ the solutions for all matrix elements given by \matV\ and
\solv\ are consistent with $\nab{\cdot A}=0$.

For $r=0$, in a similar fashion to \zero, the non zero results reduce to just
\eqn\rzero{ \eqalign{
g^+{}_{\!\! n0+}(0) = {}& g^-{}_{\!\! n0+}(0) =
N (-1)^n2^{2n+2} (\lambda+1)_{n} (\mu)_{n+1} \, , \cr
f_{n1-}(0) = {}& N (-1)^n2^{2n}(\lambda+1)_{n-1}(\mu+1)_{n-1}(\mu\lambda+n) \, , \cr
f_{n1+}(0) = {}& - N (-1)^n2^{2n+1}(\lambda+1)_{n}(\mu+1)_{n}(2n+\lambda +d) \, . \cr}
}

\newsec{Spectral Representations}

The spectral representation for two point functions encodes completely the
analyticity requirements following from locality for any quantum field theory
together with the conditions of
unitarity which reduce to positivity of the spectral weight function. Such
representations may also be found for spaces of constant curvature \Dus\ although
they are significantly less straightforward to determine in specific cases.
For two dimensional field theories on flat space
the proof of the $c$-theorem may be recast with added insight in terms of
the spectral representation for the energy momentum tensor two point function \Cap.
It is therefore natural to also consider the spectral representation for
the energy momentum tensor two point function on spaces of constant curvature.

For the scalar field $\phi$ we first compute the two point function corresponding to
summing over the intermediate states $|n \, \ell , I \r$ labelled by $\lambda$
and for which the Casimir operator has the value $\lambda(\lambda - d +1)$.
It is simplest to set $r=0$ and then only $\ell=0$ states contribute,
in a similar fashion to \Cap,  and using \zero\ we have
\eqn\ttt{
\l 0 | \phi(\tau , 0 , \hxn) \phi(0,0,\hxn) | 0 \r 
= \sum_{n=0} {f_{n0}(0)^2 \over \N_{n0}} \, e^{-\lambda_{n0}\tau} = N^2 
e^{-\lambda \tau}F (\lambda, \mu; \lambda+1-\mu; e^{-2\tau}) \, . 
}
For  $x=(\tau , 0 , \hxn), \, y = (0,0,\hxn)$ we have $\theta = \tau$.
With the appropriate choice for $N^2$ this is the standard Green function for
$-  \nab^2 + \lambda(\lambda-d+1)\rho^2$, as verified in appendix B,
\eqn\Neq{
G_\lambda(\theta) =  {\rho^{d-2}\over 2 \pi^\mu} \, 
{\Ga(\lambda) \over \Ga ( \lambda+1 -\mu )} \, 
e^{-\lambda \theta}F (\lambda, \mu; \lambda+1-\mu; e^{-2\theta}) \, . 
}
For the conformally
coupled case the two choices of $\lambda$ are $\half d$ for Dirichlet boundary
conditions and $\half(d-2)$ for the Neumann case. The spectral representation
for the scalar two point function can then be written as
\eqn\spectphi{
\l \phi(x)\phi(y)\r = \int_{{1\over 2}(d-3)}^\infty  \! \!\!\! \!\!\!\!\!
\d \lambda \, \rho_\phi(\lambda) \, G_\lambda(\theta) \, .
}

In a similar fashion to the discussion of the scalar two point function in \ttt\
we may now determine the contribution of a single spin 1 irreducible representation
to the vector two point function. Choosing the same configuration as previously
the non zero components are
\eqn\tV{\eqalign{
\l 0 | A_{\pm} (\tau , 0 , \hxn) A_{\pm} (0,0,\hxn) | 0 \r = {}&
\l 0 | A_{\pm} (\tau , 0 , \hxn) A_{\mp} (0,0,\hxn) | 0 \r = 
e^{\mp \tau}D_{\lambda} (\tau) \, , \cr
\l 0 | A_j (\tau , 0 , \hxn) A_k (0,0,\hxn) | 0 \r = {}& (\de_{jk} - \hxn_j \hxn_k)
E_{\lambda}(\tau) \, . \cr}
}
Using the results \rzero\ with \normV\ then \ttt\ adapted to this case gives
\eqn\Gone{\eqalign{
D_{\lambda} (\tau) = {}& \sum_{n=0} 
{\big ( g^\pm{}_{\!\! n0+}(0)\big)^2 \over \N_{n+}} e^{-(\lambda+2n+1)\tau} \cr
= {}& {4N^2 \over \lambda-d+2} \, e^{-(\lambda+1) \tau}\sum_{n=0}{1\over
n!}\, {(\lambda+1)_n (\mu)_{n+1}\over (\lambda + 1 - \mu)_n} \, e^{-2n\tau} \cr
={}& {4N^2\mu \over \lambda-d+2} \, e^{-(\lambda+1) \tau}
F\big (\lambda+1, \mu+1;\lambda+1-\mu; e^{-2\tau}\big ) \, , \cr}
}
and also, by summing over intermediate states $|n \, 1 \, \pm , \C \r
= |n \, 1 \, \pm , i \r \C_i$,
\eqn\Ktwo{
E_{\lambda} (\tau) =  e^{-\lambda\tau} \sum_{n=0} A_n e^{-2n\tau} \, ,
}
where
\eqn\An{
A_0 = N^2 \, , \qquad  A_{n+1} =
{\pmatrix{ f_{n{+1}\, 1 \, -}(0) & f_{n\, 1 \, +}(0)} {\underline \N}_{n1}^{\, -1}}
\down{\pmatrix{f_{n{+1}\, 1 \, -}(0) \cr  f_{n\, 1 \, +}(0)}} \, .
}
Using \rzero\ for $f_{n1\pm}(0)$ and the results of appendix C we have
\eqn\Anr{
A_n = N^2 \, {1\over n!}\, {(\lambda)_n (\mu)_{n}\over (\lambda + 1 - \mu)_n} \, 
\Big ( 1 + {2n(\lambda+n) \over \lambda\mu(\lambda - d+2)} \Big ) \, ,
}
so that
\eqn\Ktwor{
E_{\lambda} (\tau) = \bigg ( 1 + {1\over 2\lambda\mu(\lambda - d+2)} 
\Big ( {\d^2 \over \d \tau^2} - \lambda^2 \Big ) \bigg ) \, e^{-\lambda \tau}
F(\lambda, \mu; \lambda+1-\mu; e^{-2\tau}) \, .
}

The functions $D_{\lambda}$ and $E_{\lambda}$ are constrained by  conservation
equations. To obtain these and
to show the relation to the general formalism used in section 8 we may write
in general for the correlation function of two vector currents
\eqn\AAcor{
\l A_\mu(x) A_\alpha(y) \r = \hx_\mu \hy_\alpha \, D(\theta) + ( I_{\mu\alpha}
+ \hx_\mu \hy_\alpha )\,  E(\theta) \, .
} 
Choosing, as in \tV, $x=(\tau , 0 , \hxn), \, y=(0,0,\xi)$, when $\theta=\tau$,
 then the non
zero components of $\hx, \hy$ are just $\hx_\tau=1, \, \hy_\tau=-1$ and also
we have $I_{\tau\tau}=1$. Using \nabphi\ for these $x,y$ to obtain the equivalent
results in the basis used in \tV\ given by $A_a = e_a{}^\mu A_\mu$ gives then
\eqn\nbasis{
\hx_\pm = e^{\mp\tau} \, , \quad \hy_\pm = 1 \, , \quad I_{\pm\pm} = I_{\pm\mp}
= - e^{\mp\tau} \, , \quad I_{ij}= \de_{ij} - \hxn_i \hxn_j \, .
}
Applying \nbasis\ in  \AAcor\ then leads to identical expressions for each
components as in \tV\ where $D,E$ are given by just
the single spin one irreducible representation specified by $\lambda$. The 
conservation equation $\nab^\mu \l A_\mu(x) A_\alpha(y) \r = 0$ using the results
in section 5 is easily seen to give
\eqn\GGcon{
\sinh \theta \, D{}^{\prime}(\theta) + (d-1) \cosh \theta \, D(\theta) =
- (d-1) \, E(\theta) \, .
}
It is straightforward to check that  this is satisfied by \Gone\ and \Ktwor. 
By extension of \Neq\ we define
\eqn\Gn{
G_{\lambda,n}(\theta) =  {\rho^{d-2+2n}\over 2 \pi^\mu} \,
{\Ga(\lambda+n) \over \Ga ( \lambda+1 -\mu )} \,
e^{-(\lambda+n) \theta}F (\lambda+n, \mu+n ; \lambda+1-\mu; e^{-2\theta}) \, ,
}
and then the representation for a general vector two point function may then be
obtained by expressing $D$ as, if $d>3$,
\eqn\spectV{
D(\theta) = \int_{d-2}^\infty  \! \!\!\! \!\!\! 
\d \lambda \, \rho_V(\lambda) \, G_{\lambda,1}(\theta) \, ,
}
where $\rho_V(\lambda)$ is a positive weight function, with 
$E$ determined by \GGcon.

It is of interest to consider separately the contribution from $\lambda=d-2$
when we have
\eqn\Gld{\eqalign{
D_{0}(\theta) ={}& C\, G_{d-2,1}(\theta) =
C\,{\rho^d \over 2S_d} \, {\cosh \theta\over(\sinh\theta)^d} \, , \cr 
E_{0}(\theta) ={}& {C\over d-1} \, {\rho^d \over 2S_d}{\over (\sinh\theta)^d} \, . \cr}
}
With these expressions the vector two point function \AAcor\  may be written just in 
terms of the scalar $G_S$
\eqn\AAld{
\l A_\mu(x) A_\alpha(y) \r_{S} = \pr_\mu G_{S}(\theta)
{\overleftarrow \pr}{}_{\! \alpha}
\, , 
}
where
\eqn\Gdd{
G_S(\theta) = C\, {d-2\over (d-1)^2} \, G_{d-1}(\theta) \, .
}
Thus $G_S$ is proportional to the scalar Green function, as given by \Neq,
for $\lambda=d-1$ which since then $\nab^2 G_S(\theta) = 0$ for $x\ne y$
ensures that \AAld\ satisfies the conservation equation.  In this special 
case the contribution of the spin one representation intermediate states is therefore
identical with those of spin zero. This is in accord with the representation theory
since, if $\lambda=d-2$ as may be seen in (C.7), the states 
$\{| n \, 0 \, + , \C \r\}$ span an
invariant subspace with $|0 \, 0 \, + , 1 \r = \hL_{+k}|\lambda,k\r$,
annihilated by $\hL_{-i}$, the lowest weight state.
This representation formed by this subspace is identical with a spin zero 
representation corresponding to $\lambda=d-1$.
Summing over intermediate states in this space gives just the result \AAld\
with \Gdd.

Our primary interest here is of course the energy momentum tensor two point
function. The unitary representation of the isometry group $O(d-1,2)$ for a
spin two lowest weight state, $|\lambda, ij\r = |\lambda, ji \r , \,
|\lambda, ii\r=0$ which satisfies analogous conditions to \weiV, may in principle
be constructed along similar lines to the spin one case considered above and 
in appendix C, but this would be tedious to carry out.
The norm of the state $\hL_{+j}|\lambda, ij\r$ is proportional to $\lambda-d+1$ so
for unitarity it is necessary that $\lambda \ge d-1$ in this case. Although we
do not here undertake the detailed relevant calculations
the results for spin one nevertheless suggest a natural conjecture for the
contribution of the representation built on the lowest weight state
$|\lambda, ij\r$ to the energy momentum tensor two point function. 
To describe this we first decompose  the bi-tensor  $\Gamma_{\mu\nu,\alpha\beta}(x,y)$
given by \twoG, with \UU, for $x=(\tau , 0 , \hxn), \, y=(0,0,\xi)$ in the basis
given by \nbasis\ when we obtain
\eqn\nGam{ \eqalign{
\Gamma_{\pm\pm,\pm\pm} = {}& e^{\mp2\tau} (
R-4S+2T+2U+V )  \, , \cr
\Gamma_{\pm j , \pm l} = {}& e^{\mp \tau} \hde_{jl} ( S-T ) \, , \qquad
\Gamma_{ij,\pm\pm} = \hde_{ij} (U+V) \, , \cr
\Gamma_{ij,kl} = {}& \big (\hde_{ik} \hde_{jl} + \hde_{il}\hde_{jk} \big ) T 
+ \hde_{ij} \hde_{kl}V \, , \quad
\qquad \hde_{ij} = \de_{ij} - \hxn_i \hxn_j \, , \cr}
}
and other components are trivially related to those in \nGam.
With the definitions \defQ\ and \trace\ we have
\eqn\RQ{
R-4S+2T+2U+V = {d-1\over d} \, Q  +{2d-1\over d}\, P_1 +{1\over d}\, P_2 \, .
}
The conservation equations (A.2) demonstrate that $\Gamma_{\pm\pm,\pm\pm}$
determines the other components, assuming the traceless conditions \trace, just
like $D$ determines $E$ in \GGcon\ for the vector current two point function.

Based on analogy with \spectV\ we conjecture the spectral representation for 
spin 2 intermediate states can therefore be written for $Q$ in the form, with 
the definition \Gn,
\eqn\spectT{
Q(\theta) = S_d \int_{d-1}^\infty  \! \!\!\! \!\!\!
\d \lambda \, \rho_2(\lambda) \, G_{\lambda,2}(\theta) \, .
}
Of course, by virtue of the conservation equations, 
this determines the spectral representation for
the whole energy momentum tensor two point function since for intermediate
spin 2 states it is automatically traceless. The general conserved
energy momentum tensor two point function, as given by \TTsim, can then be expressed
as in \TTgen,
\eqn\TTgenH{
S_d^{\, 2}\langle T_{\mu\nu}(x) T_{\alpha \beta}(y) \rangle_{\rm con} =
\Gamma_{0,\mu\nu,\alpha\beta} (x,y) + \Gamma_{2,\mu\nu,\alpha\beta} (x,y) \, ,
}
where the spin 0 piece is determined by $F_0(\theta)$,
\eqn\TTzeroH{
\Gamma_{0,\mu\nu,\alpha\beta} (x,y) =
\big ( \nab_\mu \nab_\nu - g_{\mu\nu} \nab^2 + (d-1)\rho^2 g_{\mu\nu} \big )
F_0(\theta) \big ( {\overleftarrow \nab_{\!\alpha}}\! {\overleftarrow \nab_{\!\beta}}
- {\overleftarrow \nab}{}^2 g_{\alpha\beta} + (d-1)\rho^2 g_{\alpha\beta}\big )
\, ,
}
and the spin 2 piece is traceless and may be written as in \TTC\ in terms of 
$F_2(\theta)$.
Since $Q$ determines $\Gamma_{2,\mu\nu,\alpha\beta}$ \spectT\ implies therefore a 
spectral representation for the spin 2 part of $\l T_{\mu\nu}(x)T_{\alpha\beta}(y) \r$
determined by the positive weight function $\rho_2(\lambda)$. 
Equivalently by integrating (A.10) with $Q$ given by \spectT\ a spectral 
representation for $F_2$ may be found.  The spectral representation for the spin 0 
part $\Gamma_{0,\mu\nu,\alpha\beta}$
is also obtained directly by writing a representation analogous to \spectphi\ for 
$F_0$,
\eqn\spectTS{
F_0(\theta) =  S_d \int_{\lambda_0}^\infty   \!\!\!\!
\d \lambda \, \rho_0(\lambda) \, G_{\lambda}(\theta) \, ,
}
where $\lambda_0$ is in general restricted just by the unitarity bound $\lambda_0 >
\half(d-3)$ if $d>3$. In \Cap\ it is suggested that $\lambda_0=d$ in order to
ensure that $\int \d^d x \sqrt g \, g^{\mu\nu} T_{\mu\nu}$ is well defined, in
the coordinates corresponding to the metric in \metric, when $\sqrt{g_{H^d}}
= (\cos r)^{-d} (\sin r)^{d-2} \sqrt{g_{S^{d-2}}}$, as $x$ approaches the boundary
$r= {\pi\over 2}$, $e^\theta \sim 2\cosh \tau/ \cos r$. Directly from \TTzeroH\ we 
have
\eqn\spectTh{
S_d^{\, 2}g^{\mu\nu}(x)g^{\alpha\beta}(y)\l T_{\mu\nu}(x) T_{\alpha \beta}(y)
\r_{\rm con} = (d-1)^2 (-\nab^2 + d\rho^2)^2 \! F_0(\theta)\, ,
}
and since for $x\ne y$ this is identical with $\l \Theta (x) \Theta (y) \r$ we 
must have, neglecting possible subtractions,
\eqn\Thspec{
S_d\l \Theta (x) \Theta (y) \r = \int_{\lambda_0}^\infty   \!\!\!\!
\d \lambda \, \rho_\Theta(\lambda) \, G_{\lambda}(\theta) \, , \quad
\rho^4 (d-1)^2(\lambda+1)^2(\lambda-d)^2 \rho_0(\lambda) = \rho_\Theta(\lambda)\, .
}
so that this determines $\rho_0(\lambda)$ in terms of $\rho_\Theta(\lambda)$
except for $\lambda=d$.  As an illustrative example  we consider
free massive scalars in appendix D. However, although \TTzeroH\ is valid
in general for $\theta>0$, it may be extended to give a well defined distribution,
for test functions non zero at $x=y$, only if, for large $\lambda$, $\rho_0(\lambda) =
{\rm O}(\lambda^\alpha)$ with $\alpha<1$. Assuming \spectTS\ and \spectTh\ we have,
if $\lambda_0 \ge d$,
\eqn\intTT{ \eqalign{
\G_0 = {}& S_d
{1\over \rho^d} \int \! \d^d y \sqrt g \,
g^{\mu\nu}(x)  g^{\alpha\beta}(y)\langle T_{\mu\nu}(x) T_{\alpha \beta}(y)
\rangle_{\rm con} \cr
= {}& (d-1)^2 d^2 {1\over \rho^{d-2}} \int_{\lambda_0}^\infty   \!\!\!\!
\d \lambda \, \rho_0(\lambda) \, {1\over \lambda(\lambda-d+1)} \, , \cr}
}
which ensures that $\G_0 >0$.
The spectral representation may be modified
to allow for a large $\lambda$ behaviour with $\alpha<3$ by introducing a subtraction,
\eqn\subtr{
F_0(x,y) =  \G_0{\rho^{d-4}\over (d-1)^2 d^2}S_d \de^d (x,y) + 
S_d {1\over \rho^2} \nab^2 \!  \int_{\lambda_0}^\infty   \!\!\!\!  \d \lambda \, 
{\rho_0(\lambda)\over \lambda(\lambda-d+1)}  \, G_{\lambda}(\theta) \, ,
}
if $\lambda_0 > d-1$. Clearly this depends on the subtraction constant $\G_0$, 
for which there is then no necessary positivity constraint, as well as 
$\rho_0(\lambda)$.  In \Thspec\
it is necessary to require $\alpha <-1$ is subtractions are to be avoided.
In four dimensions the expected behaviour in renormalisable field theories away
from fixed points is $\alpha=1$, apart from powers of $\ln \lambda$, so that it is
then necessary to use \subtr. Of course unless there is a well defined representation
for $\l \Theta (x) \Theta (y) \r$ such as given by \spectTh\ and \subtr\
relations involving local contact terms like \trTtheta\ are without significance. 

An important role is played by the contribution to $F_0(\theta)$ for $\lambda=d$ 
for which $\Gamma_{0,\mu\nu,\alpha\beta}(x,y)$ satisfies the traceless conditions 
for $x\ne y$. In this case, following \FG\ since $\Gg_0=S_d G_d$,
\eqn\Fth{
\rho_0(\lambda) =  C_0 \rho^{d-2} \de(\lambda - d ) \quad
\Rightarrow \quad F_0(\theta) = C_0 \rho^{d-2} \, S_d G_{d}(\theta) \, ,
}
leads to a form for $\Gamma_{0,\mu\nu,\alpha\beta}$ which may be expressed in terms
of the result calculated in \Qd\ for $Q_0$. From the definition of $G_{\lambda,2}$
in \Gn\ we have
\eqn\Qth{
Q_0(\theta) = 4d(d+1) C_0 \rho^{d-2} \, S_d G_{d-2,2}(\theta) \, . 
}
This result shows exact agreement with the contribution expected from a spin 2 
representation,  as in \spectT, for $\lambda=d-2$. Although this is outside the 
unitarity bound it corresponds to the invariant subspace present for this value
of  $\lambda$ and which may be constructed from the spin zero lowest weight state
$\hL_{+k}\hL_{+l}|\lambda, kl\r$, which is then annihilated by $\hL_{-i}$. 

If we assume the minimal form given by \TTconH\ for conformally invariant theories
we find
\eqn\Qconf{
Q(\theta)_{\rm conformal} = C_T \Big ( {1\over s^d} + {1\over \bs^d} \Big )
= C_T \rho^{2d} \, e^{-d\theta} \big ( (1-e^{-\theta})^{-2d}  +
(1+e^{-\theta})^{-2d}  \big ) \, .
}
The spectral representation \spectT\ can be written in this case in the form
\eqn\spectQ{
Q(\theta)_{\rm conformal} = 2 C_T  \rho^{2d} \sum_{r=0}
A_r \, e^{-(d+2r)\theta} F \big ( d+2r, \mu+2; \mu+2r ; e^{-2\theta} \big ) \, .
}
By expanding both \Qconf\ and \spectQ\
in powers of $e^{-2\theta}$ we find, by matching the first ten terms,
\eqn\Ar{
A_0 = 1 \, , \qquad A_r = 2\, {(d)_{2r} (d-2)_{2r-1} \over (2r)! (\mu)_{2r-1}} \, ,
\ \ r=1,2,\dots \, .
}
For $d=3$, when $A_r = 2(r+1)(2r+1)$ if $r>0$, the hypergeometric functions may be 
reduced to elementary functions and the summation carried out explicitly. As
required $A_r>0$ for $d>2$ and $A_r=0$ if $r>0$ and $d=2$. For $r>0$ the
coefficients $A_r$ determine the appropriate $\rho_2(\lambda)$ in \spectT\ while
from \Qth\ we must have\foot{The coefficient $C_0$ was calculated directly in
\Cap\ for free scalar fields with exact agreement, with due regard for conventions,
with this result.} 
\eqn\CCT{
2^{d-1}d(d+1)(d-1)C_0 = C_T \, .
}
The remaining contribution is then given by taking
\eqn\rspec{
S_d \rho_2(\lambda)_{\rm conformal} = 4 \pi^\mu C_T \rho^{d-2} \sum_{r=1} 
{\Gamma(\mu+2r)\over \Gamma(d+2r)}A_r \, \de(\lambda - d+2-2r) \, .
}
Asymptotically $\rho_2(\lambda)_{\rm conformal} \sim (d-2)\rho C_T(\half \rho
\lambda )^{d-3}/\Gamma(d)$.

For scalar fields the simple form given by \TTconH, and hence \Qconf, is not
the whole story in the conformal limit. The extra terms present in \RSTH\ 
inserted into \defQ\ give, writing $Q(\theta)_\phi = Q(\theta)_{\rm conformal}
+ Q_1(\theta)_\phi$ with $C_T$ given by \CTphi,
\eqn\Qex{\eqalign{
Q_1(\theta)&_{\phi} =  - 4C_{T,\phi} \, d{(d-2)(d+1) \over d-1} \,
{1\over \rho^4 (s\bs)^{{1\over 2}d+1}} \cr
& = - C_{T,\phi}(\half \rho)^{d-2}  \, {(d-2)(d+1)\over d-1} \,
S_d G_{d,2}(\theta) \, .}
}
This corresponds to the contribution with only $\lambda=d$ in the spectral
representation. Added to \spectQ, and using the result \Ar, leads just to the
cancellation of the term involving $A_1 = 2d(d+1)(d-2)/(d-1)$. In particular
the coefficient $A_0$ is unaffected.

In general $C_0$ determines the leading large distance behaviour of the
energy momentum tensor two point function while $C_T$ is related to its singular
form at short distances. It is unclear whether the relation \CCT\ survives
in interacting conformal field theories.

\newsec{Implications for a Possible $c$-Theorem in Four Dimensions}

The initial stimulus for this paper was to investigate the possibility of deriving
a $c$-theorem by considering one and two point functions of the energy
momentum tensor on spaces of constant curvature. Within the framework described
here this does not seem to be feasible. A possible $C$-function need not
necessarily reproduce all the attractive features uncovered by Zamolodchikov
in two dimensions \Zam\ but may satisfy some or all the following properties:

\noindent
1) $\CC(\mu \ell    ;g)$ should be a physically measurable positive
function of the couplings
$g^i$ and some length scale $\ell   $ such that at a fixed point, 
$\beta^i(g_*) =0$, $\CC(\mu \ell    ; g_*) = \CC_*$ 
is independent of $\ell   $ and unambiguously
and universally defined in terms of the properties of the conformal field 
theory which is obtained at the fixed point. The minimal condition of
irreversibility of RG flow is that for a unitary quantum field theory in which there
are both $UV$ and $IR$ fixed points we require
\eqn\irrev{
\CC_*(UV) - \CC_*(IR) > 0 \, .
}
Ideally $\CC(\mu \ell    ;g)$ should be a function of the couplings for all
relevant and irrelevant operators in the space of cut-off quantum field theories
since its essential definition is independent of perturbation theory, $\CC=0$
should correspond to a totally trivial theory with no finite energy degrees
of freedom. It may, although
this does not seem essential, be independent of any strictly marginal couplings
whose $\beta$-functions vanish.\foot{In this case $\CC$ is just a constant equal to
its free field value in ${\cal N}=4$ SYM.} The $\CC$ function should be extensive
so that if $\{g\}$ can be separated into two distinct sets $\{g_1\},\{g_2\}$ 
corresponding to two decoupled theories then
\eqn\ext{
\CC(\mu \ell    ;g) = \CC_1(\mu \ell    ;g_1) + \CC_2(\mu \ell    ;g_2) \, .
}

\noindent
2) The $\CC$-function should obey the usual $RG$ equation expressing its independence
of the arbitrary RG scale $\mu$
\eqn\RGC{
\bigg ( \mu {\pr \over \pr \mu} + \beta^i {\pr \over \pr g^i} \bigg )
\CC(\mu \ell    ;g) = 0 \, .
}
Irreversibility of RG flow, at least in some finite domain of couplings $\{g\}$
near a fixed point, is then entailed by the requirement
\eqn\flow{
\ell    {\pr \over \pr \ell   } \CC(\mu \ell    ;g) \cases{ < 0 \ \ \hbox{on} \ \
\{g\} \setminus g_* \, , \cr
= 0 \ \ \hbox{if} \ \ g = g_* \, .}
}
With these conditions $\CC$ defines a Liapunov function in the region
$\{g\}$ for the RG flow
\eqn\Lia{
{\dot \CC}(\mu \ell    ;g_t) < 0 \quad \hbox{if} \quad g_t \ne g_* \, , \qquad
{\dot g}^i{}_{\! t} = - \beta^i(g_t) \, .
}

\noindent
3) A natural condition which clearly entails \flow\ is to require
\eqn\flowB{
\ell    {\pr \over \pr \ell   } \CC(\mu \ell    ;g) = - \GG_{ij} (\mu \ell   ; g) 
\beta^i(g)\beta^j(g) \, ,
}
where $\GG_{ij}(\mu \ell   ; g)$ is independently defined as a positive symmetric tensor
on the space of couplings and can be regarded as playing the role of a metric.
$\GG_{ij}(\mu \ell   ; g)$ should satisfy a homogeneous RG equation,
\eqn\RGG{
\mu {\pr \over \pr \mu} \GG_{ij}(\mu \ell   ; g) + (\L_\beta \GG)_{ij}(\mu \ell   ; g)
= 0 \, , \qquad (\L_\beta \GG)_{ij} = \beta^i \pr_i \GG_{ij} + \pr_i \beta^k
\GG_{kj} + \pr_j \beta^k \GG_{ik} \, ,
}
where  $\pr_i \beta^j$ is the anomalous dimension matrix for the operators 
$\O_j$. Trivially from
\RGC\ and \flowB\ we may obtain the essential Zamolodchikov equation \flowC,
\eqn\flowCC{
\beta^i(g) {\pr \over \pr g^i} \CC(\mu \ell    ;g) = \GG_{ij} (\mu \ell   ; g)
\beta^i(g)\beta^j(g) \, .
}

\noindent
4) Assuming \flowC\ the definition of $\CC$ is still ambiguous to the extent that
\eqn\amb{\eqalign{
\CC(\mu \ell    ;g) \to {\tilde \CC}(\mu \ell    ;g) = {}& \CC(\mu \ell    ;g) + 
D_{ij} (\mu \ell   ; g) \beta^i(g)\beta^j(g) \, , \cr
\GG_{ij} (\mu \ell   ; g) \to {\tilde \GG}_{ij} (\mu \ell   ; g) = {}&
\GG_{ij} (\mu \ell   ; g) + ({\cal L}_\beta D)_{ij} (\mu \ell   ; g) \, , \cr}
}
leaves \flowC\ invariant. This defines an equivalence amongst $\CC$-functions
and associated metrics which ensures that the precise value of $\ell   $ chosen 
in \flowC\ is irrelevant but of course all such functions give the same
value $\CC_*$ at a fixed point. To ensure the desired properties of irreversible 
RG flow it is only necessary that there is a $D_{ij}$ such that ${\tilde \GG}_{ij}$ 
is positive.\foot{In a perturbative context \Weyl\ we may obtain
$\CC(\mu \ell    ;g) = C(g) + \Omega_{ij}(\mu \ell    ;g)\beta^i(g)\beta^j(g)$,
$\GG_{ij} (\mu \ell    ;g) = G_{ij}(g) + (\L_\beta \Omega)_{ij} (\mu \ell    ;g)$ where
$\beta^i\pr_i C = G_{ij} \beta^i\beta^j$ and $\mu{\pr\over \pr \mu} \Omega_{ij}
+  (\L_\beta \Omega)_{ij} = - G_{ij}$. The perturbative $G_{ij}$ need not
be positive although $\GG_{ij}$ is.}
Introducing the correct terms linear in $\beta^i$, as in \defC, are in general
necessary to find a $\CC$ satisfying \flowC\ although they also do not change $\CC_*$.

\noindent
5) A stronger condition, analogous to \CGW, which implies \flowC\ is
\eqn\grad{
\pr_i  \CC(\mu \ell    ;g) =  \TT_{ij} (\mu \ell   ; g) \beta^j(g) \, , \qquad
\TT_{ij} = \GG_{ij} + \pr_i \WW_j - \pr_j \WW_i \, .
}
If $\TT_{[ij]} =0$ this ensures that the RG flow is a gradient flow. In general 
\grad\ shows that $\CC$ is independent of marginal coupling if $\TT_{ij}$
has no off diagonal pieces. Under variations as in
\amb\ then \grad\ still holds if at the same time
\eqn\Wamb{
{\tilde \WW}_j(\mu \ell    ;g) = \WW_j (\mu \ell   ; g) +
D_{jk} (\mu \ell   ; g)\beta^k(g) \, ,
}
which demonstrates that $\WW_j$ cannot be zero in general and that the assumed
form for $\TT_{[ij]}$ in \grad\ is consistent. 

To illustrate some aspects of the above we first attempt to rederive the
Zamolodchikov $c$-theorem \Zam\ (for a recapitulation see \Cbook). The basic
inequality is similarly obtained using previous results applied to the two point
function of the energy momentum tensor in two dimensions on $S^2$ 
$\l T_{\mu\nu}(x) T_{\alpha\beta}(y) \r = \Gamma_{\mu\nu,\alpha\beta}(x,y)$
as in \twoG. The positivity condition is provided by
$g^{\mu\nu}g^{\alpha\beta} \l T_{\mu\nu}(x) T_{\alpha\beta}(y) \r =
P_1 + 2P_2 \ge 0$. Using \RSTtwo{a}\ and \PPcon\ for $d=2$ we require
\eqn\Ctwo{
\rho^4 C(\theta)= 2 \sin^4 \! \half \theta (R+P_1) + f(\theta)(P_1+P_2) \, ,
}
to satisfy
\eqn\Cdash{
\rho^4 C'(\theta) = \half f'(\theta)(P_1+ 2P_2) \, .
}
This provides a differential equation for $f$ which is readily solved,
\eqn\sof{
f(\theta) = - 4 \sin^4 \! \half \theta + 4 \sin^2 \theta \, \ln \cos \half 
\theta \, .
}
As $\theta \to 0$ $f(\theta) \sim - {3\over 4} \theta^4$ so that $f'(\theta)<0$
for some finite region near $\theta=0$ ($f'(\theta)<0$ for $\theta \lesssim 2.5$)
in which then $C'(\theta) <0$. Nevertheless $C$ is in general a function
of the dimensionless $\mu/\rho$ as well as $\theta$ and the couplings $g^i$
and the $\theta$ derivative is not linked to the dependence on the RG scale $\mu$.
Hence there appears to be no demonstration of irreversibility of RG flow
for general $\rho$ from such an inequality. Only with the
assumption of a sensible flat space limit which requires $C(\theta) \sim
C(\theta\mu/\rho)$ as $\rho \to 0$ may such a result be obtained. In this case
the resulting $C$-function, which satisfies \flowB\ as well as \RGC\ with
$\ell    = \sqrt s = \theta/\rho$, is equivalent to that found by Zamolodchikov.\foot{
To make the comparison clear, using a similar notation to that in \Cbook, we may define
${F= {1\over 16}s^2 R}, \ G = {1\over 4}s^2 P_1 , \ H = s^2 (P_1 + 2P_2)$ and then
${C = 2F -G - {3\over 8}H + g(G + \quar H)}$, from \Ctwo, for 
$g(s) = 1 + (1-\quar\rho^2 s)/ \quar\rho^2 s \, \ln(1- \quar\rho^2 s) $ 
satisfies $sC'(s) = - \quar (3 - h ) H$ where
${h(s) = 1 + (1-\half\rho^2 s)/\quar\rho^2 s \,\ln (1-\quar\rho^2 s)}$. Neglecting $g,h$, 
which is justified as $s\to 0$, this is identical with the standard flat space result.}

More recently \Forte\ an alternative derivation of a $c$-theorem for general $d$ 
in terms of quantum field theories on spaces of constant negative curvature, 
where the $c$-function $C$ is defined by $g^{\mu\nu}\l T_{\mu\nu}\r = - C\rho^d$,
has been suggested which is based on equations akin to \rhoC{a,b}. Defining
\eqn\Chat{
{\hat C} = S_d C \, ,
}
\rhoC{a}\ may be written, using \intTT, as
\eqn\CTT{
\rho {\d \over \d \rho} {\hat C} = \G_0 - d  {\hat C} \, ,
}
where, if the unsubtracted representation \spectTS\ is valid, \intTT\ implies that
$\G_0>0$. If we assume $\A \propto \rho^d$, as in sections three and four for
both two and four dimensions,  then from \DC\ we have
\eqn\RGCh{
\beta^i {\pr\over \pr g^i} {\hat C} = \rho {\d \over \d \rho} {\hat C} \, .
}
In this case the right hand side of \CTT\ must vanish for $\beta^i = 0$. This
becomes more apparent by using \rhoC{b}\ to now write
\eqn\CG{
\beta^i \pr_i {\hat C} =  \G_\Theta 
- {1\over \rho^d} S_d \beta^i \big ( \pr_i \A 
+ \pr_i \beta^j \l \O_j \r \big )  \, ,  \quad 
\G_\Theta = {1\over \rho^d} S_d \int \! \d^d y \sqrt g \, 
\langle \Theta (x) \Theta (y) \rangle \, . 
}
In \Forte\ the additional terms beyond $\G_\Theta$, which is ${\rm O}(\beta^2)$,
are missing (there
is no discernible redefinition of $\hat C$ which achieves this). As it stands,
given the definition of $\hat C$, \CG\ is just an identity.
If these additional terms \CG\ are disregarded then taking $\hat C(\mu/\rho,g)$, 
which satisfies \RGC\  with $\ell   =1/\rho$, as a possible $C$-function hinges 
on supposing $\G_\Theta >0$ in order to obtain the essential inequality \flowB.
For $x\ne y$ in unitary theories $ \langle \Theta (x) \Theta (y) \rangle >0$.
However the regularisation of potential singularities in the integration defining
$\G_\Theta$ at $x=y$ is much less clear.\foot{The prescription in \Forte\ of 
subtracting
$\delta$-function contact terms is not unambiguous and does not provide a
well defined regularisation in general. Introducing additional couplings into
the quantum field theory involving curvature terms does not change the essential
argument.} It is important to recognise that $\G_\Theta$ may be defined as
a finite quantity through \CG\ but the regularisation of the integral need not
preserve any positivity conditions. These also do not apply to contact terms
that may be present in $\l \Theta (x) \Theta (y) \rangle$ (and which may be 
arbitrary although any such ambiguity cancels on both sides of \CG). In 
general $\G_\Theta$ depends on  possible subtraction constants in the spectral 
representation of the two point function whose positivity need not be entailed 
by that of the spectral weight function.  

A perhaps convincing argument as to the essential difficulties
of such an approach, independent of intricacies of the precise definition of 
$\G_\Theta$, is that in four dimensions \Tfour\ shows that we
expect $\hat C>0$ for either positive or negative curvature, and at a fixed point it
is equal to $a$ up to a factor, and then $\G_\Theta>0$ would be sufficient to prove  
the desired $c$-theorem for $\hat C$
implying \irrev, so long as the other terms on the r.h.s of \CG\ are neglected.
However in two dimensions from \Ttwo\ $\hat C$ has no definite sign although
it is proportional to the Virasoro central charge $c$ at a fixed point. 
In particular for the positive curvature case $\hat C<0$ and then $\G_\Theta>0$ 
would be
the wrong sign inequality to generate the required irreversible RG flow. 
It is nevertheless difficult to see why any putative derivation along these lines
should not apply for both positive and negative curvature.

It is worth emphasising that considering a field theory on a space of constant
non zero curvature of course introduces an extra scale, here denoted by $\rho$.
This complicates the discussion of physical consequences from the RG equations
except in some flat space limit as became apparent in the attempt to generalise
the Zamolodchikov derivation of the $c$-theorem. Another way of appreciating
the difference is that on flat space the RG equation may be regarded as implementing
broken scale invariance identities. On flat space the full conformal group
$O(d+1,1)$ is reduced, except at fixed points, to $O(d)\ltimes T_d$. The broken
generators decompose under $O(d)$ into singlet generating scale transformations and
a $d$-dimensional vector corresponding to special conformal transformations. The 
scale invariance Ward identities may still be implemented in a quantum field theory
away from fixed points if they are associated with a flow in the space of couplings 
generated by the usual $\beta$-functions. Although such linear equations for the
correlation functions do not involve any arbitrary scale $\mu$ they are equivalent to 
the standard RG equations. There are corresponding identities relevant for special
conformal transformations \conform\ but these involve the insertion of operators
which are not generated by derivatives with respect to the couplings. In the absence
of closed equations such identities are not then of much practical significance. 
If a quantum field theory is defined on $S^d$ or $H^d$, as considered in this
paper,  then the manifest symmetry group is
$O(d+1)$ or $O(d,1)$. In either case the remaining generators of $O(d+1,1)$ transform
as a $d+1$-dimensional vector. The broken conformal identities are then similar
to those for special conformal transformations on flat space and there are no
associated linear equations relating directly the $x$-dependence and the dependence
on the couplings through the $\beta$-functions.

Although, as illustrated in \Stwo, \twophi, \FFV,  and \TTconS, with \CTphi, \CTV,
\CTpsi, the results for free field theories on $S^d$ are in accord with expectation
from conformal invariance, using \defs, \tinv\ and \conII, this does not apply
to the corresponding results on $H^d$. This arises since in order to derive
the identities expressing conformal symmetry it is necessary to integrate by
parts and surface terms on the boundary cannot be dropped due to lack of
sufficient fall off of the Killing vector fields as the boundary of $H^d$ is
approached. In consequence the results are less constrained in this case.

In the end the analysis of the energy momentum tensor on spaces of constant 
curvature has not apparently led to new insight concerning a proof of a
possible $c$-theorem away from two dimensions.
Nevertheless we have rederived \aG\ which is the four dimensional analogue of
\cG\ in two dimensions which then gave \CGW, a perturbative analogue of the
$c$-theorem. This result can be shown to be directly connected with the full
$C$-function which entails irreversible RG flow \Weyl. The result \aG\ may be expected
to be related to similar equations involving three point functions at non
coincident points. In such cases positivity is no longer manifest but might
be linked to positivity conditions on the energy momentum tensor \posem. It is
interesting to note that a recent demonstration of irreversible RG flow in an
ADS/CFT context depended on a positive energy condition \Holo, albeit for
classical supergravity.

\bigskip
\noindent{\bf Acknowledgements}
\medskip
We both wish to thank Jos\'e Latorre for lengthy discussions during the 
initial stages of this work and Stefano Forte for interesting correspondence. 
We are also grateful to the British Council
for an Acciones Integradas grant. GMS would also like to acknowledge partial 
financial support from PPARC Grant GR/L56374  and EC TMR Grant ERB-FMRX-CT96-0012.

\vfill\eject

\appendix{A}{Results for Negative Curvature}

For the negative curvature case the conservation equations in \conG\ become
\eqn\conGH{\eqalign{
\!\!\! & R'-2S'+ U_2{\!}' + (d-1)(\coth \theta \, R - 2 \tanh  \half \theta \, S )
+ 2 \coth \half \theta \, S - 2\cosech  \theta \, U_2 =  0 \, , \cr
\!\!\! & S'-T'  + d \coth \theta \, S - d \tanh  \half \theta \, T -
\cosech  \theta \, U_2 = 0 \, , \cr
\!\!\! & U_1{\!}' + V' + (d-1) \coth \theta \, U_1 - 2 \cosech  \theta \, S - 2
\tanh  \half \theta \, T =  0 \, . \cr}
}
With the definition \defQ, \PPcon\ and \conQ{a,b} become
\eqn\conH{ \eqalign{
P_1\!{}' +  P_2\!{}'  = {}&  - (d-1) \coth \theta P_1 \, , \cr
Q' + d\coth \theta \, Q + {1\over d} P_1\!{}' ={}& - 2d \cosech \theta \, (S-T) \, , \cr
(S-T)' + d \coth \theta \, (S-T) = {}& - \cosech \theta \, {1\over d-1}
\big ( (Q + (d-2)(d+1)T\big) \cr
{}& +\cosech \theta \,{1\over d} \, P_1 \, . \cr}
}
For $d=2$ the equations become
\eqn\twod{\eqalign{
R' + 2\coth \half \theta \, R + P_1\!{}' ={}& 2\cosech\theta \, P_1 \, , \cr
(R-8S+8T)' + 2\tanh \half \theta \, (R-8S+8T) + P_1\!{}'
= {}& - 2 \cosech \theta \, P_1  \, , \cr}
}
and if $P_1=0$ instead of the solution in \Rtwo\ we have
\eqn\RtwoH{
R = C\, {\rho^4 \over \sinh^4 \! \half \theta} \, , \qquad R-8S+8T = 
C'\, {\rho^4 \over \cosh^4 \! \half \theta} \, .
}
The expression obtained in \TTcc\ corresponds to $C=C'$.

For the spin zero contribution given by an analogous formula to \TTzero\ we have
replacing \ABF\ and \ABzero
\eqn\ABFH{
A = \rho^2\big (F_0{\!}'' - \coth \theta \,  F_0{\!}'\big ) \, , \qquad
B = - \rho^2\big (F_0{\!}'' +  (d-2)\coth \theta \, F_0{\!}'  - (d-1) F_0 \big )\, ,
}
and
\eqnn\ABzeroH
$$\eqalignno{
\!\!\!\!\!\!\! R_0 = {}& \rho^2\big ( A'' - ( \coth \theta + 4 \cosech  \theta ) A' 
+ 4 \cosech  \theta \coth \half \theta \, A \big )\, , \cr
\!\!\!\!\!\!\! S_0 = {}& \rho^2\big ( - \cosech \theta A'+ 
\cosech \theta \coth\half \theta \, A \big )\, , 
\qquad T_0 = \rho^2 \cosech^2 \theta \, A \, , \cr
\!\!\!\!\!\!\! U_{0,1} = {}& \rho^2\big ( B'' - \coth \theta \, B'\big ) \, , \ \
U_{0,2} = - \rho^2\big ( A'' + (d-2) \coth \theta \, A' - (d-1) 
( 2\cosech^2  \theta +1 ) A \big )\, , \cr
\!\!\!\!\!\!\! V_0 ={}& - \rho^2\big ( B''+(d-2)\coth \theta \, B'-(d-1) B 
+  2\cosech^2 \theta \, A\big ) \, . & \ABzeroH \cr}
$$
Instead of \PP
\eqn\PP{
P_1 + P_2 =  \rho^2 (d-1) \big ( - \coth \theta (A' + dB') + A +dB \big ) \, ,
}
and from 
\eqn\solFH{
A + dB = - (d-1) \big ( \nab^2 F_0 - d \rho^2F_0 \big ) \, .
}
Solutions of the homogeneous equation \PP, $A+dB \propto \cosh \theta$, may now be
discarded by requiring appropriate boundary conditions and \solFH\ may be
solved to determine $F_0$.

For the spin two contribution, which satisfies the traceless conditions \trace, we
have instead of \Gres, \QG\ in this case
\eqn\GresH{\eqalign{
8 \rho^2 G = {}& \rho^4 \big (
F_2{\!}''- \coth \theta \, F_2{\!}' + 2(d-1) \tanh \half \theta \,  F_2{\!}'
+ (d-1)(d-2) \tanh^2 \! \half \theta \, F_2 \big )  \cr
={}& {4(d-1)\over (d-3)d(d+1)} \, \sinh^2 \theta \, Q \, . \cr}
}
As in \GF\ this may be simplified to the form
\eqn\GtwoH{
\rho^4 {\d^2\over \d u^2} \big ( u^{d-1} F_2)
= {16(d-1)\over (d-3)d(d+1)}\,  u^{d-1} Q \, , \qquad u = \cosh^2\half \theta \, .
}
For $P_1=P_2=0$ \conH\ demonstrates how $R,S,T$ can then be determined from $Q$.

In two dimensions if we modify the discussion in \Ctwo, \Cdash\ and \sof\ to the
negative curvature case instead of \Ctwo\ we may define
\eqn\CtwoH{
\rho^4 C(\theta)= 2 \sinh^4 \! \half \theta (R+P_1) + f(\theta)(P_1+P_2) \, ,
}
and then \Cdash\ holds if
\eqn\sofH{
f(\theta) = - 4 \sinh^4 \! \half \theta - 4 \sinh^2 \theta \, \ln \cosh \half
\theta \, .
}
In this case $f'(\theta)<0$ for all $\theta$ and $ C(\theta)$ is monotonically
decreasing.

\appendix{B}{Construction of Green Functions}

We here discuss the various Green functions for particular differential operators
which were used in the text. For homogeneous
spaces of constant curvature these all depend just on the single variable
$\theta(x,y)$ and the differential equations for $S^d$ or $H^d$
are similar although the regularity or boundary conditions in each case lead to
different solutions.

First we determine the Green function $G_\lambda(\theta)$ for the operator
${-\nab^2 \mp \lambda(\lambda-d+1)\rho^2}$ for $R\gtrless 0$.
For the negative curvature case, writing $G_\lambda(\theta)_- = e^{- \lambda \theta}
H(e^{-2\theta})$,  then ${(-\nab^2 + \lambda(\lambda-d+1) \rho^2) G_0(\theta)_- = 0}$ 
gives using \nabsq,
\eqn\hyperH{
z(1-z) H''(z) + \big ( (\lambda+1)(1-z) - \mu(1-z) \big ) H'(z)  - \lambda \mu 
H(z) = 0 \, , \qquad \mu = \half (d-1) \, , 
}
which is of standard hypergeometric form. Imposing boundedness as $\theta\to \infty$
requires the solution $\propto F(\lambda, \mu ; \lambda+1-\mu ; z)$. For 
$G_\lambda(\theta)_\pm$ to be a Green function it must have the singular behaviour as
as $\theta\to 0$, $G_\lambda(\theta)_\pm\sim (\theta/\rho)^{-d+2} / S_d (d-2)$, so as
to generate $\de^d(x,y)$ under the action of $-\nab^2$. It is easy to see that
this gives the result \Neq. For $\lambda = d$ we have for the Green function $\Gg_0$
defined by \Gzero\ and \GGG\ when $d=2$ and $d=4$,
\eqn\GH{
\Gg_0(\theta)_- = S_d G_d(\theta)_- = \rho^{d-2} \, {2^d \over d+1} 
\, z^{{1\over 2} d } F(d,\mu ; \mu + 2 ; z) \, .
}
For $d=2,4$ this can be reduced to elementary functions giving respectively
\eqn\GgH{\eqalign{
\Gg_0(\theta)_- ={}& - {1\over 2} \Big ( \big (1 + \half \rho^2 s \big ) 
\ln {s \over \bs} + 2 \Big )\, , \cr
\Gg_0(\theta)_-  = {}& {2 \over \rho^2 s \bs} +  
{\ts {3\over 4}} \rho^2 \Big ( \big (1 + \half \rho^2 s \big ) \ln {s \over \bs}
+ 2 \Big ) \, , \cr}
}
For $R>0$ we write $G_\lambda(\theta)_+ = F(w)$ for $w= \half (1+\cos \theta) $
and then the homogeneous equation becomes
\eqn\hyperS{
w(1-w)F''(w) + d(\half - w)F'(w) + \lambda(\lambda-d+1) F(w) = 0 \, .
}
By requiring a solution which is regular at $w=0$ and
imposing the required behaviour as $\theta\to 0$ the Green function becomes
\eqn\GlS{
G_\lambda(\theta)_+ = {\rho^{d-2} \over (4\pi)^{{1\over2}d}} \, 
{\Gamma(\lambda) \Gamma(- \lambda + d-1) \over \Gamma( \half d ) } \,
F ( \lambda , - \lambda + d-1 ; \half d ; w ) \, .
}
In the text  $\Gg_0$ is defined as the Green function for $-\nab^2 - d \rho^2$,
corresponding to taking $\lambda=d$. However
\eqn\limGlS{
G_{d + \epsilon} (\theta)_+ \sim  {1\over \epsilon} \, 
{\rho^{d-2}\over (4\pi)^{{1\over 2}d}} \, {\Gamma(d) \over \Gamma(\half d)} \, 
( 1 + \half w ) \quad \hbox{as} \quad \epsilon \to 0 \, ,
}
which is a reflection of the existence of normalisable eigenvectors of $-\nab^2$
with eigenvalue $d$. By subtracting this singular piece and then taking the
limit $\epsilon \to 0$ we may define
\eqn\GS{
\Gg_0(\theta)_+ 
= S_d{\rho^{d-2}\over (4\pi)^{{1\over 2}d}} \, {\Gamma(d) \over \Gamma(\half d)}
\bigg ( \sum_{n=2} {1\over n(n-1)} \, {(d)_n \over (\half d)_n} \,
w^n - 2 \Big ( 1 + {1\over d} \Big ) w \bigg ) \, . 
}
This then satisfies
\eqn\Gzero{
(- \nab^2 - d \rho^2 ) \Gg_0(\theta)_+ = S_d \de^d(x,y) -  k_d \, 
\rho^d \cos \theta\, ,
\qquad 
k_d = {d+1\over (4\pi)^{{1\over 2}d}} \, {\Gamma(d) \over \Gamma(\half d)}S_d \, .
}
Again $\Gg_0(\theta)_+$ may be found explicitly for $d=2,4$. Dropping some terms
proportional to $\cos\theta$, which satisfy the homogeneous equation, we have in
each case,
\eqn\GgS{ \eqalign{
\Gg_0(\theta)_+ = {}&- {1\over 2} \Big ( \big (1-\half \rho^2 s \big ) 
\ln \quar \rho^2 s + 1 \Big ) \, , \cr
\Gg_0(\theta)_+ = {}&  {1\over 2 s} - {\ts {3\over 4}}\rho^2 
\Big ( \big (1-\half \rho^2 s \big ) \ln \quar \rho^2 s + 1 \Big ) \, ,  \cr}
}

For the vector field the basic equation is
\eqn\VV{
-\nab^\mu \l F_{\mu\nu}(x) A_\alpha (y) \r - {1\over \xi}\,\pr_\nu \nab^\mu 
\l A_\mu(x) A_\alpha ( y) \r = g_{\mu\alpha} \de^d(x,y) \, .
}
To solve this we adapt the methods of ref. \Free\ to our notation. 
Using the definition \defhI\ we may write
\eqn\AA{
\l A_\mu(x) A_\alpha(y) \r = F(\theta) \hI_{\mu\alpha} + \pr_\mu H(\theta)
{\overleftarrow \pr_{\!\alpha}} \, ,
}
and, since as a consequence of \DhI\ $\pr_{[\mu}\hI_{\nu]\alpha} =  0$, we then have
\eqn\FA{
\rho \l F_{\mu\nu}(x) A_\alpha (y) \r = F'(\theta) ( \hx_\mu I_{\nu\alpha}
- \hx_\nu I_{\mu\alpha} ) \, .
}
In order to solve \VV\ we first require
\eqn\nabF{
- \nab^\mu \l F_{\mu\nu}(x) A_\alpha (y) \r = g_{\nu\alpha} \de^d(x,y)
+ \pr_\nu \big ( S(\theta) \hy_\alpha \big ) \, .
}
This decomposes into two equations in either case
\eqn\FS{\eqalign{
S_+={}& \sin \theta \nab^2 F_+ - \rho^2\cos \theta F'\!{}_+ \, , \qquad \,
S'\!{}_+ = \rho^2(d-1) \cosec \theta F'\!{}_+ \, , \cr
S_-={}&  \sinh \theta \nab^2 F_- - \rho^2 \cosh \theta F'\!{}_- \, , \quad 
S'\!{}_- = \rho^2 (d-1) \cosech \theta F'\!{}_- \, , \cr}
}
which may be satisfied, along with \nabF, by imposing
\eqn\eqnF{
- \nab^2 F(\theta)_\pm \pm (d-2)\rho^2  F(\theta)_\pm = \de^d(x,y) \, .
}
This is identical to the equation for the scalar Green function for $\lambda =d-2$
so that from the above solutions we have
(for $R>0$ the result is essentially given in \ITD)
\eqn\Fpm{
F(\theta)_+ = \rho^{d-2} {\Gamma(d-2)\over (4\pi)^{{1\over 2}d} \Gamma(\half d)}
F(d-2,1;\half d; w ) \, , \quad
F(\theta)_- = {\rho^{d-2}\over (d-2)S_d} \, {1\over (\sinh \theta )^{d-2}} \, .
}
The gauge dependent part in \AA\ can be found by solving, in the positive curvature
case,
\eqn\gauge{
{1\over \rho} \nab^2 H_+\! {\overleftarrow \pr_{\!\alpha}} = (1-\xi ) S_+\hy_\alpha
+ 2\rho^2 \sin\theta F_+ \hy_\alpha \, ,
}
with a similar equation if $R<0$, although knowing $H$ is unnecessary to obtain \FFV\
when in addition we use $F(2,1;2;w)= (1-w)^{-1}$.

\appendix{C}{Calculation of Norms and Spherical Harmonics}

The calculation of the norms of the basis states defined in \basis\ or \basisV\
for the scalar or vector representations follows from computing the action of
$\hL_{-i}$ on these states. In the standard fashion we use the basic commutator 
$[\hL_{-i},\hL_{+j}]= 2 \de_{ij}\hH - 2\hL_{ij}$, as well as those involving
$\hH, \ \hL_{ij}$ with $\hL_{+j}$, until they act on the lowest weight state and
we may use \wei\ or \weiV. For the scalar case, with the aid of, 
\eqn\comK{
[\hL_{-i}, \hK_+] = 4 \hL_{+i} (\hH - \mu +1) - 2 \hL_{+j} \hL_{ij} \, ,
\qquad \mu = \half(d-1) \, ,
}
we find
\eqn\Lsc{ \eqalign{
\hL_{-i} | n \, \ell , \C \r = {}&4n(\lambda-\mu+n)\, \hL_{+i} |n{-1}\, \ell, \C \r
+ 2\ell(\lambda +2n+\ell-1) | n\, \ell{-1}, \C^-{}_{\!\! i } \r \cr
= {}& 4n(\lambda-\mu+n) \, |n{-1}\, \ell{+1},  \C^+{}_{\!\!i} \r \cr
{}& + {2\ell \over \mu+\ell-1}(\lambda+n+\ell-1)(\mu+n+\ell-1)
\, | n\, \ell{-1} ,  \C^-{}_{\!\!i} \r \, , \cr}
}
after using \lplus. With the scalar products in \orthog\ and using the
hermeticity conditions \herm\ with \Lsc\ and \lplus\ again we have
\eqn\Nnl{
\l n{-1}\, \ell{+1}, \C'| \hL_{-i} | n \, \ell , \C \r = 4n(\lambda-\mu+n)\,
\N_{n{-1}\, \ell{+1}}\, \C'\!\cdot \C^+{}_{\!\!i} =
{\ell+1 \over 2(\mu+\ell)} \, \N_{n\ell}\,\C'^-{}_{\!\!\! i }\!\cdot \C \, ,
}
where
\eqn\CCC{
\C'\!\cdot \C^+{}_{\!\!i} = \C'^-{}_{\!\!\! i }\!\cdot \C = \C'{}_{\! i i_1\dots
i_\ell} \C_{i_1\dots i_\ell} \, .
}
Hence \Nnl\ gives \ relation between $\N_{n{-1}\, \ell{+1}}$ and $\N_{n\ell}$.
Similarly
\eqn\CCCL{
\l n\, \ell{-1}, \C'| \hL_{-i} | n \, \ell , \C \r = {2\ell \over \mu+\ell-1}
(\lambda+n+\ell-1)(\mu+n+\ell-1)\,  \N_{n\, \ell{-1}} \, 
\C'\!\cdot \C^-{}_{\!\!i} = \N_{n\ell} \,
\C'^+{}_{\!\!\! i }\!\cdot \C \, ,
}
with $\C'\!\cdot \C^-{}_{\!\!i}=\C'^+{}_{\!\!\! i }\!\cdot \C$ in this case,
relates $\N_{n\, \ell{-1}}$ and $\N_{n\ell}$. Solving the recurrence relations
given by \Nnl\ and \CCCL, with $\N_{00}=1$, then gives \norm.

For the vector case the results may be obtained in an analogous fashion albeit
the expressions are more lengthy. For the states defined in \basisV\ we have
\eqn\LChV{ \eqalign{
\hL_{-i} | n \, \ell , \hhC \r = {}&4n(\lambda-\mu+n)\, 
\hL_{+i} |n{-1}\, \ell, \hhC \r
+ 4n \, | n{-1} \, \ell \, + , \hhC_i \r + 2\ell \, | n \, \ell \, + , \hhC_i \r \cr
{}& + 2\ell(\lambda +2n+\ell-1)\, 
\hK_+{}^{\!\! n} \hL_{+i_1}\dots \hL_{+ i_{\ell-1}} |\lambda,k\r
\, \hhC_{ki_1\dots i_{\ell-1} i } \cr
= {}& 4n(\lambda-\mu+n) \, |n{-1}\, \ell{+1},  \hhC^+{}_{\!\!i} \r \cr
{}& + {2(\ell-1) \over \mu+\ell-1}(\lambda+n+\ell-1)(\mu+n+\ell-1)
\, | n\, \ell{-1} ,  \hhC^-{}_{\!\!i} \r \, , \cr
{}& +{2\over d+\ell-3}\Big ( 2n(\lambda +\mu +n+\ell-2) |n{-1}\, \ell \, + ,\hhC_i \r
\cr {}& \qquad\quad {} - \big ( 2(\lambda+n-1)(\mu+n-1) + \ell(\lambda +2n-1) \big) 
|n\, \ell \, - ,\hhC_i \r \Big ) \, , \cr}
}
and
\eqn\LCpV{ \eqalign{
\hL_{-i} | n \, \ell \, + , \C \r = {}&4n(\lambda-\mu+n+1)\, \hL_{+i} 
|n{-1}\, \ell \, + , \C \r + 2\ell(\lambda + 2n +\ell) 
| n\, \ell{-1}\, + , \C^-{}_{\!\! i } \r \cr
{}& + 2(\lambda-d+2)\, \hK_+{}^{\!\! n} \hL_{+i_1}\dots \hL_{+ i_{\ell}} |\lambda,i\r
\, \C_{i_1\dots i_{\ell} } \cr
= {}& 4n(\lambda-\mu+n+1) \, |n{-1}\, \ell{+1}\, + ,  \C^+{}_{\!\!i} \r \cr
{}& + {2\ell \over \mu+\ell-1}(\lambda+n+\ell)(\mu+n+\ell-1)
\, | n\, \ell{-1} \, + ,  \C^-{}_{\!\!i} \r \, , \cr
{}& + 2(\lambda -d+2) \big ( |n\, \ell{+1}\, - ,  \C^+{}_{\!\!i} \r 
- \ell \, | n \, \ell , \C_i \r \big ) \cr
{}& +{\ell(\lambda-d+2)\over ( \mu+\ell-1)(d+\ell-4)}\Big ( 
2(\mu+\ell-2) | n\, \ell{-1} \, + ,  \C^-{}_{\!\!i} \r \cr
{}& \qquad \qquad \qquad \qquad \qquad \quad {}- (\ell-1)
|n{+1}\, \ell{-1} \, - ,\C^-{}_{\!\!i} \r \Big ) \, , \cr}
}
and also
\eqnn\LCmV
$$  \eqalignno{ \!\!\!\!
\hL_{-i} | n \, \ell \, - , \C \r = {}&4n(\lambda-\mu+n)\, \hL_{+i}
|n{-1}\, \ell \, - , \C \r + 2(\ell-1)(\lambda + 2n +\ell-1)
| n\, \ell{-1}\, - , \C^-{}_{\!\! i } \r \cr
{}& -4n\,\hK_+{}^{\!\! n-1} \hL_{+i_1}\dots \hL_{+ i_{\ell}} |\lambda,i\r
\, \C_{i_1\dots i_{\ell} } +4n \, | n{-1}\, \ell{-1}\, + , \C^-{}_{\!\! i } \r \cr
= {}& 4n(\lambda-\mu+n) \, |n{-1}\, \ell{+1}\, - ,  \C^+{}_{\!\!i} \r 
+ 4n (\lambda-\mu+n+\ell)\, |n{-1}\, \ell, \C_i \r \cr
{}& + {2(\ell-1) \over \mu+\ell-1}(\lambda+n+\ell-1)(\mu+n+\ell-1)
\, | n\, \ell{-1} \, - ,  \C^-{}_{\!\!i} \r \, ,  & \LCmV\cr
{}& +{2n\over ( \mu+\ell-1)(d+\ell-4)}\Big (
(\ell-1) (\lambda - \mu + n+\ell) | n\, \ell{-1} \, - ,  \C^-{}_{\!\!i} \r \cr
{}& \qquad \qquad \qquad \ \ {}+ (d-3) (\lambda+\mu+n+2\ell-3)
|n{-1}\, \ell{-1} \, + ,\C^-{}_{\!\!i} \r \Big ) \, . \cr}
$$
With these formulae we may find the norms of the basis vectors defined in
\orthov\ and \orthoh\ by following the same procedure as in the spinless case.
First for $\hhN_{n\ell}$ we obtain
\eqn\hNnl{
\l n{-1}\, \ell{+1}, \hhC'| \hL_{-i} | n \, \ell , \hhC \r = 4n(\lambda-\mu+n)\,
\hhN_{n{-1}\, \ell{+1}}\, \hhC'\!\cdot \hhC^+{}_{\!\!i} =
{\ell\over 2(\mu+\ell)} \, \hhN_{n\ell}\,\hhC'^-{}_{\!\!\! i }\!\cdot \hhC \, ,
}
and
\eqn\hNnlm{
\l n\, \ell{-1}, \hhC'| \hL_{-i} | n \, \ell , \hhC \r = {2(\ell-1) \over \mu+\ell-1}
(\lambda+n+\ell-1)(\mu+n+\ell-1)\,  \hhN_{n\, \ell{-1}} \,
\hhC'\!\cdot \hhC^-{}_{\!\!i} = \hhN_{n\ell} \,
\hhC'^+{}_{\!\!\! i }\!\cdot \hhC \, ,
}
by using \LChV\ and \raiseV. In \hNnl\ we have the relation
\eqn\CCV{
\hhC'\!\cdot \hhC^+{}_{\!\!i} = {\ell\over \ell+1} \,
\hhC'^-{}_{\!\!\! i }\!\cdot \C = \hhC'{}_{\! j i_1\dots i_\ell i} 
\C_{i_1\dots i_\ell} \, ,
}
and similarly in \hNnlm\ with $\ell \to \ell-1$. Thus $\hhN_{n\ell}$ obeys identical
recurrence relations to $\N_{n\ell}$ and \normV\ is obtained starting from
$\hhN_{01} = 2(\lambda-1)$.

The $2\times 2$ matrix ${\underline \N}_{n\ell}$ defined by
\orthov\ may now be obtained directly with the aid algebraic relations expressing
each element in terms of  $\hhN_{n\ell}$. Using \LCmV\ and \LCpV\ we may find, with an
obvious notation for the components of  ${\underline \N}_{n\ell}$,
\eqn\Nnlpm{ \eqalign{
\l n\, \ell, \hhC| \hL_{-i} | n{+1} \, \ell \, - , \C \r = {}& 4(n+1)
(\lambda-\mu+n+\ell + 1 )\,
\hhN_{n\ell}\, \hhC\!\cdot \C{}_{i} \cr
= {}& {1\over d +\ell-3} \big ( \N_{n\ell \, +-}  - \N_{n\ell \, --} \big ) \,
\hhC_{i}\!\cdot \C \, , \cr
\l n\, \ell, \hhC| \hL_{-i} | n \, \ell \, + , \C \r = {}& -2\ell
(\lambda-d+2)\, \hhN_{n\, \ell}\, \hhC\!\cdot \C{}_{i} \cr
= {}& {1\over d +\ell-3} \big ( \N_{n\ell \, ++}  - \N_{n\ell \, -+} \big ) \,
\hhC_{i}\!\cdot \C \, , \cr}
}
where
\eqn\ChC{
\hhC_{i}\!\cdot \C = - \ell\, \hhC\!\cdot \C{}_{i} = \hhC_{ii_1 \dots i_\ell}
\C_{i_1 \dots i_\ell} \, .
}
Furthermore we have using \LChV
\eqn\Nnlpp{ \eqalign{
\l n \, \ell \, + , \C & | \hL_{-i} | n{+1} \, \ell , \hhC \r = 0 \cr
={}& {2\over d+\ell-3}\Big (2(n+1)(\lambda +\mu +n+\ell-1) \, \N_{n\ell \, ++} \, \cr
{}& \qquad\quad \qquad {} - \big ( 2(\lambda+n)(\mu+n) + \ell(\lambda +2n+1) \big)
\, \N_{n\ell \, +-} \Big ) \, \C \! \cdot \hhC_i  \, . \cr}
}
Although a further relation may be obtained by considering
$\l n{+1} \, \ell \, - , \C | \hL_{-i} | n{+1} \, \ell , \hhC \r $ \Nnlpm\ with
\ChC\ and \Nnlpp, since $\N_{n\ell \, -+} = \N_{n\ell \, +-}$, are sufficient to 
determine ${\underline \N}_{n\ell}$. Writing
\eqn\NNN{
{\underline \N}_{n\ell} = 2^{4n+\ell+1}n! (\ell-1)!(\lambda+1)_{n+\ell-1}(\mu+\ell)_n
(\lambda+1-\mu)_n \,  \pmatrix{a & b \cr b & c} \, ,
}
we then find
\eqnn\abc
$$\eqalignno{
a = {}&4(n+1) \big ( (\lambda-\mu)(\lambda+\ell-1)(\mu +\ell-1) + (n+1)((\lambda-1)
(\mu-1) + \ell (\lambda-\mu) ) \big ) \, , \cr
b = {}& 2(n+1) \ell ( \lambda + \mu + n + \ell -1) ( \lambda - d+2) \, , & \abc \cr
c= {}& \ell \big (2(\lambda+n)(\mu+n) + \ell (\lambda+2n+1) \big ) ( \lambda - d+2)
\, . \cr}
$$
The result for $\N_{n+}$ in \normV\ may be obtained by the result for
$\N_{n\ell \, ++}$ given by \NNN\ and \abc\ by setting $\ell=0$. Since we
require an expression for ${\underline \N}_{n\ell}{}^{\!\! -1}$ it is useful to
note that
\eqn\invN{
\det \pmatrix{a & b \cr b & c}  = 4 (n+1)\ell(\lambda-1)(\lambda + n + \ell)
(\mu+n+\ell)(\lambda-\mu + n+1) (d+\ell-3)(\lambda - d+2) \, ,
} 
which for a unitary representation must of course be positive.

We also summarise the essential results for the spherical harmonics defined 
as in \har\ by $Y_\ell^I (\hxn) = 
\C^I{}_{\!\!i_1 \dots i_\ell} \hxn_{i_1} \dots \hxn_{i_\ell} $ with $\C^I$ a
basis of symmetric traceless tensors of rank $\ell$ so that $\sum_I 1 =
(d-2)_{\ell-1}(d-3+2\ell)/\ell! \, $. Using
\eqn\intS{
\int_{S^{d-2}} \!\!\!\!\!\!\! \d v \, \hxn_{i_1} \dots \hxn_{i_{2\ell}} = 
{(2\ell)! \over 2^{2\ell} \ell! (\mu)_\ell} \, S_{d-1} \,
\de_{(i_1 i_2} \dots \de_{i_{2\ell-1} i_{2\ell})} \, ,
}
and \tenC\ we have
\eqn\sporth{
\int_{S^{d-2}} \!\!\!\!\!\!\! \d v \, Y_\ell^{I'} (\hxn) Y_\ell^I (\hxn)
= {\ell! \over 2^\ell (\mu)_\ell} \, S_{d-1} \, \de_{I'I} \, .
}
Furthermore we have
\eqn\comp{
\sum_I Y_\ell^I (\hxn_1) Y_\ell^I (\hxn_2) = {\ell! \over 2^\ell (\mu-1)_\ell} \,
C_\ell^{\mu-1} ( \hxn_1 \! \cdot \hxn_2) \, , \qquad
C_\ell^{\mu-1} ( 1) = {(d-3)_\ell\over \ell!} \, ,
}
with $C_\ell^{\mu-1}$ a Gegenbauer polynomial.

\appendix{D}{Results for Free Massive Scalar Fields}

Forte and Latorre \Forte\ have discussed the case of free massive scalar fields on
a space of constant negative curvature. We here re-examine this 
case, which despite being a free field theory is non trivial,
in the light of the main discussion in this paper.

For $\phi$ a free scalar field satisfying $(\Delta + m^2) \phi = 0$ then if
\eqn\mlam{
m^2 = (\lambda_\phi-\half d)(\lambda_\phi-\half d+1)\rho^2 \, ,
}
the basic $\phi$ two point function is $G_{\lambda_\phi}(\theta)$. For such free
massive fields the energy momentum tensor may still be taken to be given by
\Tphi\ but this is no longer traceless giving
\eqn\trphi{
\Theta = - m^2 \phi^2 \, .
}
In terms of the general formalism set up earlier $m^2$ may be regarded as
a coupling with the associated operator $\O_{m^2} = \half \phi^2$ and 
$\beta^{m^2} = - 2m^2$.

In order to determine the spectral representation for $\phi^2$  in this case
we make use of
\eqn\Fsq{\eqalign{
F(\lambda_1,\mu;& \lambda_1+1-\mu;e^{-2\theta}) 
F(\lambda_2,\mu;\lambda_2+1-\mu;e^{-2\theta})\cr
& = \!\sum_{n=0} B_{\lambda_1 \lambda_2 ,n}e^{-2n\theta} 
F(\lambda_1 + \lambda_2 + 2n, \mu;\lambda_1 + \lambda_2 +1-\mu+2n;e^{-2\theta}) \, ,
\cr}
}
where, in a similar fashion to  \Ar, 
\eqn\Bn{
B_{\lambda_1 \lambda_2 ,n} = {(\lambda_1)_n (\lambda_2)_n \over 
(\lambda_1+1-\mu)_n (\lambda_2+1-\mu)_n } \,{(\mu)_n
(\lambda_1 + \lambda_2 + 1 -2\mu +n)_n \over n! (\lambda_1+\lambda_2 + n -\mu)_n} \, .
}
For $d=3$, $B_{\lambda_1 \lambda_2 ,n} = 1$, when \Fsq\ is easily checked. 
For the free field $\phi$ we therefore have
\eqn\tphi{
\l \phi^2(x) \phi^2(y) \r = 2G_{\lambda_\phi}(\theta)^2 = \int_{2\lambda_\phi}^\infty
\!\!\!\!\! \d \lambda \, \rho_{\phi^2}(\lambda) \, G_\lambda(\theta) \, ,
}
where from \Fsq
\eqn\rphi{
\rho_{\phi^2}(\lambda) = {\rho^{d-2}\over \pi^\mu}\, {\Gamma(\lambda_\phi)^2 \over
\Gamma(\lambda_\phi+1-\mu)^2} \sum_{n=0} {\Gamma(2\lambda_\phi+2n+1-\mu) \over
\Gamma(2\lambda_\phi+2n)} B_{\lambda_\phi\lambda_\phi,n} \de(\lambda - 2 
\lambda_\phi - 2n) \, .
}
Asymptotically $\rho_{\phi^2}(\lambda) \sim 2(d-1)\rho (\half \rho
\lambda )^{d-3}/\Gamma(d)S_d$.
Using the relation \Thspec\  we have
\eqn\rlim{
\rho_0(\lambda)  = {m^4 S_d\over \rho^4 (d-1)^2(\lambda+1)^2(\lambda-d)^2}
\rho_{\phi^2}(\lambda) \toinf{m^2\to 0} \rho^{d-2} 
{1 \over 2^{d-1}(d-1)^2(d+1)} \, \de(\lambda -d) \, ,
}
where the limit $m^2\to 0$ may be taken either by $\lambda_\phi \to \half d$ or 
$\lambda_\phi \to \half d - 1 $ in \mlam. Comparing with \Fth\ the result for
$C_0$ is in agreement with \CCT\ where $C_T$ is given by \CTphi.

For general $d$, $G_\lambda(\theta)$ may be separated, by using standard
hypergeometric identities, into two pieces one of which contains terms of the
form $\theta^{-d+2 + 2n}$, $n=0,1,2,\dots$, and the other which is analytic in 
$\theta$. Discarding the former we may easily obtain using \Tphi
\eqn\onephi{
\l \phi^2 \r = {\rho^{d-2}\over (4\pi)^{{1\over 2} d}}\,  \Gamma(1-\half d)\,
{\Gamma(\lambda_\phi)\over \Gamma( \lambda_\phi +2 -d ) } \, , \qquad
\l T_{\mu\nu} \r = - {1\over d} m^2 g_{\mu\nu} \l \phi^2 \r \, .
}
The result for $\l T_{\mu\nu} \r$ is of course as expected from \trphi. 
With the expression in \onephi\ for $\l \phi^2 \r$ we have, noting that 
$\psi(x) = \Gamma'(x)/\Gamma(x)$,
\eqn\dphi{\eqalign{
- 2 {\pr \over \pr m^2} \l \phi^2 \r = {}& - \l \phi^2 \r \, {\psi(\lambda_\phi) -
\psi ( \lambda_\phi +2 -d ) \over (\lambda_\phi - \mu) \rho^2 } \cr
= {}& \int \! \d^d y \sqrt g \, \l \phi^2(x) \phi^2(y) \r =  
\int_{2\lambda_\phi}^\infty
\!\!\!\!\! \d \lambda \, \rho_{\phi^2}(\lambda) {1\over \lambda ( \lambda
- d +1 ) \rho^2 } \, . \cr}
}
The representation  in terms of $\rho_{\phi^2}$ may be verified directly for $d=3$
and is convergent for $d<4$ when it demonstrates $\pr \l \phi^2 \r / \pr m^2 <0$.
For $d=2$ \dphi\ gives
\eqn\ptwo{
\rho^2 \int \! \d^2 y \sqrt g \, \l \phi^2(x) \phi^2(y) \r = {1\over 2\pi}
\, {\psi'(\lambda_\phi) \over \lambda_\phi - \half} =
\int_{2\lambda_\phi}^\infty
\!\!\!\!\! \d \lambda \, \rho_{\phi^2}(\lambda) {1\over \lambda ( \lambda
- 1 )} \, .
}

As expected the one point functions in \onephi\ are  divergent when $d=2,4$. 
Subtracting the poles at $d=2,4$ as usual we have
\eqn\onephir{\eqalign{
2\pi \l \phi^2 \r_{\rm reg} \big |_{d=2} = {}&
- \ln {\rho \over \mu} - \psi(\lambda_\phi) - r \, , \cr
8\pi^2 \l \phi^2 \r_{\rm reg} \big |_{d=4} = {}& m^2 \Big (
\ln {\rho \over \mu} + \psi(\lambda_\phi - 1) + r \Big ) + \rho^2 s \, , \cr}
}
where $\mu$ is a regularisation scale and $r$, which may be absorbed into the
definition of $\mu$, and $s$ are arbitrary parameters reflecting the precise choice of
renormalisation scheme (to obtain the result for $d=4$ it is essential to
subtract a pole term $\propto m^2/\varepsilon$, $\varepsilon=4-d$, where $m^2$ is 
given by \mlam\ including $\rO(\varepsilon)$ terms).  The $\ln \rho/\mu$ terms 
reflect the mixing of the operator $\phi^2$ with $1, m^2 1$ for $d=2,4$ respectively.
By using  \consis\ we may obtain $\pr C/\pr m^2$ in terms of $\l \phi^2 \r_{\rm reg}$
and integrating this gives
\eqn\Ctf{\eqalign{
g^{\mu\nu} \l T_{\mu\nu} \r = {}& - C\rho^d = - m^2 \l \phi^2 \r_{\rm reg} + \A \, ,
\cr
2\pi\A  \big |_{d=2} = {}& - \half m^2 - {\ts{1\over 6}}c\rho^2 \, , \qquad
\quad 2\pi^2\A  \big |_{d=4} =  \half m^4  - 3a\rho^4 \, . \cr}
}
The free parameters $r$, for $d=2$, and $r,s$, for $d=4$, correspond to the
potential freedom of adding to the action terms of the form $\int \d^2 x \sqrt g m^2$
and  $\int \d^4 x \sqrt g m^4$,  $\int \d^4 x \sqrt g m^2 R $ respectively. The
undetermined integration constants $c,a$ in \Ctf, which are independent of $m^2$,
cannot be so modified and are therefore renormalisation scheme independent. The
$\ln \rho/\mu$ terms demonstrate the mixing of $T_{\mu\nu}$ with $g_{\mu\nu}m^2, \,
g_{\mu\nu}m^4$ for $d=2,\, 4$ (even for free theories and defining the energy
momentum tensor through normal ordering for instance there are such terms if
the mass used to define normal ordering is varied from the physical $m$).

Forte and Latorre effectively choose $r,s$, as well as $\mu$, by imposing the
natural decoupling condition that $\l T_{\mu\nu}\r$, and also $\l \phi^2 \r$,
 should vanish as $m^2\to \infty$.\foot{However for $d=3$ where there are
no ambiguities $4\pi \l \phi^2 \r = - (m^2 + {1\over 4} \rho^2)^{1\over 2}$ for
$R<0$ and $ 4\pi \l \phi^2 \r = - ({1\over 4} \rho^2-m^2)^{1\over 2} \cot \pi
\sqrt {{1\over 4} - m^2/\rho^2}$ if $R>0$.}
This further determines the integration constants $c,a$. With $\hat C$ defined by
\Chat\ and using the expansion of $\psi(x)$ for large $x$, the decoupling condition 
leads to
\eqn\FLC{
{\hat C}_{FL} \big |_{d=2} = {m^2 \over \rho^2} \, 2\pi\l \phi^2 \r_{FL}
+ {1\over 6} = -  {m^2 \over \rho^2} \Big ( \psi(\lambda_\phi) -
\ln {m\over \rho} - {1\over 6} \, {\rho^2 \over m^2} \Big ) \quad
\Rightarrow \ \ c=1 \, ,
}
for $d=2$ and for $d=4$
\eqn\FLCf{ \eqalign{
{\hat C}_{FL} \big |_{d=4} ={}&{m^2 \over \rho^4}\, 2\pi^2 \l \phi^2 \r_{FL}
+ {1\over 120} \cr
={}&  {m^4 \over 4 \rho^4} \Big ( \psi 
(\lambda_\phi - 1 )  - \ln {m\over \rho} - {1\over 6} \, {\rho^2 \over m^2} 
+ {1\over 30} \, {\rho^4 \over m^4}  \Big )  \quad \Rightarrow \ \
a = {\ts{1\over 360}}\, . \cr}
}
These results for $c,a$ determine $C$ in the conformal limit $m^2 \to 0$ and
are just as expected for free scalar theories in $2,4$ dimensions. This prescription
also ensures that $\A$ is independent of $m^2$, in contrast to \Ctf. Both results
for $\hat C_{FL}$ decrease monotonically to zero as $m^2$ increases from $0$ to
$\infty$.\foot{For $R>0$ a similar approach gives for $d=2$, 
$4\pi \l \phi^2\r = \ln m^2/\rho^2 - \psi(\alpha_+) - \psi(\alpha_-)$ and for $d=4$,
$16\pi^2 \l \phi^2\r = m^2 ( -  \ln m^2/\rho^2 + {1\over 3} \rho^2/m^2 +  \psi(\alpha_+)
+  \psi(\alpha_-) )$ where in both cases $\alpha_\pm = {1\over 2} \pm 
\sqrt{{1\over 4} - m^2/\rho^2}$. By integrating \consis\ with the boundary condition
that $\hat C$ vanishes as $m^2 \to \infty$ we get for $d=2$, ${\hat C} = m^2 
2\pi \l \phi^2\r/\rho^2 - {1\over 6}$ and for $d=4$, ${\hat C} = m^2 2\pi^2
 \l \phi^2\r/\rho^4 +{1\over 120}$. The integration constants are just those expected
from the trace anomaly. When $d=2$ $\hat C$ decreases monotonically with increasing 
$m^2$, despite the sign of the anomaly in this case, as a consequence of the
result $4\pi \l \phi^2 \r \sim \rho^2/m^2$ as $m^2\to 0$ with the singularity
arising from the existence of
constant normalisable zero modes for $-\nab^2$ on a sphere. When $d=4$ $\hat C$
changes sign at $m^2\approx 0.2\rho^2$ before tending to zero.}
For $d=2$ the result for the $m^2$ derivative can be expressed, in
agreement with \CTT\ and \RGCh,  as
\eqn\Cder{
- 2m^2 {\pr \over \pr m^2} {\hat C}_{FL} = -2  {\hat C}_{FL} + \G_0 \, ,
}
where explicitly
\eqn\Gphi{
\G_0 = {m^4 \over \rho^4}\,  {\psi'(\lambda_\phi) \over \lambda_\phi - \half} -
{m^2 \over \rho^2} + {1\over 3} =
4 \int_{2\lambda_\phi}^\infty
\!\!\!\!\! \d \lambda \, \rho_0(\lambda) {1\over \lambda ( \lambda
- 1 )}  \, .
}
The $\lambda$ integral, with $\rho_0$ given by \rlim, \rphi\ and \Bn\ for $d=2$,
may be verified numerically and in special cases analytically.
$\G_0$ is equal to $2\pi\rho^{-2} m^4\int \! \d^2 y \sqrt g \l \phi^2(x)\phi^2(y) \r$
with the ${\rm O}(m^2)$ and ${\rm O}(1)$ terms at large $m^2$ subtracted. For
$d=4$ the corresponding result is again in accord with the general expression
given by \CTT\ and \RGCh,
\eqn\Cderf{
- 2m^2 {\pr \over \pr m^2} {\hat C}_{FL} = -4  {\hat C}_{FL} + \G_0 \, ,
}
if we now take
\eqn\Gphif{
\G_0 = {1\over 4}\bigg(-{m^6 \over \rho^6}\,  {\psi'(\lambda_\phi-1) 
\over \lambda_\phi - {\ts{3\over 2}}} +
{m^4 \over \rho^4} - {1\over 3}{m^2 \over \rho^2} + {2\over 15} \bigg ) =
144 {1\over \rho^2} \int_{2\lambda_\phi}^\infty
\!\!\!\!\! \d \lambda \, \rho_0(\lambda) {1\over \lambda ( \lambda
- 3 )} - {1\over 15} \, .
}
The additional constant present in \Gphif\ in the expression for $\G_0$ beyond that
given by \intTT\ 
in terms of the integral over  $\rho_0(\lambda)$ is necessary since
the right hand sides of \Cder,\Cderf\ should vanish as $m^2 \to 0$ while the 
integral in this limit is restricted to $\lambda=d$, according to \rlim, 
and is determined by $C_0 \propto C_T \propto c$
while in four dimensions ${\hat C}_{FL} \propto a$. From \intTT\ and \rlim\ we would
have $\G_0|_{m^2=0} = d/(2^{d-1}(d+1))$.
\listrefs
\bye